\documentstyle[11pt,epsfig]{article}

\renewcommand\({\left(}
\renewcommand\){\right)}
\renewcommand\[{\left[}
\renewcommand\]{\right]}

\newcommand{\ra}{\rightarrow}
\def\lsim{\raise 0.4ex\hbox{$<$}\kern -0.8em\lower 0.62
ex\hbox{$\sim$}}
\def\gsim{\raise 0.4ex\hbox{$>$}\kern -0.7em\lower 0.62
ex\hbox{$\sim$}}
\newcommand{\mpl}{M_{\rm Pl}}
\newcommand{\mgut}{M_{\rm GUT}}

\newcommand{\ogw}{\Omega_{\rm gw}}
\newcommand{\hogw}{h_0^2\Omega_{\rm gw}}
\newcommand{\hc}{h_c(f)}
\newcommand{\hn}{h_n(f)}

\newcommand\eq[1]{eq.~(\ref{#1})}
\newcommand\eqs[2]{eqs.~(\ref{#1}) and (\ref{#2})}
\newcommand\eqss[3]{eqs.~(\ref{#1}), (\ref{#2}) and (\ref{#3})}

\newcommand\ee{\end{equation}}
\newcommand\be{\begin{equation}}
\newcommand\ees{\end{eqnarray}}
\newcommand\bees{\begin{eqnarray}}
\newcommand\sub[1]{_{\rm #1}}
\newcommand\su[1]{^{\rm #1}}
\def\dslash{\not{\hbox{\kern-2pt $\partial$}}}
\def\Dslash{\not{\hbox{\kern-4pt $D$}}}
\def\pslash{\not{\hbox{\kern-2.3pt $p$}}}
\begin{document}

\begin{titlepage}
\begin{flushright}
IFUP-TH 20/99\\
May 1999\\
\end{flushright}
\vspace{5mm}
\begin{center}
{\Large \bf 
Gravitational Wave Experiments and \\

\vspace{4mm}
Early Universe Cosmology\\}
\vspace{.4in}
{\large\bf  Michele Maggiore}\\
\vspace{.6 cm}
{\em INFN, sezione di Pisa, and Dipartimento di Fisica, Universit\`a
  di Pisa\\
via Buonarroti 2, I-56127 Pisa, Italy\footnote{Address from
Sept.~1, 1999 until May~31, 2000:  Theory Division, CERN, CH-1211 Geneva
23, Switzerland. }\\}
\end{center}

\vspace{.6cm}
\begin{abstract}
\noindent
Gravitational-wave experiments with
interferometers  and with
resonant masses can search for
stochastic backgrounds of gravitational waves  of cosmological
origin. We review both experimental and theoretical aspects of the search for
these backgrounds. We give a pedagogical derivation of the various
relations that characterize the response of a detector to a stochastic
background. 
We discuss the sensitivities of 
the large interferometers
under constructions (LIGO, VIRGO, GEO600, TAMA300, AIGO) 
 or planned (Avdanced LIGO, LISA) and of the
presently operating resonant bars, and we
give the sensitivities for various two-detectors correlations. 
We examine the existing
limits on the energy density in gravitational waves 
from nucleosynthesis, COBE and
pulsars, and their effects on  theoretical predictions.
We discuss general theoretical principles for order-of-magnitude
estimates of cosmological production mechanisms, and 
then we turn to specific theoretical predictions from inflation,
string cosmology, phase transitions, cosmic strings and other
mechanisms. 
We finally compare with the stochastic backgrounds of astrophysical origin.

\end{abstract}

\vspace{1.6cm}

\centerline{ To appear in {\it Physics Reports}}

\end{titlepage}
                                                                         
\tableofcontents
\newpage

\section{Introduction}\label{s1}

A large effort is presently under way in the scientific community for
the construction of the first generation of large scale
interferometers for gravitation wave (GW) 
detection. The two LIGO detectors are under
construction in the US, and the Italian-French interferometer VIRGO is
presently under construction near Pisa, Italy. The detectors are
expected to start taking data around the year 2002, with a
sensitivity where we might find  signals from known astrophysical
sources. These detectors are also expected to
evolve into second generation experiments, with might better
sensitivities. At the same time, two other somewhat smaller interferometers
are being built:
GEO600 in Germany and TAMA300 in Japan, and they will be important
also for developing the techniques needed by second generation
experiments. 
To cover the low-frequency region, where many interesting signals are
expected, but which on the Earth is unaccessible because of seismic noise,
it is planned to send an interferometer  into space: this
extraordinary experiment is LISA, which has been selected by the
European Space Agency has a cornerstone mission in his future science
program Horizons 2000 Plus, 
and might flight somewhere around 2010-2020, with a
sensitivity levels where many sources are virtually guaranteed. 
Finally, although at a lower sensitivity level, cryogenic resonant
bars are in operation since 1990, and more sensitive resonant
masses, e.g. of spherical shape, are under study.

A possible target of these experiments is a stochastic background of
GWs of cosmological origin. The detection of such a background would
have a profound impact on early Universe cosmology and on high-energy
physics, opening up a new window and exploring very early times in the
evolution of the Universe, and correspondingly high energies,
that will never be accessible by other means. 
The basic reason why relic GWs carry such a unique information is
that  particles which decoupled  from the primordial
plasma at  time $t\sim t\sub{dec}$, when the Universe had
a temperature $T_{\rm dec}$, give a `snapshot' of 
the state of the Universe at
$T\sim T_{\rm dec}$. All informations on the Universe 
when the particle was still in thermal equilibrium
has instead  been obliterated by the successive interactions. 
The weaker the interaction of a particle, the higher is the energy
scale when they drop out of thermal equilibrium. Hence, GWs can probe
deeper into the very  early Universe.

To be more quantitative, 
the condition for thermal equilibrium is that the rate $\Gamma$
of the processes that mantain equilibrium be larger than the 
rate of expansion of the Universe, as measured by the Hubble
parameter $H$. 
The rate (in units $\hbar =c=1$) is given by $\Gamma =n\sigma |v|$ 
where $n$ is the number density of the particle in question,
and for massless or light particles in equilibrium at a temperature $T$,
$n\sim T^3$;  $|v|\sim 1$ is the typical velocity and
$\sigma$ is the cross-section of the process. 
Consider for instance the weakly interacting neutrinos. In this
case the equilibrium is mantained, e.g., by electron-neutrino
scattering, and at energies below the $W$ mass
 $\sigma\sim G_F^2\langle E^2\rangle
\sim G_F^2T^2$ where $G_F$ is the
 Fermi constant and $\langle E^2\rangle$ is the average energy squared.
 The Hubble parameter during the radiation
dominated era is related to the temperature by $H\sim
 T^2/\mpl$. Therefore
\be
\left(\frac{\Gamma}{H}\right)_{\rm neutrino}
\sim \frac{G_F^2T^5}{T^2/\mpl}\simeq\left(
\frac{T}{1\rm MeV}\right)^3\, .
\ee
Even the weakly interacting neutrinos, therefore, cannot carry
informations on the state of the Universe at temperatures larger than
approximately 1 MeV.
If we repeat the above computation for gravitons, the Fermi constant
$G_F$ is replaced by Newton constant $G=1/\mpl^2$,
where $\mpl \sim 10^{19}$ GeV is the Planck mass.
At energies below $\mpl$, 
\be
\left(\frac{\Gamma}{H}\right)_{\rm graviton}
\sim \left(\frac{T}{\mpl}\right)^3\, .
\ee
The gravitons are therefore decoupled below the Planck scale.
It follows that relic gravitational waves  are a potential source of
informations on very high-energy physics. Gravitational waves
produced in the very early Universe have not lost memory of 
the conditions in which they have been produced, as it happened to
all other particles, but still retain in their spectrum, typical
frequency and intensity, important informations on the state of the
very early Universe,  and therefore on 
physics at correspondingly high energies, 
which cannot be accessed experimentally in any
other way.
It is also clear that the property of gravitational waves that
makes them so interesting, 
i.e. their extremely small cross section, is also
responsible for the difficulties of the experimental
detection.

In this review we will examine a number of experimental and
theoretical aspects of the search for a stochastic background of GWs.
The paper is organized as follows. We start from the experimental
side; in sects.~2 to 4 we present the basic concepts for
characterizing a stochastic background and for describing
the detector. This seems to be a subject where everybody has its own
definitions and notations, and we made an
effort to clean up many formulas appearing in the literature,  to
present them in what we consider the clearest form, and to give the
most straightforward derivation of various relations. 

In sect.~5 we examine the various detectors under construction or
already operating; we
show the sensitivity curves of LIGO, VIRGO, GEO600, Advanced LIGO,
LISA and NAUTILUS (kindly provided by the various collaborations), and
we discuss the limiting sources of
noise and the perspectives for improvement in the near and mid-term
future. 

To detect a stochastic background with Earth-based detectors, it turns
out that it is mandatory to correlate two or more detectors. The
sensitivities that can be
reached with two-detectors correlations are discussed in sect.~6. We
will show graphs for the overlap reduction functions between VIRGO and
the other major detectors, and we will give the estimate of the
minimum detectable value of the signal.

In sect.~7 we
examine the existing bounds on the energy density in GWs coming
from nucleosynthesis, COBE, msec pulsars, binary pulsar orbits, pulsar
arrays.

Starting from sect.~8 we  move toward more theoretical issues.
First of all, we discuss general principles for order of magnitude
estimate of the characteristic frequency and intensity  of the
stochastic background produced by  cosmological mechanisms.
With the very limited
experimental informations that we have on the very high energy
region,  $\mgut \lsim E\lsim \mpl$, and on the history of the very
early Universe,
it is unlikely that  theorists will be able to
foresee all the interesting sources of relic  stochastic backgrounds,
let alone to compute their spectra. This is particularly
clear in the Planckian or
string theory domain where, even if we succeed in predicting some
interesting physical effects, in general we cannot
 compute them reliably. So, despite the large efforts that have
been devoted to understanding possible sources, it is still quite
possible that, if a relic background of gravitational waves will be
detected, its physical origin will be a surprise.
In this case it is useful to understand what features of a theoretical
computation 
 have some general validity, and what are specific to a given
model, separating for instance kinematical from dynamical effects.
The results of this section provide a sort of benchmark, against which
we can compare the results from the specific models discussed in
sect.~9 and~10.

In sect.~9 we discuss  the most general cosmological
mechanism for GW production, namely the amplification of vacuum
fluctuations. We examine both the case of standard inflation, and a
more recent model, derived from string theory, known as pre-big-bang
cosmology. 
In sect.~\ref{other} we examine a number of other
production mechanisms.

Finally, in sect.~\ref{astro} we discuss the stochastic 
background due to many unresolved astrophysical sources, which from
the point of view of the cosmological background is a `noise' that
can compete with the signal (although, of course, it is very
interesting in its own right).

\section{Characterization of stochastic backgrounds of GWs}

A stochastic background of GWs of cosmological origin is expected to
be isotropic, stationary  and unpolarized. Its main property will be 
therefore its
frequency spectrum. There are different useful characterization of the
spectrum: (1) in terms of a (normalized)  energy density per unit
logarithmic  interval of frequency, $\hogw (f)$; (2) in terms  of the 
spectral density of the ensemble
avarage of the Fourier component of the metric, $S_h(f)$; (3) 
or  in terms of a
characteristic amplitude of the stochastic background, $\hc$. In this
section we examine the relation between these quantities, that
constitute the most common description used by the theorist.

On the other hand, the experimentalist
expresses the sensitivity of the apparatus in terms of a
strain sensitivity with dimension Hz$^{-1/2}$, or thinks in terms of the
dimensionless amplitude for the GW, which includes some form of binning. 
The relation 
between  these quantities and the variables $S_h(f),\hogw (f),\hc$
will be discussed in  sect.~3.

\subsection{The energy density $\ogw (f)$}

The intensity of  a stochastic background of gravitational waves (GWs)
can be characterized by the dimensionless quantity
\be
\ogw (f)=\frac{1}{\rho_c}\,\frac{d\rho_{\rm gw}}{d\log f}\, ,
\ee
where $\rho_{\rm gw}$ is the energy density of the stochastic
background of gravitational waves, $f$ is the frequency ($\omega =2\pi
f$) and $\rho_c$ is the present value of the
critical energy density for closing the Universe. In terms of the present
value of the Hubble constant $H_0$, the critical density is given by
\be\label{rhoc}
\rho_c =\frac{3H_0^2}{8\pi G}\, .
\ee
The  value of $H_0$ is usually written as $H_0=h_0\times 100 $
km/(sec--Mpc), where $h_0$ parametrizes the existing experimental
uncertainty.  Ref.~\cite{PDG} gives a value
 $0.5<h_0<0.85$. In the last
few years there has been a constant trend toward lower values of
$h_0$ and typical estimates are now in the range $0.55<h_0<0.60$
or, more conservatively, $0.50<h_0<0.65$. For instance
ref.~\cite{San}, using the method of type IA supernovae,
 gives two slightly different estimates
$h_0=0.56\pm 0.04$ and $h_0=0.58\pm 0.04$.
Ref.~\cite{Tri}, with the same method,
finds $h_0=0.60\pm 0.05$ and ref.~\cite{KK}, using a gravitational
lens, finds $h_0=0.51\pm 0.14$. 
The spread of values obtained gives an idea of
the systematic errors involved.

It is not very convenient to normalize $\rho_{\rm gw}$
to a quantity, $\rho_c$,
which is uncertain: this uncertainty would appear
in all the subsequent formulas, although it has nothing to do with the 
uncertainties on the GW background itself. 
Therefore, we
rather characterize the stochastic GW background with
the quantity $\hogw (f)$, which is independent of $h_0$. All theoretical
computations of primordial GW spectra are actually  computations of 
$d\rho_{\rm gw}/d\log f$ and  are  independent of the
uncertainty on $H_0$. Therefore the result of these computations is
 expressed in terms of $\hogw$, rather than of
$\ogw$.\footnote{This simple point has occasionally been missed in the
literature, where one can find the statement that, for small values of
$H_0$, $\ogw$ is larger and therefore easier to detect. Of course, it
is larger only because it has been normalized using a smaller
quantity.} 

\subsection{The spectral density $S_h(f)$ and
the characteristic amplitudes $\hc$}\label{2.2}

To understand the effect of the stochastic background on a detector,
we need however to think in terms of amplitudes of GWs.
A stochastic GW at a given point $\vec{x}=0$ can be expanded, in the
transverse traceless gauge, as\footnote{Our convention for the Fourier 
transform are 
$\tilde{g}(f)=\int_{-\infty}^{\infty}dt\, \exp\{2\pi ift\}g(t)$,
so that $g(t)=\int_{-\infty}^{\infty}df\, \exp\{-2\pi ift\}\tilde{g}(f)$.}
\be\label{hab}
h_{ab}(t)=\sum_{A=+,\times}\int_{-\infty}^{\infty}df\int d\hat{\Omega}
\,\tilde{h}_A(f,\hat{\Omega})e^{-2\pi ift}e_{ab}^A(\hat{\Omega})\, ,
\ee
where $\tilde{h}_A(-f,\hat{\Omega})=\tilde{h}_A^*(f,\hat{\Omega})$. 
$\hat{\Omega}$ is a unit vector representing the direction of
propagation of the wave and $d\hat{\Omega}=d\cos\theta d\phi$.
The polarization tensors can be written as
\bees\label{6}
e_{ab}^+(\hat{\Omega})&=&\hat{m}_a\hat{m}_b-\hat{n}_a\hat{n}_b\, ,\nonumber\\
e_{ab}^{\times}(\hat{\Omega})&=&\hat{m}_a\hat{n}_b+\hat{n}_a\hat{m}_b\, ,
\ees
with $\hat{m},\hat{n}$ unit vectors ortogonal to 
$\hat{\Omega}$ and to each other. With these definitions,
\be
e^A_{ab}(\hat{\Omega})e^{A',ab}(\hat{\Omega})=2\delta^{AA'}\, . 
\ee
For a
stochastic background,  assumed to be isotropic, unpolarized and
stationary (see~\cite{All,AR} for a discussion of these assumptions)
the ensemble average of the Fourier amplitudes
can be written as
\be\label{ave}
\langle \tilde{h}_A^*(f,\hat{\Omega})
\tilde{h}_{A'}(f',\hat{\Omega}')\rangle =
\delta (f-f')\frac{1}{4\pi}
\delta^2(\hat{\Omega},\hat{\Omega}')\delta_{AA'}
\frac{1}{2}S_h(f)\, ,
\ee
where $\delta^2(\hat{\Omega},\hat{\Omega}')=\delta (\phi -\phi ')
\delta (\cos\theta -\cos\theta ')$. The {\em spectral density}
 $S_h(f)$ defined by
the above equation
 has dimensions Hz$^{-1}$ and  satisfies $S_h(f)=S_h(-f)$. 
The factor 1/2 is conventionally inserted  in the definition of
$S_h$ in order to compensate for the fact
that the integration variable $f$
in eq.~(\ref{hab}) ranges between $-\infty$ and
$+\infty$ rather than over the physical domain $0\leq f<\infty$.
The factor $1/(4\pi )$ is a choice of normalization such that
\be\label{norm}
\int d\hat{\Omega}d\hat{\Omega}'\,
\langle \tilde{h}_A^*(f,\hat{\Omega})
\tilde{h}_{A'}(f',\hat{\Omega}')\rangle =
\delta (f-f')\delta_{AA'}
\frac{1}{2}S_h(f)\, .
\ee
Using eqs.~(\ref{hab},\ref{ave}) we get
\be\label{hc1}
\langle h_{ab}(t)h^{ab}(t)\rangle =
2\int_{-\infty}^{\infty}df\, S_h(f)=
4\int_{f=0}^{f=\infty}d(\log f)\,\, fS_h(f)\, .
\ee
We  now define the characteristic amplitude $h_c(f)$ from
\be\label{hc2}
\langle h_{ab}(t)h^{ab}(t)\rangle =
2 \int_{f=0}^{f=\infty}d(\log f)\,\, h_c^2(f)\, .
\ee
Note that $h_c(f)$ is dimensionless, and represents a characteristic
value of the amplitude, per unit logarithmic interval of frequency.
The factor of two on the right-hand side of eq.~(\ref{hc2}) is
part of our  definition, and is motivated by the fact that the
left-hand side is made up of two  contributions,
given by
$\langle \tilde{h}_+^*\tilde{h}_{+}\rangle$ and
$\langle \tilde{h}_{\times}^*\tilde{h}_{\times}\rangle$. In a
unpolarized background these contributions are equal, while 
the mixed term $\langle \tilde{h}_+^*\tilde{h}_{\times}\rangle$
vanishes, \eq{ave}.

Comparing eqs.~(\ref{hc1}) and
(\ref{hc2}), we get
\be\label{hc3}
h_c^2(f)= 2fS_h(f)\, .
\ee
We  now relate $\hc$ and $\hogw (f)$. The starting point is the
expression for the energy density of gravitational waves,
given by the $00$-component of the energy-momentum tensor. 
The energy-momentum tensor of a GW cannot be localized inside a
single wavelength
(see e.g.~ref.\cite{MTW}, sects.~20.4 and 35.7 for a careful
discussion) but it can be defined  with a spatial averaging over
several wavelengths:
\be\label{rho1}
\rho_{\rm gw}=\frac{1}{32\pi G}\langle \dot{h}_{ab}\dot{h}^{ab}
\rangle\, .
\ee
For a stochastic background, the spatial average over a few wavelengths
is the same as a time average at a given point, 
which, in Fourier space, is the 
ensemble average performed using eq.~(\ref{ave}). We therefore
insert eq.~(\ref{hab}) into eq.~(\ref{rho1}) and use
eq.~(\ref{ave}). The result is 
\be\label{rho2}
\rho_{\rm gw}=\frac{4}{32\pi G}
\int_{f=0}^{f=\infty}d(\log f)\,\, f (2\pi f)^2S_h(f)\, ,
\ee
so that 
\be\label{rho3}
\frac{d\rho_{\rm gw}}{d\log f}=\frac{\pi}{2G}f^3S_h(f)\, .
\ee
Comparing eqs.~(\ref{rho3}) and (\ref{hc3}) we get the important
relation
\be\label{rho4}
\frac{d\rho_{\rm gw}}{d\log f}=\frac{\pi}{4G}f^2h_c^2(f)\, ,
\ee
or, dividing by the critical density $\rho_c$,
\be\label{ogwhc}
\ogw (f)=\frac{2\pi^2}{3H_0^2} f^2h_c^2(f)\, .
\ee
Using eq.~(\ref{hc3}) we can also write
\be\label{ooo}
\ogw (f)=\frac{4\pi^2}{3H_0^2} f^3S_h(f)\, . 
\ee
Some of the equations that we will find below are very naturally
expressed in terms of $\hogw (f)$. This will be true in particular for
all theoretical predictions. Conversely, all equations involving the
signal-to-noise ratio and other issues related to the detection are
much more transparent when written in terms of
$S_h(f)$. Eq.~(\ref{ooo}) will be our basic formula for moving
between the two descriptions. 

Inserting the numerical
value of $H_0$, \eq{ogwhc} gives
\be\label{rho5}
h_c(f)\simeq 1.263\times 10^{-18}\,
\left(\frac{{\rm 1 Hz}}{f}\right)\sqrt{\hogw (f)}\,\, .
\ee
Actually, $\hc$ is not yet the most useful dimensionless quantity to
use for the comparison with experiments. In fact, any experiment
involves some form of binning over the frequency. In a total
observation time $T$, the resolution in frequency is $\Delta f=1/T$,
so one does not observe $\hogw (f)$ but rather 
\be
\int_f^{f+\Delta f}d(\log f)\,\, \hogw (f)\simeq \frac{\Delta f}{f}\hogw
(f)\, ,
\ee
and, since $\hogw (f)\sim h_c^2(f)$, it is convenient to define
\be\label{rho6}
h_c(f,\Delta f)=\hc \(\frac{\Delta f}{f}\)^{1/2}\, .
\ee
Using $1/(1\, {\rm yr})\simeq 3.17\times 10^{-8}$ Hz
 as a reference value for $\Delta f$, and
$10^{-6}$ as a reference value for $\hogw$, \eqs{rho5}{rho6} give
\be\label{hrms}
h_c(f,\Delta f)\simeq 2.249\times 10^{-25}
\left(\frac{{\rm 1 Hz}}{f}\right)^{3/2} 
\(\frac{\hogw (f)}{10^{-6}}\)^{1/2}
\(\frac{\Delta f}{3.17\times 10^{-8}\,{\rm Hz}}\)^{1/2}\, ,
\ee
or, with reference value that will be more useful for LISA,
\be\label{hrms2}
h_c(f,\Delta f)=7.111\times 10^{-22}
\left(\frac{{\rm 1 mHz}}{f}\right)^{3/2} 
\(\frac{\hogw (f)}{10^{-8}}\)^{1/2}
\(\frac{\Delta f}{3.17\times 10^{-8}\,{\rm Hz}}\)^{1/2}\, .
\ee
Finally, we mention
another useful formula which expresses $\hogw (f)$ in terms of the
number of gravitons  per cell of the phase space,
$n(\vec{x},\vec{k})$. For an isotropic stochastic background 
$n(\vec{x},\vec{k})=n_f$ depends only on the frequency
$f=|\vec{k}|/(2\pi)$, 
and 
\be
\rho_{\rm gw}=2\int \frac{d^3k}{(2\pi)^3}\,  2\pi f n_f\,  =
16\pi^2\int_0^{\infty}
d(\log f)\, f^4n_f\, . 
\ee
Therefore  
\be
\frac{d\rho_{\rm gw}}{d\log f} = 16\pi^2n_ff^4\, , 
\ee
and
\be\label{37}
\hogw (f) \simeq 3.6\left( \frac{n_f}{10^{37}}\right)
\left(\frac{f}{\rm 1 kHz}\right)^4\, .
\ee
This formula is useful in particular when one computes the production
of a stochastic background of GW due to amplification of vacuum
fluctuations, since the computation of the Bogoliubov coefficients
(see sect.~\ref{Bogo}) gives directly $n_f$.

As we will discuss below, to be observable at the LIGO/VIRGO
interferometers, we should have at least $\hogw\sim 10^{-6}$ between
1 Hz and 1 kHz, corresponding to $n_f$ of order
$10^{31}$ at 1 kHz and $n_f\sim 10^{43}$ at 1 Hz. A detectable
stochastic GW background is therefore exceedingly classical, $n_k\gg 1$.

\section{The response of a single detector}

\subsection{Detector tensor and detector pattern functions}\label{Dpf}

The quantities $\ogw (f),\hc, S_h(f)$ discussed above are all equivalent
characterization of the stochastic background of GWs, 
and have nothing to do with the
detector used. Now we must  make contact 
with what happens in a  detector.
The total output of the detector $S(t)$ is in general 
of the form
\be
S(t)=s(t)+n(t)
\ee
where $n(t)$ is the noise and $s(t)$ is the contribution to the output
due to the 
gravitational wave. For instance,
for an interferometer with 
equal arms of length $L$ in the $x-y$ plane, 
$s(t)=(\delta L_x(t)-\delta L_y(t))/L$,
where $\delta L_{x,y}$ are the displacements  produced by the GW.
The relation between the scalar output $s(t)$ and the gravitational
wave $h_{ab}(t)$ in the transverse-traceless gauge 
has the general form~\cite{For,Fla}
\be
s(t)=D^{ab}h_{ab}(t)\, .
\ee
$D^{ab}$ is known as the {\em detector tensor}.

Using \eq{hab}, we get
\be\label{27}
s(t)=\sum_{A=+,\times}\int_{-\infty}^{\infty}df\int d\hat{\Omega}
\,\tilde{h}_A(f,\hat{\Omega})e^{-2\pi ift}
D^{ab} e_{ab}^A(\hat{\Omega})\, .
\ee
It is therefore convenient to define the {\em detector pattern
functions} $F_{A}(\hat{\Omega} )$,
\be\label{defF}
F_{A}(\hat{\Omega} )=D^{ab} e_{ab}^A(\hat{\Omega})\, 
\ee
so that
\be\label{28}
s(t)=\sum_{A=+,\times}\int_{-\infty}^{\infty}df\int d\hat{\Omega}
\,\tilde{h}_A(f,\hat{\Omega})F_A(\hat{\Omega})e^{-2\pi ift}\, ,
\ee
and the Fourier transform of the signal, $\tilde{s}(f)$, is
\be
\tilde{s}(f)=\sum_{A=+,\times}\int d\hat{\Omega}
\,\tilde{h}_A(f,\hat{\Omega})F_A(\hat{\Omega})\, .
\ee
The pattern functions $F_A$
depend on the direction $\hat{\Omega}=(\theta ,\phi )$ of arrival
of the wave. Furthermore, they depend on an angle $\psi$ 
(hidden in $e_{ab}^A$) which
describes a rotation in the plane orthogonal to $\hat{\Omega}$, i.e. in the
plane spanned by the vectors $\hat{m},\hat{n}$, see \eq{6}.
Once we have made a definite choice for the vectors $\hat{m},\hat{n}$,
we have chosen the axes with respect to which the  $+$ and $\times$
polarizations are defined. For an astrophysical source there can be
a natural choice of axes, with respect to which the radiation
has an especially simple form. Instead, for a unpolarized
stochastic background, there is no privileged choice of basis,
and the angle $\psi$ must cancel from the final result, as we will
indeed check below.

For a stochastic background the average of $s(t)$ vanishes and, if we
have only one detector, the best we can do is to
consider the average of $s^2(t)$. Using \eqss{28}{defF}{ave},
\be\label{FSh}
\langle s^2(t)\rangle =F\, \int_{-\infty}^{\infty}df\, \frac{1}{2}
S_h(f)=
F\, \int_{0}^{\infty}df\, 
S_h(f)\, ,
\ee
where
\be\label{FF}
F\equiv
\int\frac{d\hat{\Omega}}{4\pi}\sum_{A=+,\times} F^A(\hat{\Omega},\psi )
F^A(\hat{\Omega},\psi )\, .
\ee
Note that, while the value of $\langle s^2(t)\rangle$ is obtained
{\em summing} over all stochastic waves coming from all directions 
$\hat{\Omega}$, the factor $1/(4\pi )$ in \eq{ave} produces an {\em
  average} of $F_+^2+F_{\times}^2$ over the directions.
The factor $F$ gives a measure of the loss of sensitivity due to the
fact that the stochastic waves come from all directions, compared to the
sensitivity of the detector for waves coming from the optimal direction.

To compute the pattern functions, we need the explicit expressions for 
the vectors $\hat{m},\hat{n}$  that enters the definition of \eq{6}.
We use  polar coordinates  centered
on  the detector. Then
\be
\hat{\Omega}=(\sin\theta\sin\phi ,\sin\theta\cos\phi ,\cos\theta)
\ee
and a possible choice for $\hat{m},\hat{n}$ is
\be
\hat{n}=(-\cos\theta\sin\phi ,-\cos\theta\cos\phi ,\sin\theta )\, ,
\hspace{5mm}
\hat{m}=(-\cos\phi ,\sin\phi ,0)\, .
\ee
The most general choice is obtained with the rotation
\bees
\hat{n}&\ra &\hat{n}\cos\psi +\hat{m}\sin\psi\, ,\nonumber\\
\hat{m}&\ra &-\hat{n}\sin\psi +\hat{m}\cos\psi\, ,
\ees
so that 
\bees
\hat{n}&=&(-\cos\theta\sin\phi\cos\psi -cos\phi\sin\psi ,
-\cos\theta\cos\phi\cos\psi +\sin\phi\sin\psi ,
\sin\theta\cos\psi )\, ,\nonumber\\
\hat{m}&=&(-\cos\phi\cos\psi +\cos\theta\sin\phi\sin\psi ,
\sin\phi\cos\psi +\cos\theta\cos\phi\sin\psi ,
-\sin\theta\sin\psi)\, .
\ees
We  now compute the explicit expression for $F^A$ and $F$ in the
most interesting cases.

\subsubsection{Interferometers}
For an interferometer with arms along the $\hat{u}$
and $\hat{v}$ directions (not necessarily orthogonal)
\be\label{Dint}
D^{ab}=\frac{1}{2}\( \hat{u}^a\hat{u}^b-\hat{v}^a\hat{v}^b\) \, .
\ee
Using the definition of $F_A$, \eq{defF}, together with 
\eq{6} for $e^{A}_{ab}$ and 
with the above expressions for
$\hat{m},\hat{n}$ one obtains, restricting now to an interferometer with
perpendicular arms~\cite{For2,ST,Th},
\bees\label{Fint}
F_+(\theta, \phi ,\psi)&=&\frac{1}{2}(1+\cos^2\theta )\cos 2\phi\cos
2\psi - \cos\theta\sin 2\phi\sin 2\psi\, ,\nonumber\\
F_{\times}(\theta, \phi ,\psi)&=&\frac{1}{2}(1+\cos^2\theta )\cos 2\phi\sin
2\psi + \cos\theta\sin 2\phi\cos 2\psi\, ,
\ees
The factor $F$ is then given by
\be\label{Fval}
F\equiv\int\frac{d\hat{\Omega}}{4\pi}\sum_{A=+,\times} F^A(\hat{\Omega},\psi )
F^A(\hat{\Omega},\psi )=\frac{2}{5}\, .
\ee
Note also that $\langle s^2(t)\rangle$ depends on the $F_A$ only
through the combination
$F_+^2+F_{\times}^2$ which is independent of the angle
$\psi$, as can be checked from the explicit expressions, in agreement
with the argument discussed above. The same will be true for resonant
bars and spheres. Therefore below we will often
use the notation $F_A(\hat{\Omega})$ instead of
$F_A(\hat{\Omega},\psi )$.

It is interesting to see how these results are modified if the arms
are not perpendicular. If $\alpha$ is the angle between
the two arms, we find
\bees
F_+(\theta, \phi ,\psi)&=&(\sin\alpha )\[
\frac{1}{2}(1+\cos^2\theta )\sin (\alpha +2\phi )\cos 2\psi +
\cos\theta\cos (\alpha +2\phi )\sin 2\psi\]\, ,\nonumber\\
F_{\times}(\theta, \phi ,\psi)&=&(\sin\alpha )\[
\frac{1}{2}(1+\cos^2\theta )\sin (\alpha +2\phi )\sin 2\psi -
\cos\theta\cos (\alpha +2\phi )\cos 2\psi\]\, .
\ees
Therefore
\be\label{2/5sin}
F=\frac{2}{5}\sin^2\alpha\, .
\ee
For $\alpha =\pi/2$ we recover  the results
(\ref{Fint},\ref{Fval}) and the sensitivity is maximized, 
while for $\alpha =0$ the sensitivity of
the interferometric detection of course vanishes.

\subsubsection{Cylindrical bars}
For a cylindrical bar with axis along the direction $\hat{l}$, 
one has instead
\be\label{Dbar}
D^{ab}=\hat{l}^a\hat{l}^b\, .
\ee
Since $D^{ab}$ is contracted with the traceless tensor
$h_{ab}$, it is  defined only apart from terms $\sim \delta^{ab}$,
so that we can equivalently use the
traceless tensor
\be
D^{ab}=\hat{l}^a\hat{l}^b-\frac{1}{3}\delta^{ab}\, .
\ee
To compute $F_A$ it is convenient, compared to the case of the
interferometer, to perform
a redefinition $\psi\ra\psi +\pi /2$, and define $\theta$
as the polar angle measured from the bar direction $\hat{l}$. (of
course, when we will discuss interferomer-bar correlations in
sect.~\ref{3.5}, we will be careful to use the same definitions in the
two cases). Then
\bees
F_+(\theta, \phi ,\psi)&=&\sin^2\theta\cos 2\psi\, ,\nonumber\\
\label{Fbar}
F_{\times}(\theta, \phi ,\psi)&=&\sin^2\theta\sin 2\psi\, .
\ees
and the angular factor $F$ is
\be
F\equiv\int\frac{d\hat{\Omega}}{4\pi}\sum_{A=+,\times} F^A(\hat{\Omega},\psi )
F^A(\hat{\Omega},\psi )=\frac{8}{15}\, .
\ee

\subsubsection{Spherical detectors}\label{spheres}
Finally, it is interesting to give also the result for a detector of
spherical geometry.
For a spherical resonant-mass detector 
there are 
five detection channels~\cite{TIGA}
corresponding to the five degenerates quadrupole modes. The functions
$D^{ab}$ and $F_A$ for each of these channel are given in 
ref.~\cite{ZM}: one defines the real quadrupole spherical harmonics as
\bees\label{harm}
Y_0   &\equiv& Y_{20}\nonumber\\
Y_{1c}&\equiv& \frac{1}{\sqrt{2}}(Y_{2,-1}-Y_{2,+1})\nonumber\\
Y_{1s}&\equiv& \frac{i}{\sqrt{2}}(Y_{2,-1}+Y_{2,+1})\\
Y_{2c}&\equiv& \frac{1}{\sqrt{2}}(Y_{2,-2}+Y_{2,+2})\nonumber\\
Y_{2s}&\equiv& \frac{i}{\sqrt{2}}(Y_{2,-2}-Y_{2,+2})\, .\nonumber
\ees
The pattern functions associated to the $0,1c,1s,2c,2s$ channels are
\bees\label{sphere}
F_{+,0}(\theta ,\phi )=\frac{\sqrt{3}}{2}\sin^2\theta\, ,
&\hspace{5mm}&
F_{\times ,0}(\theta ,\phi )=0\nonumber\\
F_{+,1c}(\theta ,\phi )=\frac{1}{2}\sin 2\theta\cos\phi\, ,
&\hspace{5mm}&
F_{\times ,1c}(\theta ,\phi )=-\sin\theta\sin\phi\nonumber\\
F_{+,1s}(\theta ,\phi )=-\frac{1}{2}\sin 2\theta\sin\phi\, ,
&\hspace{5mm}&
F_{\times ,1s}(\theta ,\phi )=-\sin\theta\cos\phi\\
F_{+,2c}(\theta ,\phi )=\frac{1}{2}(1+\cos^2\theta )\cos 2\phi\, ,
&\hspace{5mm}&
F_{\times ,2c}(\theta ,\phi )=-\cos\theta\sin 2\phi\nonumber\\
F_{+,2s}(\theta ,\phi )=-\frac{1}{2}(1+\cos^2\theta )\sin 2\phi\, ,
&\hspace{5mm}&
F_{\times ,2s}(\theta ,\phi )=-\cos\theta\cos 2\phi\nonumber\, .
\ees
Here the angle  $\psi$ has been set to zero; as discussed above,
this corresponds to a
choice of the axes with respect to which the polarization states are
defined. With this choice,
at $\theta =0$, only $F_{A,2s},F_{A,2c}$ are non-vanishing, 
consistently with the fact  GWs have elicities $\pm 2$.
The same value as for interferometers, $F=2/5$, is obtained for the sphere 
in the  channel $m=0$, as well as for the channels
$m=\pm 1$ and $m= \pm 2$.

\subsection{The strain sensitivity $\tilde{h}_f$}

The ensemble average of the Fourier components  of the noise satisfies
\be\label{Sn1}
\langle \tilde{n}^*(f)\tilde{n}(f')\rangle =
\delta(f-f')\frac{1}{2}S_n(f)\, .
\ee
The above equation defines the
functions $S_n(f)$, with $S_n(-f)=S_n(f)$ and dimensions Hz$^{-1}$. 
The factor $1/2$ is again conventionally inserted in the definition
so that the total noise power is obtained integrating $S_n(f)$ over the
physical range $0\leq f<\infty$, rather than from $-\infty$ to
$\infty$,
\be\label{n2}
\langle n^2(t)\rangle = \int_0^{\infty}df\, S_n(f)\, .
\ee
The function $S_n$ is known as the square spectral noise
density.\footnote{Unfortunately there is not much agreement
about notations in the literature. The square spectral noise density,
that we denote by $S_n(f)$ following e.g. ref.~\cite{Fla},
is called $P(f)$ in ref.~\cite{All}. Other authors
use  the notation $S_h(f)$, which we instead reserve 
for the spectral density of the signal. To make things worse, $S_n$ is
sometime defined with or without the factor 1/2 in eq.~(\ref{Sn1}).}

Equivalently, the noise level of the  detector  is  measured by
the {\em strain sensitivity} $\tilde{h}_f$
\be
\tilde{h}_f\equiv\sqrt{S_n(f)}\, , 
\ee
where now $f>0$. Note that $\tilde{h}_f$ 
 is linear in the noise and has dimensions
Hz$^{-1/2}$.

Comparing  eqs.~(\ref{n2}) and eq.~(\ref{FSh}), we see that 
in a single detector a
stochastic background will  manifesting itself as an eccess noise, and
will be observable at a frequency $f$ if
\be
 S_h(f)\, > \, \frac{1}{F} S_n(f)\, . 
\ee
where $F$ is the angular efficency factor, \eq{FF}.
Using eq.~(\ref{ooo}) 
and $S_n(f)=\tilde{h}_f^2$, we can express this result in terms of
the minimum detectable value of $\hogw$, as
\be\label{single}
\hogw^{\rm min}(f)\simeq \frac{1}{F}\cdot 10^{-2}
\left(\frac{f}{\rm 100 Hz}\right)^3
\left(\frac{\tilde{h}_f}{10^{-22}{\rm Hz}^{-1/2}}
\right)^2\, .
\ee
In sect.~4 we will use this result to compute the sensitivity of
various single detectors to a stochastic background.

\section{Two-detectors correlation}

\subsection{The overlap reduction functions $\gamma (f), \Gamma (f)$}
To detect a stochastic GW background the optimal strategy consists in
performing a correlation between two (or more) detectors, since, as we
will see, the
signal  is expected to be far too low to exceed the noise
level in any existing or planned 
single detector (with the exception of the planned 
space interferometer LISA).

The strategy for correlating two detectors has been discussed
in refs.~\cite{Mic,Chr,Fla,VCCO,All}.
We  write the output $S_i(t)$ of the $i-$th detector as $S_i(t)=s_i(t)+n_i(t)$,
where  $i=1,2$ labels the detector, and
we have to face the situation 
in which the GW  signal $s_i$ is much smaller than the noise $n_i$.
At a generic point $\vec{x}$ we rewrite \eq{hab} as
\be\label{hab2}
h_{ab}(t,\vec{x})=\sum_{A=+,\times}\int_{-\infty}^{\infty}df\int d\hat{\Omega}
\,\tilde{h}_A(f,\hat{\Omega})e^{-2\pi if(t-\hat{\Omega}\cdot\vec{x}/c)}
e_{ab}^A(\hat{\Omega})\, ,
\ee
so that
\be\label{ht2}
s_i(t)=\sum_{A=+,\times}\int_{-\infty}^{\infty}df\int d\hat{\Omega}
\,\tilde{h}_A(f,\hat{\Omega})e^{-2\pi if(t-\hat{\Omega}\cdot\vec{x}_i/c)}
F_i^A(\hat{\Omega} )\, ,
\ee
where $F_i^A$ are the pattern functions of the $i$-th detector. 
The Fourier transform of the scalar  signal in the i-th
detector, $\tilde{s}_i(f)$, is then
related to $\tilde{h}_A(f,\hat{\Omega})$ by
\be\label{tildes}
\tilde{s}_i(f)=\sum_{A=+,\times}\int d\hat{\Omega}
\,\tilde{h}_A(f,\hat{\Omega})e^{2\pi if\hat{\Omega}\cdot\vec{x}_i/c}
F_i^A(\hat{\Omega} )\, .
\ee
We then correlate the two outputs defining
\be\label{S}
S_{12}=\int_{-T/2}^{T/2}dt\int_{-T/2}^{T/2}dt'\,
S_1(t)S_2(t')Q(t-t')\, ,
\ee
where $T$ is the total integration time (e.g. one year) and $Q$ a
real filter function. The simplest choice would be $Q(t-t')=\delta
(t-t')$, while the optimal choice will be discussed in
sect.~\ref{filter}. At any rate,
$Q(t-t')$ falls rapidly to zero for large $|t-t'|$; then, taking the
limit of large $T$,  \eq{S} gives
\be
S_{12}=\int_{-\infty}^{+\infty}df\, 
\tilde{S}_1^*(f)\tilde{S}_2(f)\tilde{Q}(f)\, .
\ee
Taking the ensemble average, the contribution to 
 $\langle S_{12}\rangle $ 
from the GW background is
\bees
\langle s_{12}\rangle
&\equiv &\int_{-\infty}^{+\infty}df\, 
\langle \tilde{s}_1^*(f)\tilde{s}_2(f)\rangle\, \tilde{Q}(f)=
\int_{-\infty}^{+\infty}df\, \int d\hat{\Omega} d\hat{\Omega}'
e^{ \frac{2\pi if}{c}
(\hat{\Omega}\cdot\vec{x}_1-\hat{\Omega}'\cdot\vec{x}_2) }
\times
\nonumber\\
& &\times\sum_{A,A'} 
F_1^A(\hat{\Omega})F_2^{A'}(\hat{\Omega}')\,
\langle \tilde{h}_A^*(f,\hat{\Omega})\tilde{h}_{A'}(f,\hat{\Omega}')
\rangle \, \tilde{Q}(f)=
\nonumber\\
&= &T\int_{-\infty}^{\infty} df\, 
\frac{1}{2}S_h(f)\Gamma (f)\tilde{Q}(f)\, ,
\label{sysrq}
\ees
where we have used \eq{ave} and
$\delta (0)=\int_{-T/2}^{T/2}dt=T$.
In the last line we have defined
\be\label{Gamma}
\Gamma (f) \equiv \int \frac{d\hat{\Omega}}{4\pi}\,
\[ \sum_A F^A_1(\hat{\Omega} )F^A_2(\hat{\Omega} )\]
\exp\left\{ 2\pi if\hat{\Omega}\cdot\frac{\Delta\vec{x}}{c}\right\}\, ,
\ee
where $\Delta \vec{x}$ is the separation between the two detectors.
Note that $\sum_A F^A_1F^A_2$ is independent of $\psi$ even if the two
detectors are of different type, e.g. an interferometer and a bar, as
can be checked from the explicit expressions, and
therefore again we have not written explicitly the $\psi$ dependence. 

We introduce  also
\be\label{FF12}
F_{12}\equiv
\int\frac{d\hat{\Omega}}{4\pi}\sum_A F^A_1(\hat{\Omega} )
F^A_2(\hat{\Omega} )|_{\rm aligned}\, ,
\ee
where the subscript means that we must compute $F_{12}$ taking the two
detectors to be perfectly aligned, rather than with their actual
orientation. 
Of course if the two
detectors are of the same type, e.g. two interferometers or two
cylindrical bars,  $F_{12}$ is the same as the constant
$F$ defined in eq.~(\ref{FF}).

The {\em overlap reduction function} $\gamma(f)$ is defined 
by~\cite{Chr,Fla} 
\be\label{gammaGamma}
\gamma(f)=\frac{\Gamma(f)}{F_{12}}\, .
\ee
This normalization 
is useful in the case of two interferometers, since
$F_{12}=2/5$ already takes into account the reduction in sensitivity due to
the angular pattern, already present in the case of one
interferometer, and therefore $\gamma (f)$ separately
takes into account the effect
of the separation $\Delta\vec{x}$ between the interferometers, and of
their relative orientation. With this definition, $\gamma (f)=1$ if the
separation $\Delta x =0$ and if the detectors are perfectly aligned.

This normalization is instead impossible when one considers the
correlation between an interferometer and a resonant sphere, since in
this case for some modes of the sphere 
$F_{12}=0$, as we will see below. Then one simply uses 
$\Gamma (f)$, which is the quantity that enters directly \eq{sysrq}.
Furthermore, the use of $\Gamma (f)$ is more convenient when we want
to write equations that hold independently of what detectors
(interferometers, bars, or spheres) are
used in the correlation.

\subsection{Optimal filtering}\label{filter}
As we have seen above, in order to correlate the output of two detectors we
consider the combination
\be
S_{12}=\int_{-\infty}^{+\infty}df\, 
\tilde{S}_1^*(f)\tilde{S}_2(f)\tilde{Q}(f)\, .
\ee
We now find the optimal choise of the filter function
$\tilde{Q}(f)$,~\cite{Mic,Chr,Fla,VCCO,All} 
following the discussion given in ref.~\cite{All}. 
Since the spectral densities are defined, in the unphysical region
$f<0$, by $S_h(f)=S_h(-f),S_n(f)=S_n(-f)$, we also require
$\tilde{Q}(f)=\tilde{Q}(-f)$ and $\tilde{Q}(f)$ real (so that 
$\tilde{Q}(f)=\tilde{Q}^*(-f)$ and $Q(t)$ is real). 

We consider 
the variation of $S_{12}$ from its average value,
\be
N\equiv S_{12}-\langle S_{12}\rangle\, .
\ee
By definition $\langle N\rangle =0$, while
\bees
\langle N^2\rangle &=&\langle S_{12}^2\rangle - \langle
S_{12}\rangle^2=
\int_{-\infty}^{\infty}dfdf'\, \tilde{Q}(f)\tilde{Q}^*(f')
\times\nonumber\\
& &\times\left[
\langle \tilde{S}_1^*(f)\tilde{S}_2(f)\tilde{S}_1(f')\tilde{S}^*_2(f')
\rangle -
\langle \tilde{S}_1^*(f)\tilde{S}_2(f)\rangle
\langle \tilde{S}_2^*(f')\tilde{S}_1(f')\rangle 
\right]\, .
\ees
We are now interested in the case $n_i\gg h_i$. Then
$S_i\simeq n_i$; if the noise in a single detector
has a Gaussian distribution and the  noise in the two detectors are
uncorrelated, then 
this becomes
\be
\langle N^2\rangle \simeq 
\int_{-\infty}^{\infty}dfdf'\, \tilde{Q}(f)\tilde{Q}^*(f')
\langle \tilde{S}_1^*(f)\tilde{S}_1(f')\rangle
\langle \tilde{S}_2^*(f')\tilde{S}_2(f)\rangle 
\ee
and
\be
\langle \tilde{S}_i^*(f)\tilde{S}_i(f')\rangle\simeq
\langle \tilde{n}_i^*(f)\tilde{n}_i(f')\rangle =
\delta (f-f')\frac{1}{2}S_n^{(i)}(f)\, .
\ee
Using $\delta (0)=T$, one therefore gets
\be\label{N2}
\langle N^2\rangle =\frac{T}{4}
\int_{-\infty}^{\infty}df\, |\tilde{Q}(f)|^2 S_n^2(f)\, ,
\ee
where we have defined
\be
S_n(f)=\left[ S_n^{(1)}(f)S_n^{(2)}(f)\right]^{1/2}\, .
\ee
We now define the signal-to-noise ratio (SNR),
\be
{\rm SNR} = \left[ \frac{\langle s_{12}\rangle}{\langle N^2\rangle^{1/2}}
\right]^{1/2}\, .
\ee
Note that we have taken an overall  square root in the
definition of the SNR since $s_{12}$ is quadratic in the signal, and so
the SNR  is linear in the signal
(this differs from the convention used in ref.~\cite{All}, where it is
quadratic); $\langle s_{12}\rangle$ is 
the contribution to $\langle S_{12}\rangle$ from the GW, 
given in \eq{sysrq}. 

We now look for the function $\tilde{Q}(f)$  that
maximizes the SNR. For two arbitrary complex functions $A(f),B(f)$ one
defines  the (positive definite) scalar product~\cite{All}
\be
(A,B)=\int_{-\infty}^{\infty}df\, A^*(f)B(f)S_n^2(f)\, .
\ee
Then
\be
\langle N^2\rangle =\frac{T}{4}(\tilde{Q},\tilde{Q})
\ee
and, from \eq{sysrq},
\be
\langle s_{12}\rangle
=\frac{T}{2}\(\tilde{Q},\frac{S_h\Gamma}{S_n^2}\)\, .
\ee
Then we have to maximize
\be
({\rm SNR})^4= \frac{\langle s_{12}\rangle^2}{\langle N^2\rangle}
=T\(\tilde{Q},\frac{\Gamma S_h}{S_n^2}\)^2\frac{1}{(\tilde{Q},\tilde{Q})}
\, .
\ee
The solution of this variational problem is standard,
\be\label{optimal}
\tilde{Q}(f)=c \frac{\Gamma (f)S_h(f)}{S_n^2(f)}
\ee
with $c$ an arbitrary normalization constant. Note that, since we used
$\Gamma (f)$ rather then $\gamma (f)$, this formula holds true
independently of the detectors used in the correlations.

It is also important to observe that the optimal filter depends on the signal
that we are looking for, since $S_h(f)$ enters \eq{optimal}. This
means that one should perform the data analysis 
considering  a set of possible filters, chosen either in order to
encompass a range of typical behaviours, e.g. chosing
$S_h(f)\sim f^{\alpha}$ for different values of the parameter
$\alpha$, or using the  theoretical predictions discussed in later
chapters. 
Of course this point is not relevant for correlations
involving resonant bars, because of their narrow bandwidth.

\subsection{The SNR for  two-detectors correlations}\label{3.5}
We can now write down the value of the SNR obtained with the optimal
filter:
\be
({\rm SNR})^4 = T\(\frac{\Gamma S_h}{S_n^2},\frac{\Gamma S_h}{S_n^2}\)\, ,
\ee
or
\be\label{SNR2}
{\rm SNR}=\left[ 2 T\int_0^{\infty}df\, 
 \Gamma^2(f)
\frac{S_h^2(f)}{S_n^2(f)}
\right]^{1/4}\, .
\ee
In particular, for two interferometers
$\Gamma (f)=(2/5)\gamma (f)$
and
\be\label{SNR3}
({\rm SNR})_{\rm intf-intf}=\left[ \frac{8}{25} T\int_0^{\infty}df\, 
 \gamma^2(f)
\frac{S_h^2(f)}{S_n^2(f)}
\right]^{1/4}\, .
\ee
For two cylindrical bars, $\Gamma (f)=(8/15)\gamma (f)$ and
\be\label{SNR3b}
({\rm SNR})_{\rm bar-bar}=\left[ \frac{128}{225} T\int_0^{\infty}df\, 
 \gamma^2(f)
\frac{S_h^2(f)}{S_n^2(f)}
\right]^{1/4}\, .
\ee
For the correlation between  an interferometer and a cylindrical bar,
again $\Gamma (f)=(2/5)\gamma (f)$, so that we get again \eq{SNR3}.

In these formulas,  
the effect due to the fact that the two detectors  will have in
general a different orientation is fully taken into account into $\gamma
(f)$. If however one wants to have a quick order of magnitude
estimate of the effect due to the misalignement without performing 
each time  numerically the integral that defines $\gamma (f)$, one can 
note, from \eq{Gamma}, that in the limit 
$2\pi f d \ll 1$, where $d=|\Delta \vec{x}|$ is the distance between
the detectors,
\be
\Gamma (f) \simeq \int \frac{d\hat{\Omega}}{4\pi}\,
\[ \sum_A F^A_1(\hat{\Omega} )F^A_2(\hat{\Omega} )\]\, ,
\ee
where of course the pattern functions $F^A$ relative to the two
detectors are computed with their actual orientation. 
Therefore in the limit $2\pi f d \ra 0$ the ratio between 
the value of $\Gamma $ that takes into account the actual orientation
and the value for aligned detectors is equal to the ratio between 
$F_{12}$ computed with the actual orientation and $F_{12}$ for aligned
detectors. 

In the definition of $\gamma$ enters $F_{12}$  computed for aligned
detectors, \eq{FF12}.
Then  a simple estimate  
of the effect due to the misalignement is
obtained substituting in \eq{SNR2} $\Gamma (f) \simeq F_{12}\gamma (f)$
where $F_{12}$ is now computed for the actual orientation of the
detectors and $\gamma (f)$ is the overlap reduction function for the
perfectly aligned detectors.

We apply this procedure to  estimate the effect of misalignement
on the SNR, in the 
correlation between an interferometer and a cylindrical bar.
In the limit $2\pi f d\ll 1$ we can also neglect the effect of
the Earth curvature. Then, in the frame where the interferometer arms
are along the $x,y$ axes, the bar also lies in the $(x,y)$
plane and its direction  is in general
given by $\hat{l}= (\sin\alpha ,\cos\alpha ,0)$; 
the angle  $\alpha$ measures the misalignement between
the interferometer $y$ axis and the bar.  In this
frame, computing  the detector pattern function of the bar we get
\bees
F_+^{\rm (bar)}&=&-\cos^2\theta\cos^2 (\phi -\alpha )\cos 2\psi
+\sin^2 (\phi -\alpha)\cos 2\psi
+\cos\theta\sin 2(\phi -\alpha)\sin 2\psi\nonumber\\
F_{\times}\su{(bar)}&=&-\cos^2\theta\cos^2 (\phi -\alpha )\sin 2\psi
+\sin^2 (\phi -\alpha)\sin 2\psi
-\cos\theta\sin 2(\phi -\alpha)\cos 2\psi\nonumber\\
& &
\ees
To compare with \eq{Fbar} note that here $\theta$ is measured from the
$z$ axis, and the bar is in the  $(x,y)$ plane, while in \eq{Fbar}
$\theta$ is measured for the direction of the bar.
For the interferometer the functions $F_A$ are given in \eq{Fint}. 
A straightforward computation shows that 
\bees
\sum_A F_A\su{(bar)}F_A\su{(intf)}&=&
\frac{1}{2}(1+\cos^2\theta )\cos 2\phi 
\[ \sin^2(\phi -\alpha)-\cos^2\theta\cos^2(\phi -\alpha) \]
\nonumber\\
& & -\cos^2\theta\sin 2(\phi -\alpha)\sin 2\phi\, .
\ees
Note that again the dependence on the angle $\psi$ has cancelled. Then 
\be
F_{12}=\int \frac{d\hat{\Omega}}{4\pi}\,
\sum_A F_A\su{(bar)}F_A\su{(intf)}=
-\frac{2}{5}\cos 2\alpha\, .
\ee
The overall sign of $F_{12}$ is irrelevant since $\Gamma
(f)$ enters quadratically in the SNR. Therefore we get
\be
{\rm SNR} (\alpha) \simeq |\cos 2\alpha|^{1/2} {\rm SNR} (0)\, .
\ee
Of course, the SNR is maximum if the bar is aligned along the $y$ axis
($\varphi =0$) of along the $x$ axis ($\varphi =\pi /2$); the
correlation becomes totally ineffective when $\alpha =\pi/4$.

Finally, it is interesting to consider the correlation between an
interferometer and a sphere~\cite{dad}. Using \eq{sphere}, and \eq{Fint} with
$\psi =0$, one finds in particular that for the correlation with the
$m=0$ channel  of the sphere,
\be
\sum_A F_A\su{(sph)}F_A\su{(intf)}=\frac{\sqrt{3}}{4}\sin^2\theta
(1+\cos^2\theta )\cos 2\phi\, ,
\ee
and therefore $F_{12}$ vanishes because of the integral over
$\phi$. Similarly for the channel $1s,1c$ one gets an integral over
$\cos\phi\cos 2\phi$, which again vanishes.
In this case of course $\gamma (f)$ cannot be defined and one
simply uses $\Gamma (f)$.
Instead, for each of the two  modes $2s,2c$, $F_{12}$ is just one half of the
value for the interferometer-interferometer correlation, so that
summing over all modes of the sphere one gets $F_{12}=2/5$.

\subsection{The characteristic noise $\hn$ of correlated detectors}\label{3.6}

We have seen in  section~\ref{2.2} that
there is a very natural definition of the characteristic
amplitude of the signal, given by $h_c(f)$,
which contains all the informations on the physical effects, and is
independent of the apparatus. We will now associate to $h_c(f)$ 
a corresponding noise amplitude $h_n(f)$, that embodies
 all the informations
on the apparatus, defining  $\hc /h_n(f)$ in terms of the optimal SNR.

If, in the integral giving the optimal SNR,
 eq.~(\ref{SNR2}), we consider
only a range of frequencies $\Delta f$ such that the integrand is
approximately constant, we can write
\be\label{SNR4}
{\rm SNR}\simeq  
\left[ 2T\Delta f
\frac{\Gamma^2(f)S_h^2(f)}{S_n^2(f)}\right]^{1/4}=
\left[\frac{T\Delta f\Gamma^2(f)h_c^4(f)}{2 f^2S_n^2(f)}\right]^{1/4}\, ,
\ee
where we have used \eq{hc3}.
The right-hand side of eq.~(\ref{SNR4}) is proportional to $h_c(f)$,
and we can therefore define $h_n(f)$
equating the right-hand side of eq.~(\ref{SNR4}) to
$h_c(f)/h_n(f)$, so that 
\be\label{hn}
h_n(f)\equiv\frac{1}{(\frac{1}{2}T\Delta f)^{1/4}}\,
\left(\frac{fS_n(f)}{\Gamma (f)}\right)^{1/2}\, .
\ee
In the case  of  two interferometers approximately at the same site,
\be
\Gamma (f)\simeq
\int \frac{d\hat{\Omega}}{4\pi}\,
\[ \sum_A F^A(\hat{\Omega},\psi )F^A(\hat{\Omega},\psi
)\]=F\, ,
\ee
and we recover eq.~(66) of ref.~\cite{Th},
\be
h_n(f)=\frac{1}{(\frac{1}{2}T\Delta f)^{1/4}}\,
\left[\frac{fS_n(f)}{F}\right]^{1/2}\, .
\ee
For  comparison,  note that the quantity
denoted $S_h(f)$ in ref.~\cite{Th} is called here $S_n(f)/2$, and
$F=\langle F_+^2\rangle +\langle F_{\times}^2\rangle=2\langle
F_+^2\rangle$. 

From the derivation of eq.~(\ref{hn})
we can better understand  the limitations implicit
in the use of  $h_n(f)$. It gives a measure of the noise level only
under the approximation that leads from eq.~(\ref{SNR2}), which is
exact (at least in the limit $h_i\ll n_i$),
to eq.~(\ref{SNR4}). This means that $\Delta f$ must be small enough 
compared to the scale on which the integrand in eq.~(\ref{SNR2}) changes,
so that $\Gamma (f)S_h(f)/S_n(f)$ must be approximately constant.
In a large bandwidth this is non trivial, and of course depends also on
the form of the signal; for instance, if $\hogw (f)$ is flat,
as in many examples that we will find in later chapters, 
then $S_h(f)\sim 1/f^3$.
For accurate estimates of the SNR at a wideband detector there is no
substitute for a numerical integration of eq.~(\ref{SNR2}). However,
for  order of magnitude estimates, eq.~(\ref{rho5}) for $\hc$ together
eq.~(\ref{hn}) for $h_n(f)$ are simpler to use, and they have the
advantage of clearly separating the physical effect, which is
described by $\hc$, from the properties of the detectors, that enter
only in $h_n(f)$.

Eq.~(\ref{hn}) also shows  very clearly the advantage of correlating two
detectors compared with the use of a single detector. With a single
detector, the minimum  observable signal, at SNR=1,
is given by the condition
$S_h(f)\geq S_n(f)/F$. This means, from
eq.~(\ref{hc3}), a minimum detectable value for $h_c(f)$ given by
\be
h_{\rm min}^{\rm 1d}=\left(\frac{2fS_n(f)}{F}\right)^{1/2}\, .
\ee
The superscript 1d reminds
that this quantity refers to a single detector. From eq.~(\ref{hn}) we
find for the minimum detectable value with two interferometers 
in coincidence, 
$h_{\rm min}^{\rm 2d}=\hn$.
For close detectors of the same type, $\Gamma (f)\simeq F$, so that
\be\label{gain}
h_{\rm min}^{\rm 2d}(f)\simeq 
\frac{1}{(\frac{1}{2}T\Delta f)^{1/4}}\,\frac{1}{\sqrt{2}}
h_{\rm min}^{\rm 1d}(f)\simeq
1.12\times 10^{-2}h_{\rm min}^{\rm 1d}(f)\,
\left(\frac{\rm 1\, Hz}{\Delta f}\right)^{1/4}
\left(\frac{\rm 1\, yr}{T}\right)^{1/4}
\, .
\ee
Of course, the  reduction factor in the noise level is larger if
the integration time is larger, 
and if we increase the bandwidth $\Delta f$
over which a useful coincidence 
is possible.  Note that $\hogw$ is quadratic
in $h_c(f)$, so that an improvement in sensitivity by two orders of
magnitudes in $h_c$ means four orders of magnitude in $\hogw$.

\section{Detectors in operation, under construction, or planned}
\label{detectors}

In this section we will discuss the characteristic of the
detectors existing, under construction, or planned, and we will
examine the sensitivities to stochastic backgrounds
that can be obtained using them as
single detectors. The results will allow us to fully appreciate the
importance of correlating two or more detectors, 
at least as far as stochastic
backgrounds are concerned.

\subsection{The first generation of large 
interferometers: LIGO, VIRGO, GEO600,  TAMA300, AIGO}

\begin{figure}
\centering
\includegraphics[width=1.2\linewidth,angle=270]{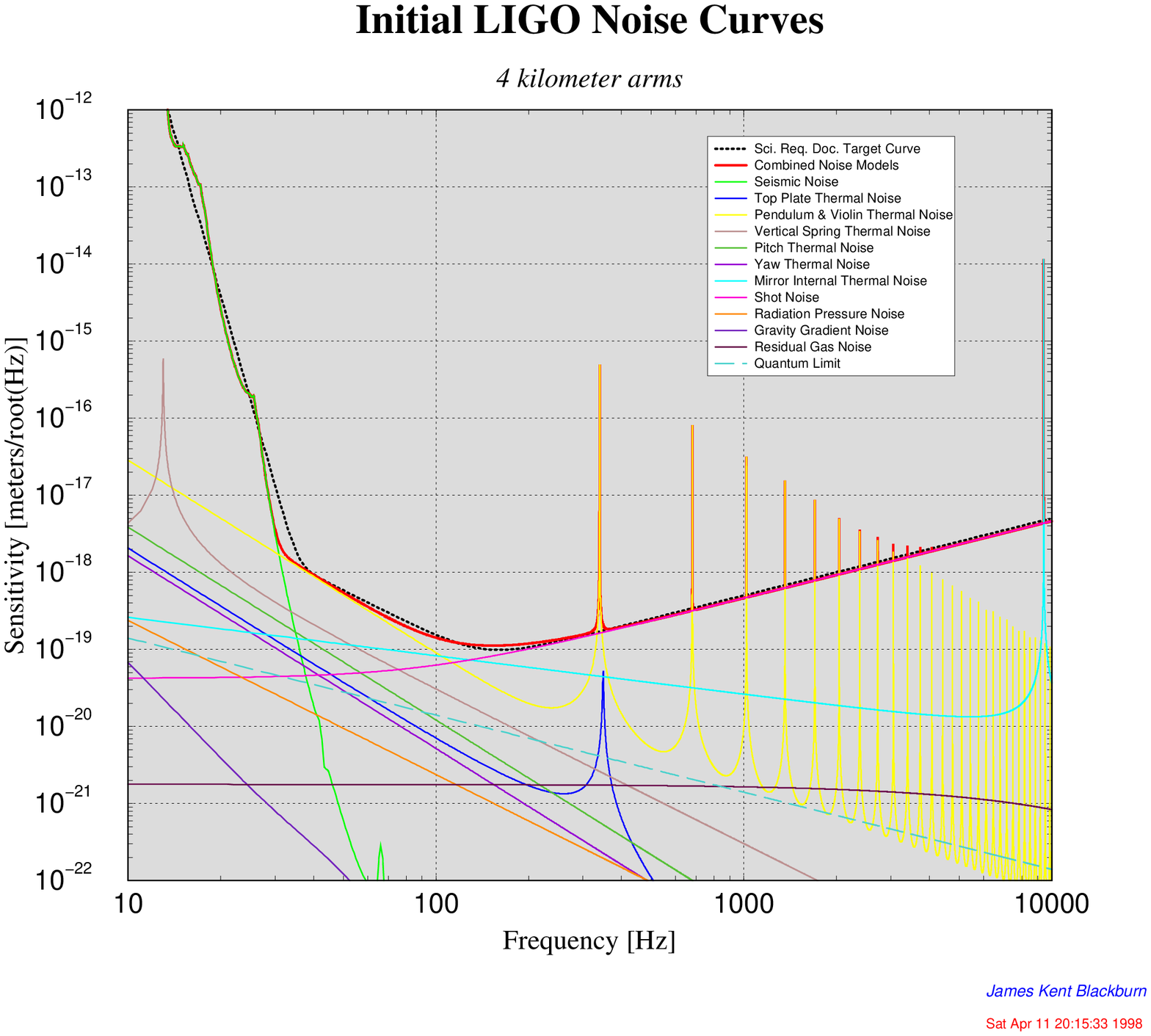}
\caption{The planned LIGO sensitivity curve, in meters$/\sqrt{{\rm Hz}}$
(courtesy of the LIGO collaboration).}
\label{4km_dhs}
\end{figure}

A great effort is being presently devoted to the construction of large
scale interferometers. The Laser Interferometer Gravitational-Wave
Observatory (LIGO) is being developed by an MIT-Caltech collaboration.
LIGO \cite{LIGO} consists of two widely separated interferometers, in
Hanford, Washington, and in Livingston, Louisiana. The commissioning
of the detectors will begin in 2000, and the first data run is
expected to begin in 2002~\cite{Bar}. The arms of both detectors will
be 4~km long; in Hanford there  will be  also a second 2~km
interferometer implemented in the same vacuum system. The sensitivity
of the 4~km interferometers is shown in fig.~\ref{4km_dhs}.
The sensitivity is here given in meters per root Hz; to obtain
$\tilde{h}_f$, measured in $1/\sqrt{{\rm Hz}}$, one must divide by the
arm length in meters, i.e. by 4000. The sensitivity in
$1/\sqrt{\rm Hz}$ is shown in fig.~\ref{ligoII}, 
next subsection, together with
the sensitivity of advanced LIGO.
The LIGO interferometer  already has a 40m prototype at Caltech, which
has been used to study sensitivity, optics, control, and even 
to do some
work on data analysis~\cite{40m}. 

\begin{figure}
\centering
\includegraphics[width=0.55\linewidth,angle=270]{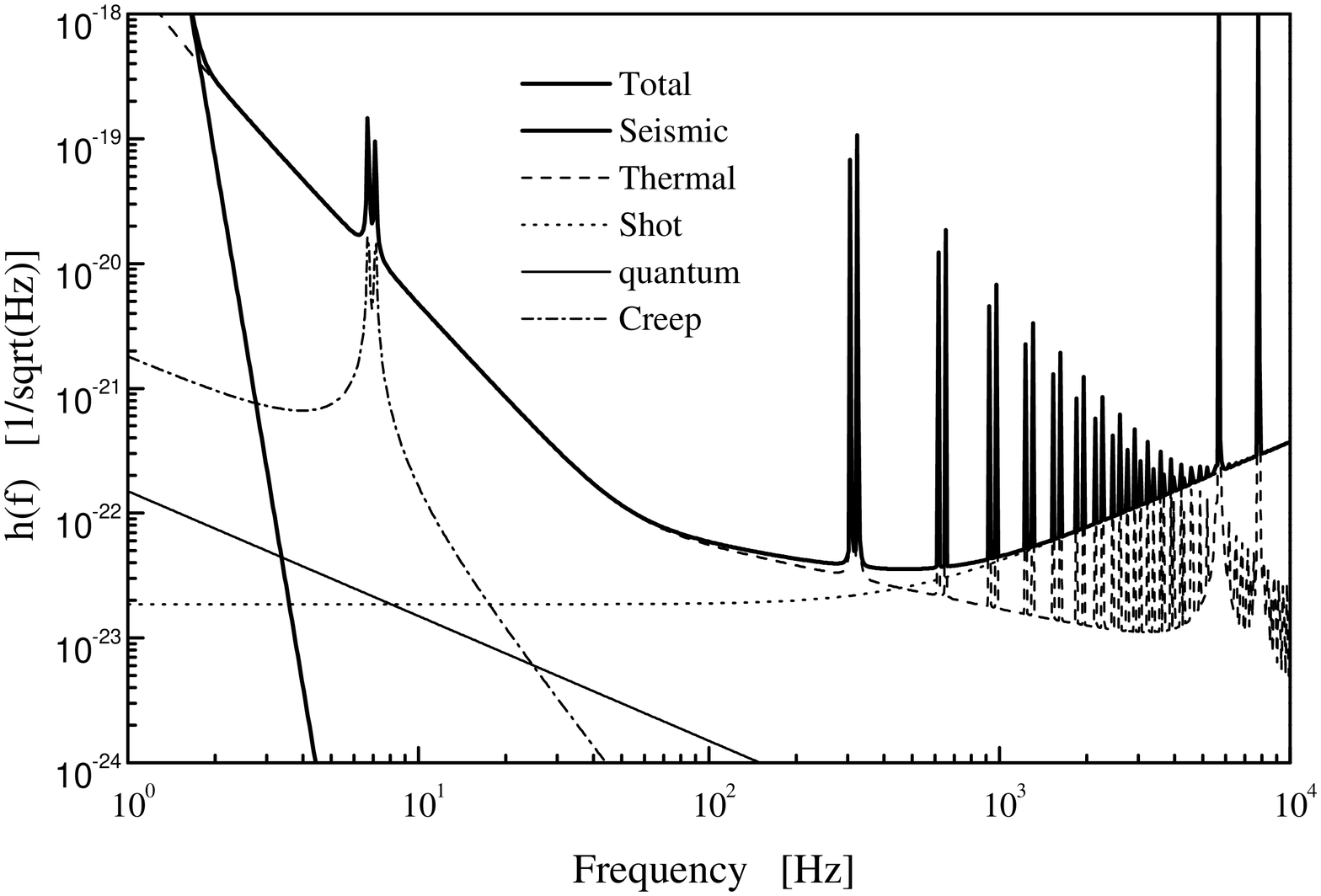}
\caption{The planned VIRGO sensitivity curve (from ref.~[16]).}
\label{virsens}
\end{figure}

A comparable interferometer is VIRGO~\cite{Car,finaldesign,Bri} 
which is being built by 
an Italian-French collaboration supported by INFN and
CNRS, with 3 km arm length, and
is presently under construction in Cascina, near Pisa. These
experiments are carried out by very large collaborations, comparable to
the collaborations of particle physics experiments. For instance,
VIRGO involves about 200 people.
Fig.~\ref{virsens} shows, in terms of $\tilde{h}_f$, 
the expected sensitivity of the VIRGO interferometer.
By the year
2000 it should be possible to have the first data from a 7 meters prototype,
which will be used to test the suspensions, vacuum tube, etc.

Interferometers are wide-band detectors, that will cover the region
between a few Hertz up to approximately a few
kHz. Figs.~(\ref{4km_dhs},\ref{virsens}) show
separately the main noise sources. At very low frequencies, $f<2$~Hz
for VIRGO and $f<40$~Hz for LIGO, the
seismic noise dominates and sets the lower limit to the frequency
band. Above a few Hz, the VIRGO superattenuator
reduces the seismic noise to a neglegible level.
Then, up to a few hundreds Hz, thermal noise
dominates, and finally the laser shot noise takes over.
The various spikes are mechanical resonances due to the thermal noise:
first, in fig.~\ref{virsens}, 
the high frequency tail of the pendulum mode (in this region
LIGO is still dominated by seismic noise), then  a series
of  narrow resonances due to
the violin modes of the wires; finally,
 the  two rightmost spikes in the figures are the low
frequency tail of the
resonances due to internal modes of the two mirrors. All resonances
appear in pairs, at frequencies close but non-degenerate, 
because the two mirrors have different masses. The width of these
spikes is of the order of fractions of Hz.
Note that the clustering of the various resonances in the kHz region
is due to the logarithmic scale in
figs.~(\ref{4km_dhs},\ref{virsens}). 
The resonances are actually
evenly spaced and narrow, so that in this frequency range the
relevant curve is mostly the one given by the shot noise. 

\begin{figure}
\centering
\includegraphics[width=0.8\linewidth,angle=0]{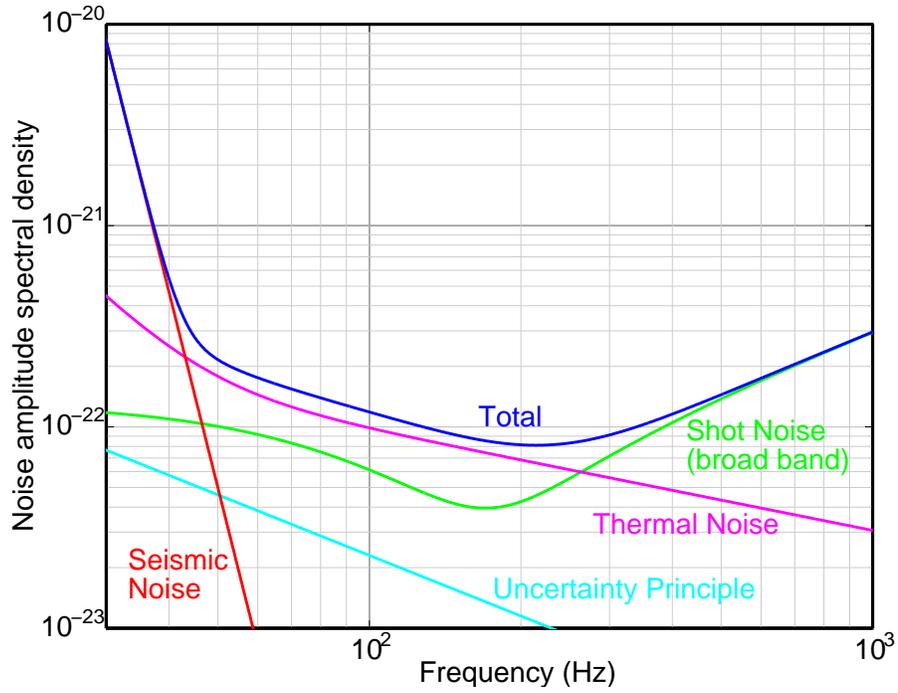}
\caption{The planned GEO sensitivity curve, without narrow-banding
 (courtesy of the GEO collaboration).}
\label{wideband}
\end{figure}

\begin{figure}
\centering
\includegraphics[width=0.8\linewidth,angle=0]{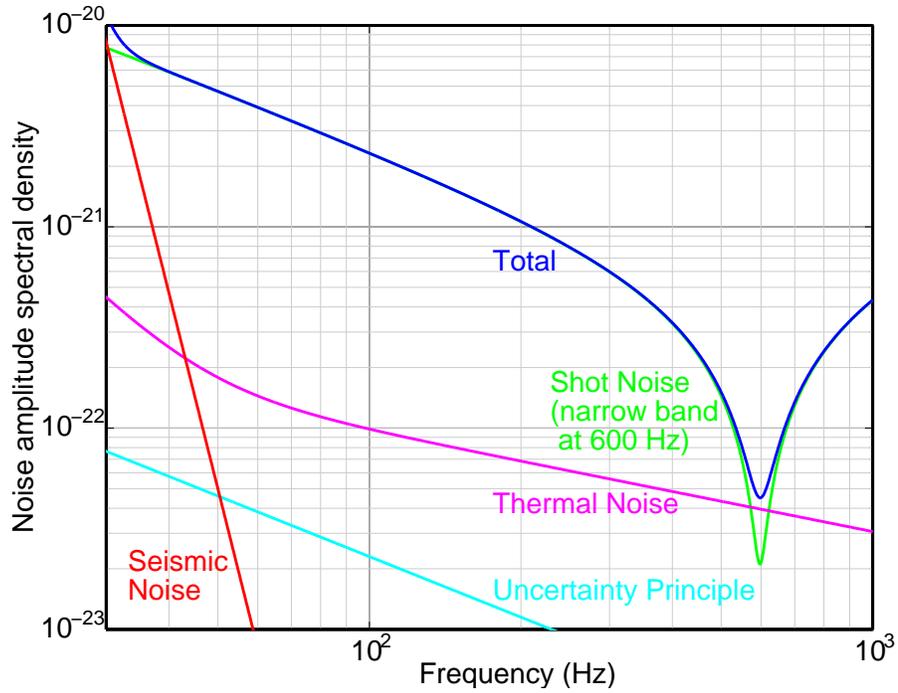}
\caption{The planned GEO sensitivity curve, with narrow-banding
at 600 Hz (courtesy of the GEO collaboration).}
\label{narrowband}
\end{figure}

Under construction is also
GEO600~\cite{GEO}, a collaboration between the
Max-Planck-Institut f\"{u}r Quantenoptik in Garching and the 
University of Glasgow. It is being built near Hannover, with arms 600
meters in length. GEO600 is somewhat smaller than LIGO and VIRGO, but
will use techniques that will be important for the advanced
detectors. In particular, the signal recycling of GEO600 provides an
opportunity to change the spectral characteristics of the detector
response, especially those due to the shot noise limitation, therefore
in the high frequency range of the available bandwidth. By choosing
low or high mirror reflectivities for the signal-recycling mirror, one
can use the recycling either to distribute the
improvement in the sensitivity on a wideband, or
to improve it even further on a narrow band. This is the so-called
narrow-banding of an interferometer.
The sensitivity of GEO600 broad-band is shown in
fig.~\ref{wideband}, and with narrow-banding at 600 Hz
in fig.~\ref{narrowband}.

The frequency of maximum sensitivity is tunable to the desired value by
shifting the signal-recycling mirror, and thus changing the resonance
frequency of the signal-recycling cavity.
In its tunable narrowband mode, GEO600 might well be able to set the
most stringent limits of all for a while.

TAMA300 (Japan)~\cite{Tama} is the other large interferometer. It is a
five year project (1995-2000) that aims to develop the advanced
techniques that will be needed for second generation experiments, and
catch GWs that may occur by chance within out local group of
galaxies. Its has 300 meters arm length. It is hoped that the project
will evolve into the proposed Laser Gravitational Radiation Telescope
(LGRT), which should be located near the SuperKamiokande detector.

Furthermore, Australian scientists have  joined forces to
form the Australian Consortium for Interferometric Gravitational
Astronomy (ACIGA)~\cite{Sand}. 
A design study and research with particular emphasis on an
Australian detector is nearing completion. The detector, AIGO, will be
located north of Perth. Its position in the southern emisphere will
greatly increase the baseline of the worldwide array of detectors, and
it will be  close to the resonant bar NIOBE so that correlations can
be performed.
The detector will use
sapphire optics and sapphire test masses, a material that LIGO plans
to use only in its advanced stage, see sect.~\ref{advL}

All these detectors will be used in a
worldwide network to increase the sensitivity and the reliability of a
detection.

Using the values of $\tilde{h}_f$ given in the figures
and the value  $1/F =2.5$ for interferometers, 
 we see  from \eq{single} that
VIRGO, GEO600 or any of the two LIGOs , used as  single detectors, can
reach a minimum detectable value for $\hogw$
of order $10^{-2}$ or at most a few times
$10^{-3}$, at $f=$ 100Hz. 
Unfortunately, at this level current theoretical expectations
exclude the possibility of a cosmological signal, as we will discuss
in sect.~\ref{boundsect} and \ref{gen}. As we will see, an interesting
sensitivity level for $\hogw$ should be at least
of order $10^{-6}$. To reach such a level with a single
 interferometer we need, e.g., 
\be
\tilde{h}_f (f=100\, {\rm Hz})< 10^{-24}\,\, {\rm
  Hz}^{-1/2}\, ,
\ee
or 
\be
\tilde{h}_f (f=1\, {\rm kHz})<   3\times 10^{-26}\, {\rm
  Hz}^{-1/2}\, .
\ee
We see from the figures
that such small  values of $\tilde{h}_f$ are
very far from the sensitivity of  first generation interferometers,
and are in fact
even well below the  limitation due to quantum noise.
Of course, one should stress that 
theoretical prejudices, however well founded, are no
substitute for a real measurement, and that even a  negative result
at the level $\hogw\sim 10^{-2}$ would be interesting.

\subsection{Advanced LIGO}\label{advL}

It is important to have in mind that the interferometers discussed in
the previous section  are the first
generation of large scale interferometers, and in this sense they
represent really a pioneering effort. At the level of sensitivity
that they will reach, they have no guaranteed source of detection. 
However, they will open the way 
to second generation interferometers, with much better sensitivity. 
GEO600 and TAMA300 are important for testing the technique that will
be needed for second generation experiments, while the larger
interferometers VIRGO and LIGOs should evolve into second generation
experiments. 

\begin{figure}
\centering
\includegraphics[width=\linewidth,angle=0]{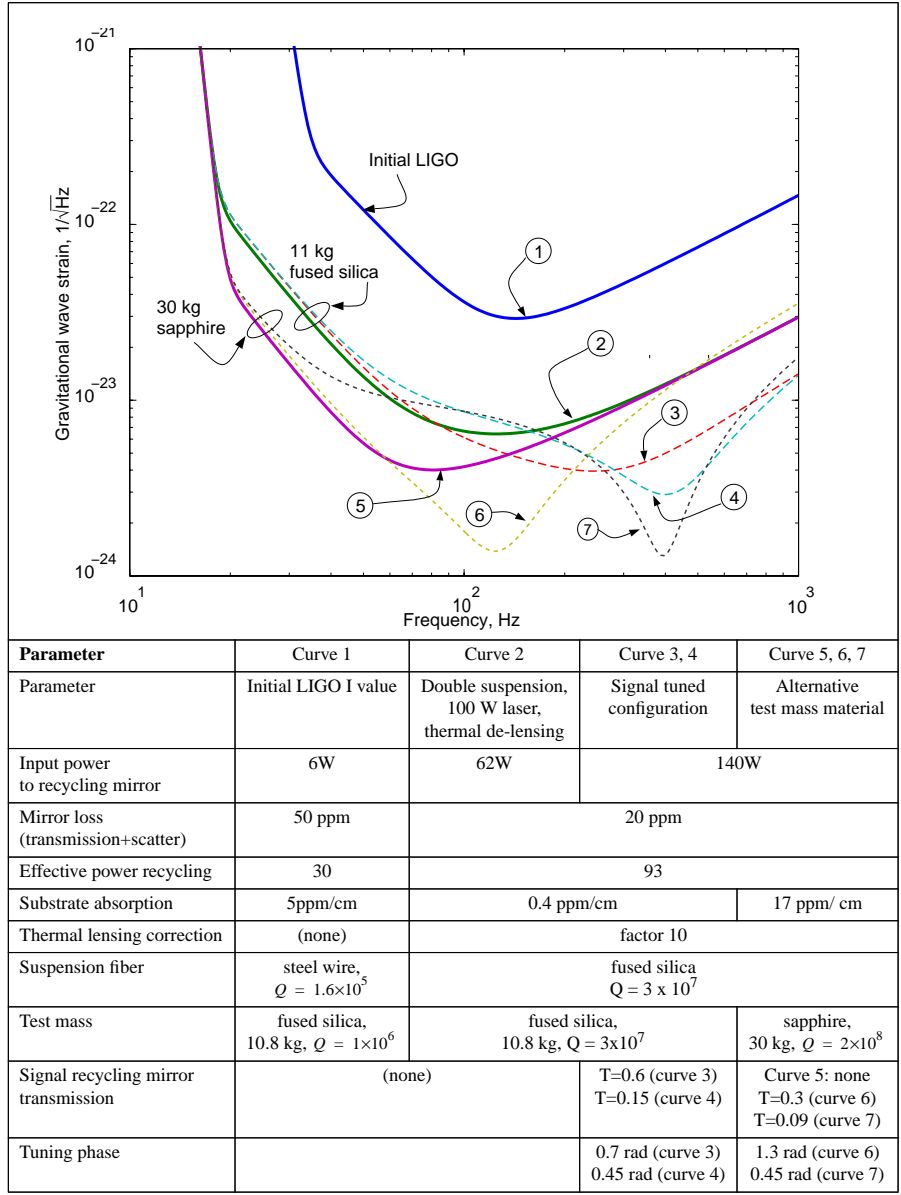}
\caption{The planned advanced LIGO sensitivity curves, and the
  parameters relative to the various improvements, from
  ref.~\cite{LSC} 
(courtesy of the LIGO collaboration).}
\label{ligoII}
\end{figure}

The LIGO collaboration has  presented a recommended program
for research and developement, that will lead to the Advanced LIGO
project~\cite{LSC}. The results are shown in fig.~\ref{ligoII},
together with the changes in a number of parameters that are
responsible for the various improvements shown in the figure. 

The improvement program is divided into three stages: LIGOII near
term, LIGOII medium term and LIGOIII. 

Curve 1 in fig.~\ref{ligoII} shows the sensitivity of LIGOI, while
LIGOII near term 
is shown in curve~2. The changes leading to LIGOII near term should
be incorporated by the end of 2004, and are based on engeneering
developents of existing technology or modest stretches from present
day systems. They would result in significant improvements of the
sensitivity, and are also necessary to gain full advantage of
subsequent changes. In particular, a significant reduction in thermal
noise will be obtained using fused silica fibers in the test mass
suspension; a moderate improvement in seismic isolation will move the
seismic noise below this new thermal noise floor, and an increase in
the laser power to $\sim 100$ watts results in a decrease of shot noise.

The goal of the medium term program are more
ambitious technically, and it is estimated that they could be
incorporated in LIGO by the end of 2006. They include signal
recycling. As discussed above, 
this could be used either to obtain a broad-band
redistribution of sensitivity (curve 3), or to  improve the
sensitivity in some narrow band (curve 4). 

Another possible improvement  is the change of the test masses from
fused silica to sapphire or other similar materials.  Sapphire has
higher density and higher quality factor, and this should result in a
significant reduction in thermal noise. For an optical configuration
without recycling, the sensitivity is given by curve~5. With signal
recycling, possible responses are shown in curves 6 and 7. These
changes require significant advances in material technology, and so
are considered possible but ambitious for LIGOII.

Fig.~\ref{ligoII} does not show performances for LIGOIII detectors,
which are expected to provide possibly another factor of 10 of
improvement in $\tilde{h}_f$, but of course are quite difficult to
predict reliably at this stage.

The overall improvement of LIGOII can be seen to be, depending on the
frequency, one or two orders of magnitude in $\tilde{h}_f$. This is
quite impressive, since
two order of magnitudes in $\tilde{h}_f$ means four order of
magnitudes in $\hogw (f)$ and therefore an extremely interesting
sensitivity, even for a single detector, without correlations.

\subsection{The space interferometer LISA}

The space interferometer LISA~\cite{LISA,LISA1,Hou}  
was   proposed
to the European Space Agency (ESA) in 1993,
in the framework of ESA's long-term space science program
Horizon 2000. The original proposal involved using laser interferometry
beteween test masses in four drag-free spacecrafts placed in a
Heliocentric orbit. In turn, this led to a proposal for a six
spacecrafts mission, which has been selected by the
European Space Agency has a cornerstone mission in its future science
program Horizon 2000 Plus. This implies that in principle the mission
is approved and that funding for industrial studies and technology
developement  is provided right away. The launch year however depends
on the availability of fundings. With reasonable estimates 
it is then expected that it will not be
launched before 2017, and possibly as late as 2023.
In Feb. 1997 the LISA team and ESA's Fundamental Physics Advisory
Group proposed to carry out LISA in collaboration with NASA. 
If approved, this
could make possible to launch it between 2005-2010.
The design has also been
 somewhat simplified,
with three drag-free spacecrafts. The
spacecrafts will be in a Heliocentric orbit, at a distance of 1 AU
from the Sun, 20 degrees behind the Earth. This design has lower
costs, and it is likely to be adopted for future studies relevant to
the project~\cite{Hou}. The mission is planned for two years, but it
could last up to 10 years without exausting on-board supplies. 

LISA has three arms, first of all for redundancy.
Thus,  it can be
thought of as two interferometers sharing a common arm. Of course,
this means that the two interferometers will have a common
noise. However, most signals are expected to have a signal-to-noise
ratio so high that the noise will be neglegible. Then, the output
from the two interferometers can be used to obtain extra informations
on the polarization and direction of a GW. For the stochastic
background, the third arm will help to discriminate  backgrounds
as those produced by binaries or by cosmological effects from anomalous
instrumental noise~\cite{LISA}. 

Goint into space, one is not limited
anymore by seismic and gravity-gradient noises; 
LISA could then explore the very low frequency
domain, $10^{-4}$~Hz $<f<1$~Hz. At the same time, there is also the
possibility of a very long path length (the mirrors will be freely
floating into the spacecrafts at distances of $5\times 10^{6}$ km from
each other!), 
so that the requirements on
the position measurment noise can be relaxed. The goal is to 
reach a strain sensitivity~\cite{LISA}
\be\label{lisastrain}
\tilde{h}_f=4\times  10^{-21} \,\, {\rm Hz}^{-1/2}
\ee
at $f=1$ mHz. At this level, one expects first of all signals from
galactic binary
sources, extra-galactic supermassive black holes binaries and
super-massive black hole formation.

Concerning the stochastic background, eq.~(\ref{single}) shows clearly the
advantage of going to low frequency:  the factor $f^3$ in 
\eq{single} gives an extremely low value for the minimum detectable
value of $\hogw$. The sensitivity of \eq{lisastrain} would correspond to
\be\label{-12}
\hogw (f=1\,{\rm mHz})\simeq 1\times 10^{-12}\, .
\ee
Since LISA cannot be correlated with any other detector, to have some
confidence in the result it is necessary to have a SNR sufficiently
large. Certainly one cannot work at SNR=1.65, as we will do when
considering the correlation between two detectors. The standard choice 
made by the LISA collaboration is SNR=5, and \eq{-12} refers to this
choice. 
Note that the minimum detectable value of $\hogw$ is proportional to 
(SNR)$^2$, since this SNR refers to the amplitude, and $\hogw
(f)\sim h_c^2(f)$.  Furthermore, \eq{-12}
takes into account  the angles between the arms, $\alpha
=60^o$, and the effect of the motion of LISA, which together
results in a loss
of sensitivity by a factor approximately equal to $1/\sqrt{5}$.
Eq.~(\ref{-12}) shows that LISA could reach a truly remarkable
sensitivity in $\hogw$. 

LISA will have its best sensitivity between 3 and 30 mHz. Above 30
mHz,  the sensitivity degrades because the wavelength of the GW
becomes shorter  than twice the arm-length of $5\times 10^6$ km. At
low frequencies, instead, the noise curve rises because of spurious
forces on the test masses. At some frequency below 0.1 mHz the
accelerometer noise will increase rapidly, and the instrumental
uncertainty would increase even more rapidly with decreasing
frequency, setting a lower limit to the frequency band of LISA.

\begin{figure}
\centering
\includegraphics[width=\linewidth,angle=0]{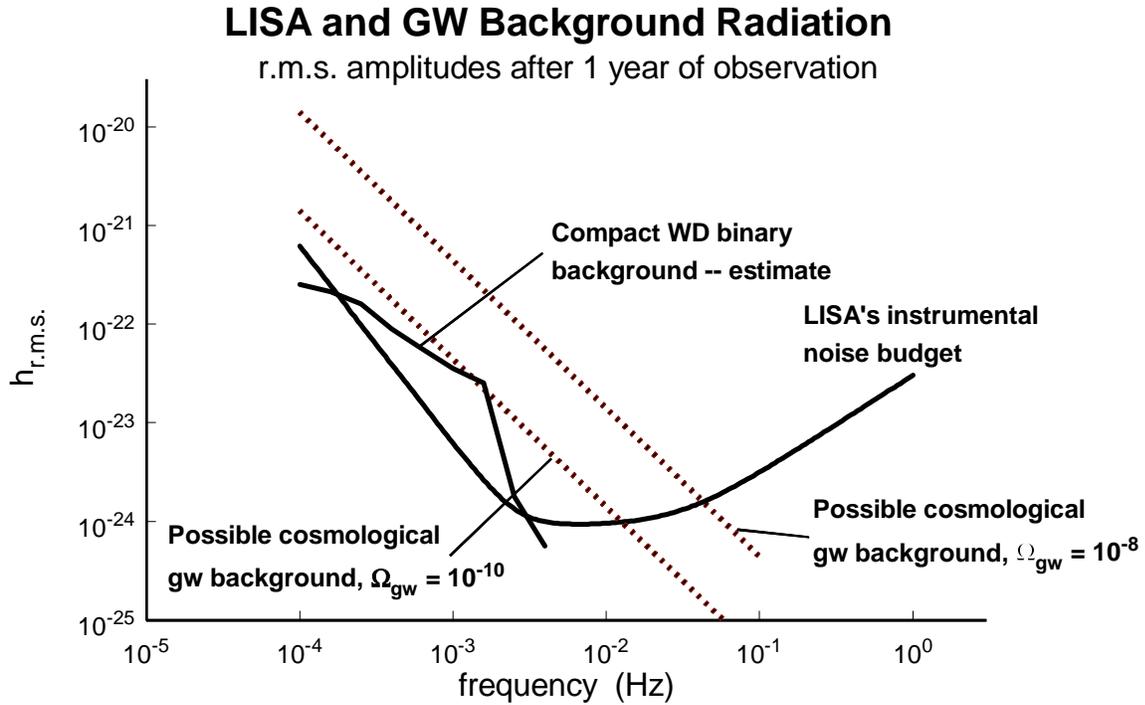}
\caption{The sensitivity of LISA to a stochastic background of GWs
  after one year of observation, and SNR=5
(from ref.~\cite{LISA}; courtesy of the LISA collaboration).}
\label{lisa}
\end{figure}

\begin{figure}
\centering
\includegraphics[width=0.55\linewidth,angle=270]{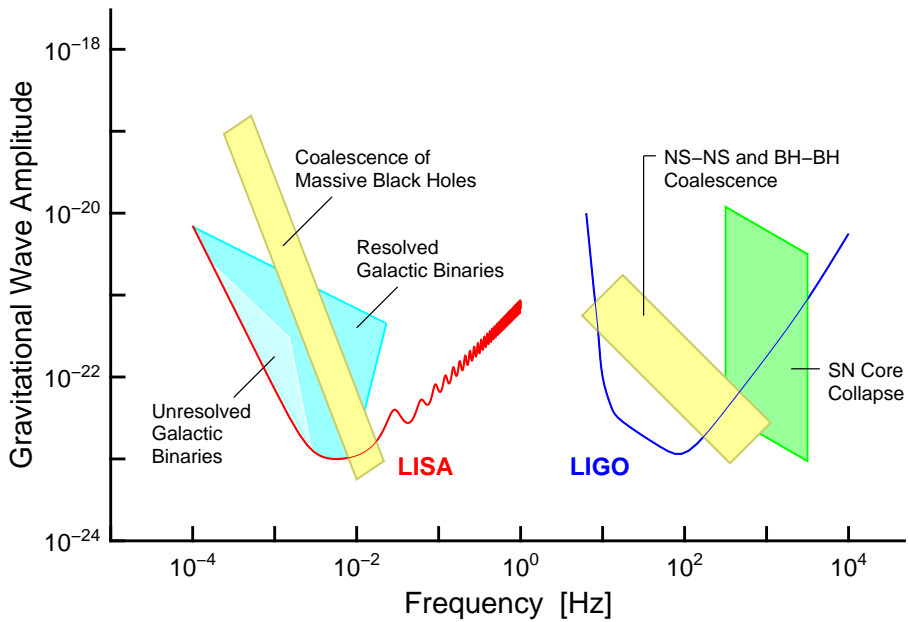}
\caption{The sensitivity of LISA for periodic sources, 
and of advanced LIGO for burst sources, together with some
  expected astrophysical signals
(from ref.~\cite{LISA}; courtesy of the LISA collaboration).}
\label{lisa2}
\end{figure}

Below a few mHz it is expected also a stochastic
background due to compact white-dwarf binaries, that could cover a
cosmological background. The sensitivity 
curve of LISA to a stochastic background, together with the
estimated white-dwarf binaries background, is shown in
fig.~\ref{lisa}. On the vertical axis is shown 
$h\sub{r.m.s.}\equiv
h_c(f,\Delta f=3\times 10^{-8}\, {\rm Hz})$,
defined in \eq{hrms2}, which is written in~\cite{LISA} in the
equivalent form
\be
h_c (f,\Delta f= 3\times 10^{-8}\,{\rm Hz} )
= 5.5\times 10^{-22}\(\frac{\ogw}{10^{-8}} \)^{1/2}
\( \frac{1\, {\rm mHz}}{f}\)^{3/2}
\( \frac{H_0}{75\,{\rm km\, s}^{-1}\,{\rm Mpc}^{-1}} \)\, .
\ee
Two lines of constant $\ogw$, equal to $10^{-10}$ and $10^{-8}$, are
also shown in the figure. 

To compare with interferometers, fig.~\ref{lisa2} shows the
sensitivity to the GW amplitude for both LISA and the advanced
LIGO, together with a number of signals expected from astrophysical
sources.  It is apparent from the figure that ground-based and
space-borne interferometers are complementary, and together can cover
a large range of frequencies, and, in case of a  detection of a
cosmological signal, together they can give crucial spectral informations.

The sensitivities shown are presented by the collaboration as
conservatives because~\cite{LISA}
\begin{enumerate}
\item The errors have been calculated realistically, including all
  substantial error sources that have been thought of since early
  studies of drag-free systems, and since the first was flown over 25
  years ago, and in most cases (except shot noise) the error allowance
  is considerably larger than the expected size of the error and is
  more likely an upper bound.
\item LISA is likely to have a significantly longer lifetime than one
  year, which is the value used to compute these sensitivities.
\item The sensitivity shown refers only to one interferometer. Using
  three arms could increase the SNR by perhaps 20\% .
\end{enumerate}

An interesting feature of LISA is that, as it rotates in its orbit,
its sensitivity to different directions changes. Then, LISA can test
the isotropy of a stochastic background. This can be quite important
to separate a cosmological signal from a stochastic background of
galactic origin, which is likely to be concentrated on the galactic
plane. By comparing two 3-month stretches of data, LISA should have no
difficulty in identifying this effect. Furthermore, if the mission
lasts 10yr, LISA will be close to the sensitivity needed to detect the
dipole anisotropy in a cosmological background due to the motion of the
solar system. If the GW background turned out not to have the same
dipole anisotropy as the cosmic microwave background, one would have
found evidence for anisotropic cosmological models.

\subsection{Resonant bars: NAUTILUS, EXPLORER, AURIGA, ALLEGRO, NIOBE}
\begin{figure}
\centering
\includegraphics[width=0.8\linewidth,angle=0]{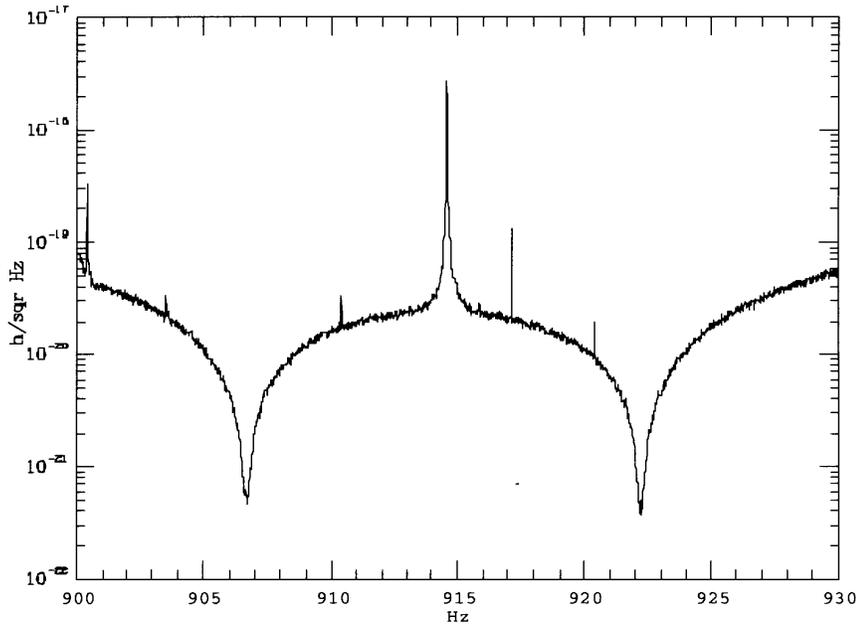}
\caption{The sensitivity curve of NAUTILUS (courtesy of the NAUTILUS
  collaboration) }
\label{10_8_98}
\end{figure}

Cryogenic resonant antennas have been taking date since 1990
(see e.g. ref.~\cite{Piz} for review).
Fig.~\ref{10_8_98}
 shows the sensitivity of NAUTILUS~\cite{Nau1,Nau2,Nau,Nau0}, 
the  resonant bar
located in Frascati, near Rome. It is an ultracryogenic detector,
operating at a temperature  $T=0.1$ K. This figure is based on a two hour
run, but the behaviour of the apparatus is by now quite stationary
over a few days,
with a duty cycle limited to 85\% by cryogenic operations.
Compared to the interferometers, we see that bars are narrow-band
detectors, and work at two resonances. The bandwidth is limited
basically by the noise in the amplifier.
The value of the resonance
can be slightly tuned, at the level of a few Hz, working on the
electronics. In the figure, the resonances are approximately
at 907 Hz and 922 Hz, with half-height bandwiths of about 1~Hz. 
At these frequencies the
strain sensitivity is 
\be
\tilde{h}_f\simeq 5\times 10^{-22}\, ,
\ee
about a factor of 5 higher than
the sensitivity of the interferometers at the same frequencies. 

With a thermodynamic temperature of 0.1~K, the experimental data are
in very good agreement with a gaussian distribution, whose variance
gives an effective temperature $T\sub{eff}=4.1$~mK. The effective
temperature $T\sub{eff}$ is related to the thermodynamic temperature
$T$ by $T\sub{eff}\simeq 4T\Gamma^{1/2}$, where $\Gamma$ is inversely
proportional to the quality factor $Q$ of the resonant mode, so that
$\Gamma\ll 1$. 

With improvements in the electronics,
the collaboration plans to reach in a few years the target sensitivity 
\be\label{sensbar}
\tilde{h}_f\simeq 8.6\times 10^{-23}\, {\rm Hz}^{-1/2}\, ,
\ee
and 
$T\sub{eff}\sim 10\mu$K, over a bandwidth of 10 Hz~\cite{Nau2,Nau}.

The cryogenic resonant bar EXPLORER~\cite{Exp} 
is in operation at Cern since
1990, cooled at a temperature $T=2.9$~K,  and
has  taken  data at approximately the same frequencies as NAUTILUS, 
905 and 921 Hz; its sensitivity is approximately the same as
NAUTILUS. 

Similar sensitivities are obtained  
by ALLEGRO, (Louisiana)~\cite{Allegro,Ham}. 
ALLEGRO has been taking data since 1991. The resonances are at 897 and
920 Hz. 

The resonant bar NIOBE~\cite{Niobe1,Niobe2,Niobe3,Niobe4} 
is located in Perth, Australia, and
it has operated since 1993. 
The resonant frequencies are at  694.6 Hz
and 713 Hz. It is made of
niobium, instead of aluminium as the other bars, since it has a higher
mechanical quality factor. This has allowed to reach a noise
temperature of less then 2mK, while the typical noise temperature is
around 3mK. Improvement in the transducer system
should allow to lower $T\sub{eff}$ to a few microkelvin, in a
bandwidth of 70~Hz. This would allow to reach a sensitivity comparable
to interferometers over a large bandwidth, approximately 650-750
Hz~\cite{Niobe4}.

AURIGA~\cite{Cer,Aur1,Aur2} is located in Legnaro, near Padua, Italy.  It
began full operations in 1997. 
The bar is cooled at 0.2K, and the
effective temperature is $T\sub{eff}=7$~mK. The resonances are at 911
and 929 Hz. The spectral sensitivity reached~\cite{Aur2} 
is  similar to that
of NAUTILUS.

All these bars are designed to operate in a well coordinated way,
in order  to improve the chances of reliable 
detection\footnote{There is an {\em International Gravitational
    Event Collaboration}, an agreement between the bar detector
  experiments presently in operation, signed at CERN on  July 4,
  1997.}.

From \eq{sensbar} we see that, without performing correlations
between different detectors, the  sensitivity that 
one could get from resonant bars is between at most  
$\hogw^{\rm min}(f)\sim O(1)$ and $\hogw^{\rm min}(f)\sim O(10)$
 and therefore quite far from a level where one can expect a
cosmological signal.
In sect.~\ref{5} we will discuss the level that can be reached with
correlations between bars, and between a bar and an interferometer.

\subsection{Some projects at a preliminary stage}

A number of other projects for gravitational wave detection have been
put forward and are  presently at the stage of  testing  prototypes.

An especially interesting project involves
resonant mass detectors with  truncated icosahedron
geometry~\cite{TIGA,MJ1,MJ2,CLO,BBCFL,Stev}. 
A spherical detector, in fact, has a
larger mass compared to a bar with the same resonance frequency, and
therefore a larger cross section. Furthermore, it has the informations about
the directions and polarization of a GW that could only be obtained
with 5 different resonant bars. The problem of attaching mechanical
resonators to the detector suggests a truncated icosahedral geometry,
rather than a sphere.

A prototype called TIGA (Truncated Icosahedral Gravitational Wave
Antenna) has been built at the Lousiana State University~\cite{MJ1}.  
It has its first resonant mode at 3.2 kHz.
The Rome group has started the SFERA project~\cite{SPHERA}.
A working group has been formed  to
to carry out studies and
measurements in order to define a project of a large spherical 
detector, 40 to
100 tons of mass, competitive with large interferometers but 
with complementary features. A correlation between an
interferometer  and a sphere  could in particular have
quite interesting sensitivities for the stochastic
background.

These spheres  could reach 
approximately one order of
magnitude better in the sensitivity $\tilde{h}_f$~\cite{CLO,Lob}
compared to resonant bars.
Furthermore, they have different channels, see \eqs{harm}{sphere}, that
are sensitive to different spin content~\cite{For2,Wag,TIGA}. This would
allow  to discriminate between different theories of gravity,
separating for instance the 
effects of scalar GWs, which could originate from a
Brans-Dicke theory~\cite{Wag1,Wag2,Lob,BCFF,BBCFL,ANic,MN}. Scalars fields
originating from string theory, as the dilaton and the moduli of 
compactifications, are however a much more difficult target:
to manifest
themselves as coherent
scalar GWs they should be extremely light, $m< 10^{-12}$
eV~\cite{MN}. 
Actually, there is the possibility to keep the dilaton and moduli fields
extremely light or even massless,  
 with a mechanism that has been proposed 
by Damour and Polyakov \cite{DP}. In fact, 
assuming some
form of universality in the string loop corrections, it is possible to
stabilize a massless 
dilaton during the cosmological evolution, at a value where
it is essentially decoupled from the matter sector. In this case,
however, the dilaton becomes  decoupled also from the detector, 
since the dimensionless coupling of the dilaton to matter 
($\alpha$ in the notation of~\cite{DP}) is smaller
than $10^{-7}$ (see also~\cite{DE2}). Such a dilaton
would then be  unobservable at VIRGO, although it
could still produce a number of small deviations from General
Relativity which might in principle be observable improving by several
orders of magnitude the experimental tests of the equivalence
principle~\cite{Dam}.

Recently, hollow spheres have been proposed in ref.~\cite{Coc3}.
The theoretical study suggests a very interesting value for
the strain sensitivity, even of order
\be
\tilde{h}_f\sim\, {\rm a\, few}\times 10^{-24}\, .
\ee
The resonance frequency, depending on the material used and 
other parameters, can be between approximately 200 Hz and 1-2 kHz, and
the bandwidth can be of order 20~Hz.

Another idea which is discussed is an array of resonant
masses~\cite{FraP,Fra,Fin}, each with a different frequency
$f$, and  a bandwidth $\Delta f\sim f/10$,
which together would cover the region from below the kHz up to a few
kHz.

Although not of immediate applicability to GW search, we finally mention
the  proposal  of using two coupled 
superconducting microwave cavities to detect very small
displacements.  The idea goes back to the works 
\cite{PPR,PR,PRBP,Cav,IPPR}. 
These detectors have actually been
constructed in ref.~\cite{RRM}, using two coupled  cavities with
frequencies of order of 10GHz; the coupling induces a shift in the
levels of order of 1~MHz, and transitions between these levels could
in principle  be induced by  GWs. However, \eq{single} shows clearly
the problem with working at such a high frequency, $f\sim 1$ MHz: to
reach an interesting value for $\hogw (f)$ of order $10^{-6}$, one
needs $\tilde{h}_f\sim 10^{-30}$ Hz$^{-1/2}$, 
very far from present technology. 
There is however the proposal~\cite{BGPP} to repeat the experiment
with improved sensitivity, so to reach
$\tilde{h}_f\sim 6\times 10^{-24}$ Hz$^{-1/2}$ at $f=1$ MHz, and, if
this should work, then one would try to lower the resonance frequency
down to, possibly, a few kHz. This second step would mean to move
toward much larger cavities, say of the kind used at LEP.

\section{Sensitivity of various two-detectors correlations}\label{5}
In the previous section we have seen that with a single detector 
(and with the exception of LISA) we
cannot reach sensitivities interesting for
cosmological backgrounds of GWs. On the other hand, we have found in
sects.~(\ref{3.5}) and (\ref{3.6}) that many orders of magnitudes can be
gained correlating two detectors. In this section we examine the
sensitivities for various two-detector correlations. 

It is also important to stress that performing multiple detector
correlations is crucial not only for improving the sensitivity, but
also because in a GW detector there are in general noises which are
non-gaussian, and can only be eliminated correlating two or more
detectors. An example of a non-gaussian noise in an interferometer 
is the creep, i.e. the sudden energy release in the
superattenuator chain, and most importantly in the wires holding the
mirrors, and due to accumulated internal stresses in the
material. Indeed, it is feared that this and similar non-stochastic
noises might turn out to be the effective sensitivity limit of
interferometers, possibly exceeding the design
sensitivity~\cite{DeS}. Some sources of non-stochastic noises have
been identified, and there are technique for neutralizing or
minimizing them. However, it is impossible to guarantee that all
possible sources of non-stochastic noises have properly been taken
into account, and only the cross correlation between different
detectors can provide a reliable GW detection. 

\subsection{An ideal two-interferometers correlation}\label{ideal}

We now discuss the sensitivities that could be obtained correlating
the major interferometers.\footnote{ 
Correlations between two interferometers have
already been carried out  using prototypes operated by the groups 
in Glasgow and at the
Max Planck Institute for Quantum Optics, with an effective coicident
observing period of 62 hours~\cite{Nic}. Although the  sensitivity  of
course is not yet significant, they demonstrate the possibility of
making long-term coincident observations with interferometers.}

In order to understand what is the best result that could be obtained
with present interferometer technology, 
we first consider the
sensitivity that could be obtained if one of the interferometers under
constructions 
were correlated with a second identical interferometer
located at a few tens of kilometers from the first, and with the same
orientation. This
distance would be optimal from the point of view of the stochastic
background, since it should be sufficient to decorrelate local noises
like, e.g., the seismic noise and local electromagnetic disturbances,
but still the two interferometers would
be close enough so that the overlap reduction
function does not cut off the high
frequency range.  For this exercise
we use the data for the sensitivity of VIRGO, fig.~\ref{virsens}.

Let us first give a rough estimate of 
the sensitivity using $h_c(f),h_n(f)$.
From fig.~\ref{virsens} we see that we can take, for our estimate, 
$\tilde{h}_f\sim 10^{-22} {\rm Hz}^{-1/2}$ over a bandwidth
$\Delta f\sim$ 1 kHz. Using $T=$ 1~yr, eq.~(\ref{hn}) gives
\be
h_n(f)\sim 4.5\times 10^{-24}
\left(\frac{f}{100{\rm Hz}}\right)^{1/2}
\left(\frac{\tilde{h}_f}{10^{-22} {\rm Hz}^{-1/2}}\right)\, .
\ee
We require for instance  SNR=1.65 (this corresponds to $90\%$
confidence level;  a more precise discussion of the
statistical significance, including the effect of the false alarm rate
can be found in ref.~\cite{AR}). Then the minimum detectable value of
$\hc$ is $h_c\su{min}(f)=1.65 h_n(f)$, and
from \eq{rho5} we get an estimate for the minimum detectable value
of $\hogw (f)$,
\be\label{min}
\hogw^{\rm min}(f)\sim 3\times 10^{-7}
\left(\frac{f}{100{\rm Hz}}\right)^{3}
\left(\frac{\tilde{h}_f}{10^{-22} {\rm Hz}^{-1/2}}\right)^2\, .
\ee
This suggests that correlating two VIRGO interferometers we can detect
a relic spectrum with 
$\hogw ({\rm 100 Hz})\sim 3\times 10^{-7}$ at SNR=1.65, or
$1\times 10^{-7}$ at SNR=1.
Compared to the case of a single interferometer with SNR=1, 
eq.~(\ref{single}),
we gain five orders of magnitude.
As already
discussed, to obtain a precise numerical value one must however resort
to eq.~(\ref{SNR3}). This involves an integral over all frequencies,
(that replaces the somewhat arbitrary choice of $\Delta f$ made above)
and depends on the functional form of $\hogw (f)$. If for instance
$\hogw (f)$ is independent of the frequency, using the numerical
values of $\tilde{h}_f$
plotted in fig.~\ref{virsens}
and performing the numerical integral, we have found for the minimum
detectable value of $\hogw$
\be\label{min2}
\hogw\su{min}\simeq 2 \times 10^{-7}\left(\frac{\rm
SNR}{1.65}\right)^2
\left(\frac{1\rm yr}{T}\right)^{1/2}
\hspace*{15mm} (\hogw (f)={\rm const.})\, .
\ee
This number is quite consistent with the approximate
estimate~(\ref{min}), and with the value $2\times 10^{-7}$
reported in ref.~\cite{CS}. Stretching the parameters to
SNR=1 ($68\%$ c.l.) and $T=4$ years, the value goes down
at $(3-4)\times 10^{-8}$. This might be considered an absolute
(and quite optimistic) upper bound on
the capabilities of first-generation experiments.
 
It is interesting to note that the main contribution to the integral
comes from the region $f<$ 100 Hz. In fact, neglecting the
contribution to the integral of the region $f>$ 100 Hz, the result for
$\hogw^{\rm min}$ changes only by approximately $2\%$. Also, the lower
part of the accessible frequency range is not crucial. Restricting for
instance to the region 20 Hz $\leq f \leq$ 200 Hz, the sensitivity on
$\hogw$ degrades  by less than  $1\%$, while restricting
to the region 30 Hz $\leq f \leq$ 100 Hz, the sensitivity on
$\hogw$ degrades  by approximately $10\%$.
Then, from fig.~\ref{virsens} we conclude 
that by far the most important source of noise for the
measurement of a flat stochastic background is the 
thermal noise. In particular, {\em the sensitivity to a 
flat stochastic background
is limited basically by the  mirror thermal
noise}, which dominates in the region 40 Hz$\lsim f\lsim 200$ Hz, while
the  pendulum thermal noise  dominates below approximately 
40 Hz.

The sensitivity depends however  on the functional form of 
$\ogw (f)$. Suppose for instance that
in the VIRGO frequency band we can  approximate the signal as
\be
\ogw (f)=\ogw ({\rm 1kHz})\left(\frac{f}{\rm 1 kHz}\right)^{\alpha}
\, .
\ee
For $\alpha =1$ we find that
the spectrum is detectable at SNR=1.65 if $\hogw ({\rm 1kHz})\simeq
3.6\times 10^{-6}$. For $\alpha =-1$ we find (taking $f=$ 5Hz as
lower limit in the integration)
$\hogw ({\rm 1kHz})\simeq 6\times 10^{-9}$. Note however that in
this case, since $\alpha <0$, the spectrum is peaked at low
frequencies, and  $\hogw (5{\rm Hz})\simeq 1\times 10^{-6}$. So,
both for increasing or decreasing spectra, to be detectable
$\hogw$ must have a peak value, within the VIRGO band, of order
a few $\times 10^{-6}$ 
in the case $\alpha =\pm 1$, while a constant spectrum can
be detected at the level $2\times 10^{-7}$.
Clearly, for detecting 
increasing (decreasing) spectra,  the upper (lower) part
of the frequency band becomes more important, and this is the reason
why the sensitivity  degrades  compared to flat spectra, since 
for increasing or decreasing spectra the maximum of the signal is at
the edges of the accessible frequency band, where the interferometer
sensitivity is worse.

\subsection{LIGO-LIGO}
Let us now see what can be done with existing interferometers. 
The two LIGO detectors are under construction at a
large distance from each other, $d\sim 3000$ km. This choice optimizes
the possibility of detecting the direction of arrival of GWs from
astrophysical sources, but it is not optimal from the point
of view of the stochastic background, since the overlap reduction
function cuts  off the integrand in eq.~(\ref{SNR2})  at a frequency
of the order of $1/(2\pi d)$. The overlap function $\gamma (f)$ for
the LIGO-LIGO correlation has been computed in ref.~\cite{Fla}.
and it has its first zero
at $f\simeq 64$~Hz. Furthermore, the arms of the two
detectors are not exactly parallel, and therefore $|\gamma (0)|=0.89$
rather than 1.

The sensitivity to a stochastic background  for the LIGO-LIGO correlation
has been computed in refs.~\cite{Mic,Chr,Fla,All,AR}. The result,
as we have discussed, depends on the functional form of $\hogw$. For 
$\ogw (f)$ independent of $f$, the minimum detectable value is
\be\label{LigoLigo}
\hogw\simeq 5\times 10^{-6}
\ee
for the initial LIGO. However, we have seen that second generation
interferometers could result in much better sensitivities.
The sensitivity of the correlation between two advanced LIGO is
estimated in ref.~\cite{All} to be
\be
\hogw\simeq 5\times 10^{-11}\, , 
\ee
which is an extremely interesting level.
These numbers are given at $90\%$ c.l. in~\cite{All}, and a 
detailed analysis of the statistical significance is given in~\cite{AR}.

\subsection{VIRGO-LIGO, VIRGO-GEO, VIRGO-TAMA}

\begin{figure}
\centering
\includegraphics[width=\linewidth,angle=0]{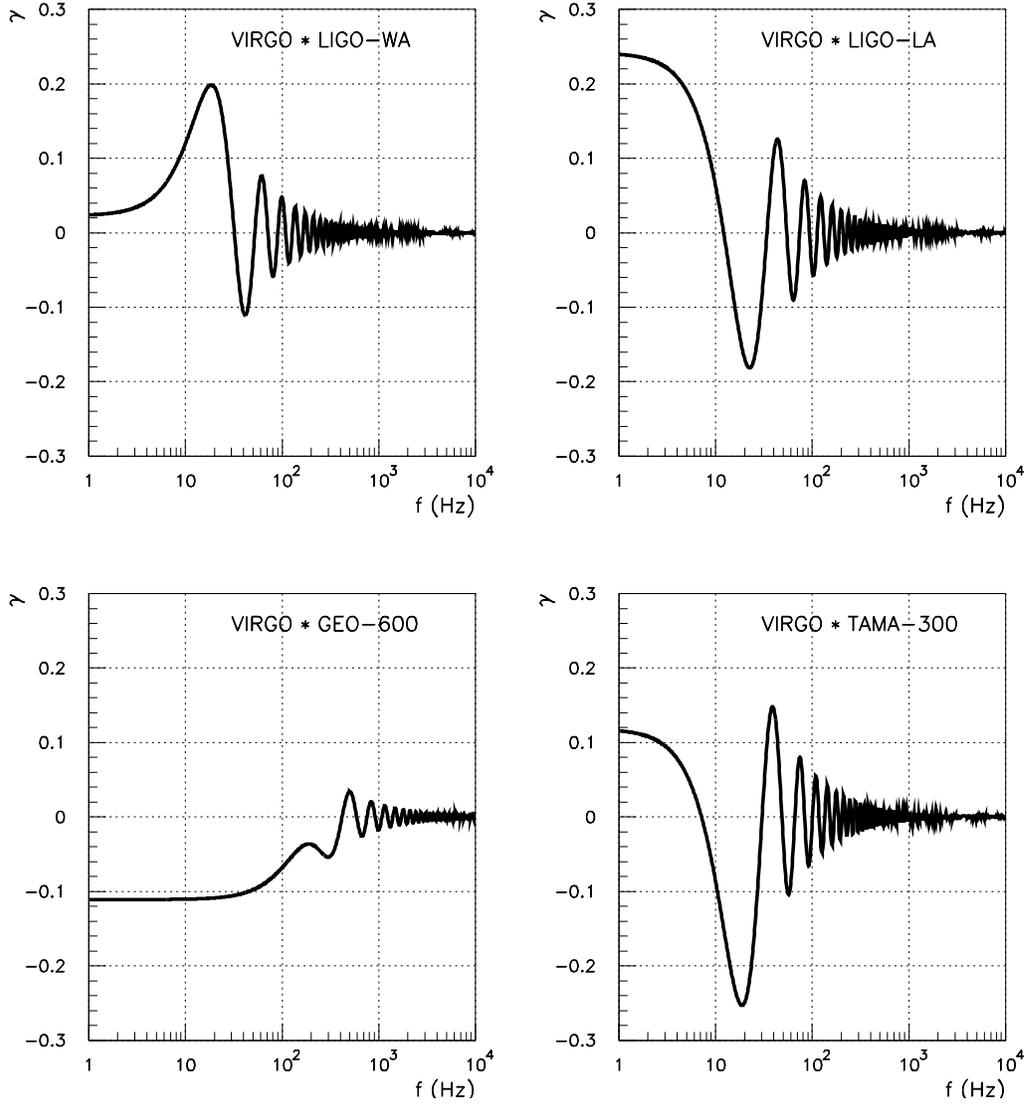}
\caption{The overlap reduction functions $\gamma (f)$
for the correlation of VIRGO
  with the other major interferometers (from ref.~\cite{dad}).}
\label{dad1}
\end{figure}

In Fig.~\ref{dad1}  we show the overlap reduction
functions for the correlation of VIRGO with the other major
interferometers. Using these functions, one can compute numerically
the integral in  \eq{SNR3} and obtain the minimum detectable value of
$\hogw$. These values are 
shown in Table~1 (from ref.~\cite{dad}), together with
the value for LIGO-LIGO computed
in ref.~\cite{All}. All these numbers are at
 90\% confidence level.

\begin{table}
\begin{center}
\begin{tabular}{|l|l|}\hline
correlation      & $\hogw$             \\ \hline\hline
LIGO-WA*LIGO-LA  & $5\times 10^{-6}$ \\ \hline
VIRGO*LIGO-LA    & $4\times 10^{-6}$ \\ \hline
VIRGO*LIGO-WA    & $5\times 10^{-6}$ \\ \hline
VIRGO*GEO600     & $5.6\times 10^{-6}$ \\ \hline
VIRGO*TAMA300    & $1\times 10^{-4}$ \\ \hline
\end{tabular}
\caption{The sensitivity of various two-interferometers correlation.}
\end{center}
\end{table}

We see that the correlation between VIRGO and any of
the two LIGO is  suppressed by the overlap reduction function
above, say, 30-40 Hz. 
In the case of the VIRGO-GEO correlation, instead,
$\gamma (f)$ cuts the integrand only above say 200  Hz. However, the
sensitivity of GEO is below 200 Hz is
lower then LIGO, so that the result are basically
the same as for VIRGO-LIGO, see Table~1.

To obtain the precise numbers for the sensitivity, of course one has
to perform the integral in \eq{SNR3}. However, it is easy to have an
understanding of the numbers that come out.
As we have found  in sect.~\ref{ideal},
the minimum detectable value of $\hogw$, using two identical VIRGO
detectors, 
degrades only by less than 10\% if, from the full VIRGO bandwidth,
we restrict to  the region 20 Hz $<f<$ 100 Hz; in this region
$|\gamma (f)|$ for the VIRGO-GEO correlation
is constant to a good accuracy, and of order 0.1. 
From \eqs{SNR3}{ooo} we see that the minimum detectable value of
$\hogw$ scales with $\gamma (f)$ as
$|\gamma (f)|^{-1}$. Therefore for the VIRGO-GEO correlation we get
about a factor of 10 worse than the ideal result~(\ref{min2}).
Comparing figs.~\ref{virsens} and \ref{narrowband}, 
and recalling that $\hogw\su{min}\sim\tilde{h}_f^2$, we see
that we lose about
another factor of 3 in $\hogw$ compared to an ideal 
VIRGO-VIRGO correlation, so that one
gets the estimate
\be
\hogw^{\rm min}\sim 6 \times 10^{-6}\left(\frac{\rm
SNR}{1.65}\right)^2
\left(\frac{1\rm yr}{T}\right)^{1/2}
\hspace*{15mm} (\hogw (f)={\rm const.})\, ,
\ee
quite in agreement with the result of the more accurate computation
shown in Table~1.

The situation with the VIRGO-TAMA300 correlation is instead  
worse, as can be seen from the overlap reduction function and from the
value in Table~1.

\subsection{VIRGO-Resonant mass and Resonant mass-Resonant mass}

The correlation between two resonant bars
 and between a bar and an interferometer
has been considered  in  refs.~\cite{VCCO,CS,ALS,APP,AFPR,dad}. 

\begin{figure}
\centering
\includegraphics[width=\linewidth,angle=0]{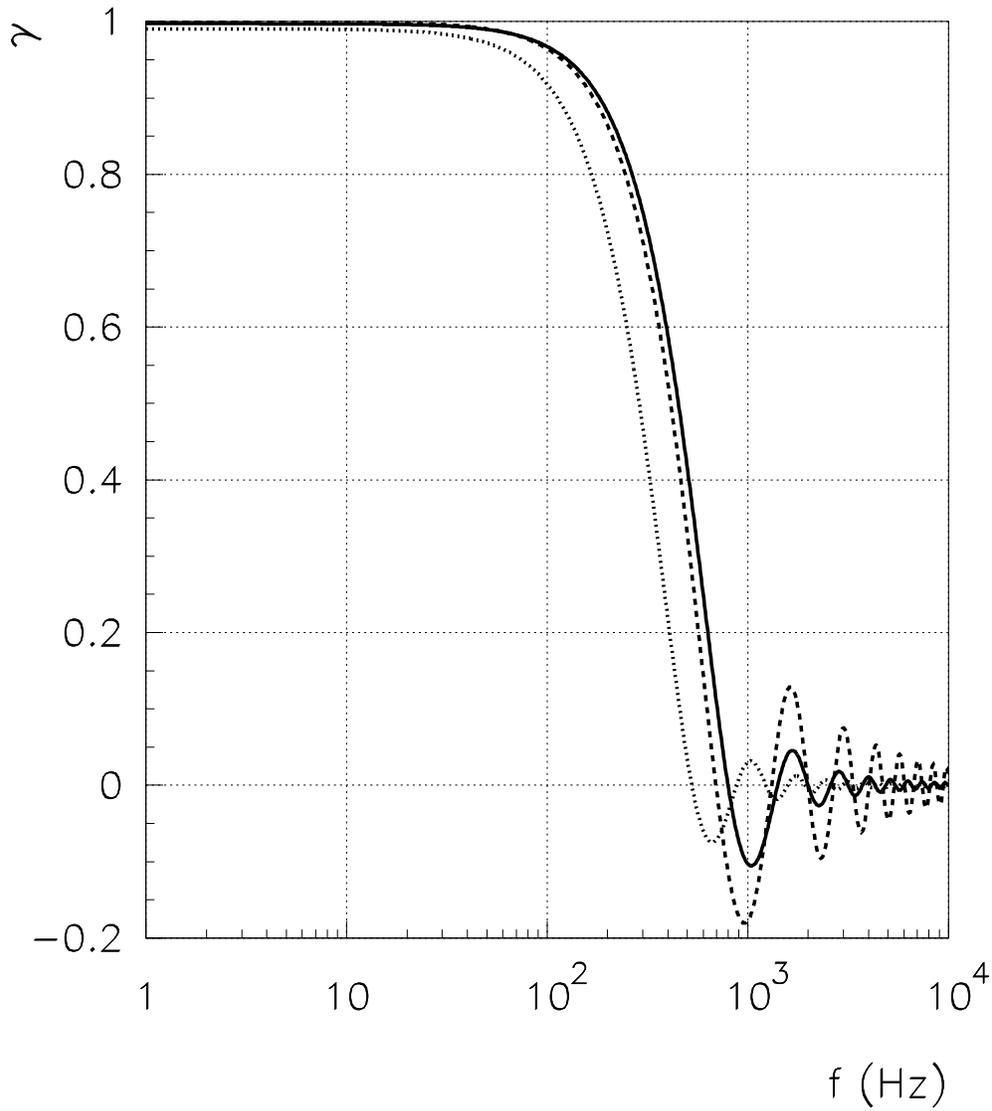}
\caption{The overlap reduction function $\gamma (f)$ for the
  correlation of VIRGO with NAUTILUS (solid line), AURIGA (dashed
  line) and EXPLORER (dotted line) in the case in which the bars have
  been reoriented so to be aligned with VIRGO (from ref.~\cite{dad}).}
\label{dad2}
\end{figure}

Fig.~\ref{dad2} shows the overlap reduction functions for the
correlation between VIRGO and one of the three resonant bars NAUTILUS,
EXPLORER, AURIGA. These overlap reduction functions have been computed
assuming that the bars have been reoriented so to achieved the maximum
correlation with VIRGO, which is technically feasible. 

One  should also note that there is the danger that one of the spikes due to
  the violin modes in the VIRGO sensitivity curve comes close to a
  resonant frequency of the bar; surprisingly, this apparent unlikely
  event is just what happens with the data used to draw 
fig.~\ref{virsens}; in
  fact, according to these data there is a pair of violin modes at  922.6
  Hz and 977.2 Hz. The first one happens to fall  exactly at
  the resonance at 922 Hz of NAUTILUS!
However, these data for the violin modes are not yet final. For
instance, the VIRGO collaboration is presently considering the
possibility of  using silica instead of steel for the wires, which
would change the position of the resonances. Furthermore, the
resonance frequency of the  bars can be tuned within a few Hz
with the electronics; this would be quite sufficient, since the
violine modes have a very high Q and so are much narrower than one
Hz.

The minumum detectable values for $\hogw$
for some bar-bar and bar-interferometer correlations
are given in  Table 2  (from ref.~\cite{dad}),
for one year of observation and 90\% confidence
level (SNR=1.65).

\begin{table}
\begin{center}
\begin{tabular}{|l|l|}\hline
correlation     & $\hogw$             \\ \hline\hline
VIRGO*AURIGA    & $4\times 10^{-4}$ \\ \hline
VIRGO*NAUTILUS  & $7\times 10^{-4}$ \\ \hline
AURIGA*NAUTILUS & $5\times 10^{-4}$ \\ \hline
\end{tabular}
\caption{The sensitivity of the correlation of VIRGO with resonant
  bars, and of two resonant bars
(from ref.~\cite{dad}), but setting   the confidence level to 90\% ).
}
\end{center}
\end{table}

A three detectors correlation
AURIGA-NAUTILUS-VIRGO, with present orientations, would reach 
\be
\hogw\simeq 3\times 10^{-4}\, ,
\ee
at 90\% c.l., while with optimal orientation, 
\be
\hogw\simeq 1.6\times 10^{-4}\, . 
\ee
Although the improvement in
sensitivity in a bar-bar-interferometer correlation is not large
compared to a bar-bar or bar-interferometer 
correlation,  a three detectors
correlation would be important in ruling out spurious
effects~\cite{VCCO}.

In the case of correlations involving resonant bars, obviously
there is no issue
of optimal filtering, since they are narrow-band detector, and
the sensitivity does not depend on the shape of the spectrum.
As we have seen in sect.~\ref{detectors},
using resonant optical techniques, it is possible
to improve the sensitivity of interferometers at special values of the
frequency, at the expense of their broad-band sensitivity. Since bars
have a narrow band anyway, narrow-banding the interferometer improves
the sensitivity of a bar-interferometer correlation by about one order
of magnitude~\cite{CS}. Thus, the limit of bar-bar-interferometer
correlation, with narrow banding of the interferometer, is of order 
$\hogw\sim {\rm a\, few}\times 10^{-5}$.

Cross-correlation experiments have already been performed
using NAUTILUS and EXPLORER~\cite{Nau3}. The bar are oriented so that
they are parallel, and then the overlap reduction function $\Gamma
(f)$ results in a reduction of sensitivity of about a factor of 6, 
compared to ideal same site detectors. The detectors are tuned at the
same resonance frequency $f=907.2$ Hz, with an overlapping band $\pm
0.05$ Hz, and the overlapped data cover a period of approximately 12
hours. Of course, with such a short coincidence time, 
the bound on $\hogw$ is not yet
significant. Running for one year, it is expected to reach $\hogw
<1$. Cross-correlations, searching for bursts, have also been done
between ALLEGRO and EXPLORER~\cite{AllExp}

It is also important to observe that, at the level of sensitivity that
we are discussing, the effect of cosmic rays on the resonant detectors
can be relevant~\cite{AP,Coc5,MaPa,Ast,Piz2}. For a stochastic
background, cosmic rays at sea level degrade the power spectrum
sensitivity of a single resonant mass detector 
by a factor $(20 {\rm mK}+T_{\rm  eff})/T_{\rm eff}$~\cite{Piz2}.
For the coincidence between
two resonant masses it would then be useful to place one of the two
detectors underground (not much is gained placing both detectors
underground).  In the coincidence between a resonant mass and
an interferometer, instead, this is  not necessary, since the
interferometer is much less sensitive to cosmic rays.

While resonant bars have been taking data for years, spherical
detectors are at the moment still at the stage of theoretical studies
(although prototypes might be built in the near future), but could reach
extremely interesting sensitivities. In particular, two spheres with a
diameter of 3 meters, made of Al5056, and located at the same site,
could reach a sensitivity $\hogw\sim 4\times
10^{-7}$~\cite{VCCO}. This figure  improves using a more dense
material or increasing the sphere diameter, but it might be 
difficult to build a heavier sphere. Another very promising
possibility is given by hollow spheres~\cite{Coc3}. The theoretical
studies of ref.~\cite{Coc3} suggest that correlating two hollow
spheres one could reach the
\be
\hogw\simeq 10^{-9}\times\(\frac{f}{200\,{\rm Hz}}\)^3
\(\frac{\tilde{h_f}}{10^{-24}\,{\rm Hz}^{-1/2}}\)^2
\(\frac{20\,{\rm Hz}}{\Delta f}\)^{1/2}
\(\frac{10^7\, {\rm s}}{T}\)^{1/2}\, .
\ee
With the value of $\tilde{h}_f$ and $\Delta f$ suggested by
ref.~\cite{Coc3},
it could be possible to reach an extremely interesting value,
$\hogw\sim 10^{-9}$.

\clearpage

\section{Bounds on $\hogw (f)$}\label{boundsect}

In this section we discuss various experimental bounds on $\hogw
(f)$. We will be interested not only on the bounds at  values of $f$
where interferometers or resonant bars can operate, but also at all
possible frequencies.
The reason will become apparent when we will
 discuss the spectra from various specific cosmological
mechanisms for the production of relic GWs.
These spectra depend of course on the parameters of the cosmological
model. Often the  frequency dependence  is, to a first approximation,
completely determined,
but the overall value of $\hogw$  depends on some
parameters of the
model. 
In some case, and especially for the amplification of vacuum
fluctuations (sect.~\ref{Bogo}) 
these spectra  extend over a huge range of
frequencies, ranging from frequencies as small as $10^{-18}$
Hz (corresponding to wavelength of the order of the present Hubble
radius of the Universe) up to possibly the GHz region. It is therefore
important to see what are the experimental constraint, at any
frequency, on $\hogw$, since they automatically imply bounds on the
parameters of the model that enter in
 the spectrum, and therefore on its the value at frequencies
accessible to interferometers or resonant masses.

\begin{figure}
\centering
\includegraphics[width=\linewidth,angle=270]{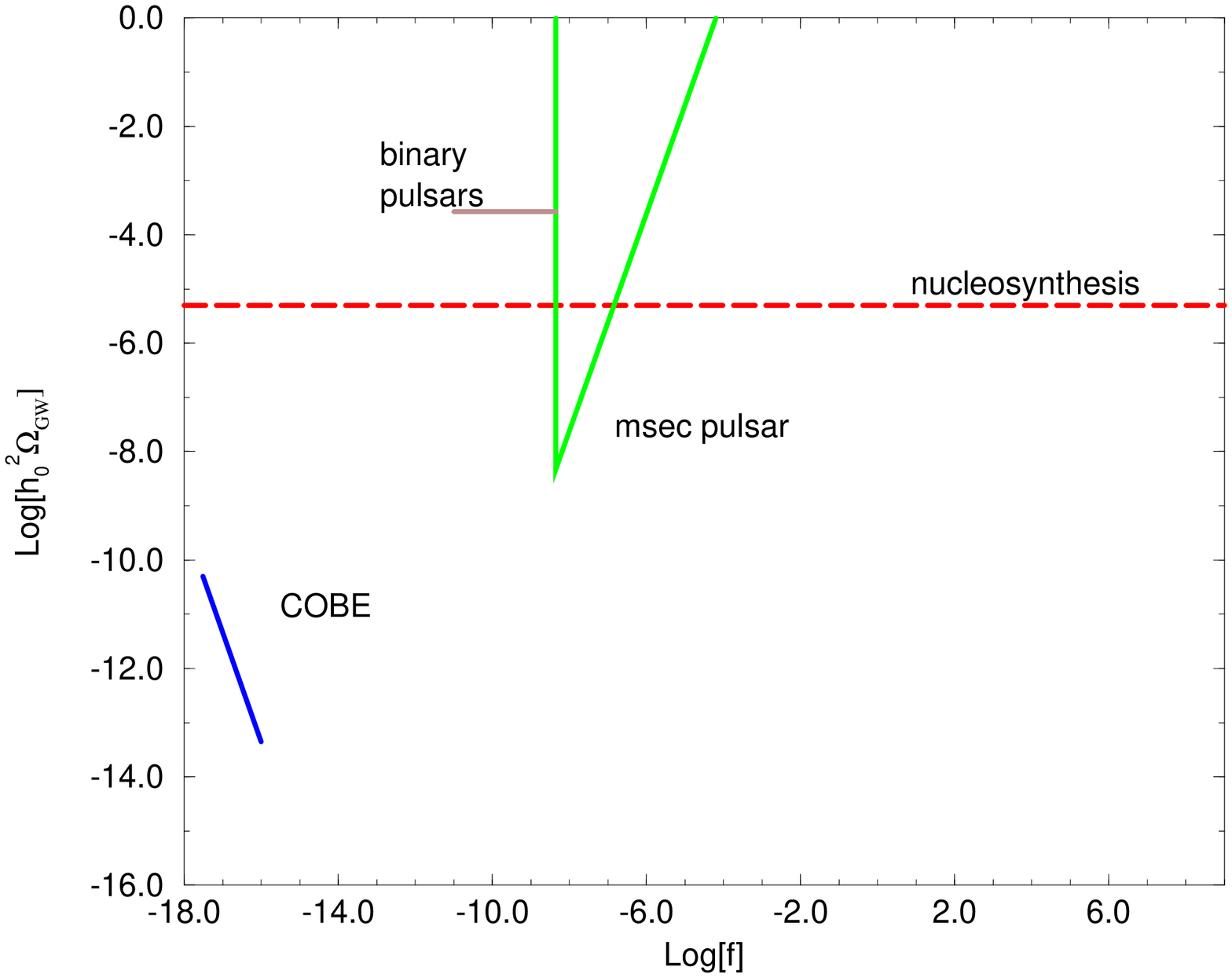}
\caption{The various bounds on $\hogw (f)$ discussed in the text.}
\label{bounds}
\end{figure}

The various limits discussed in this section are summarized in
fig.~\ref{bounds}.

\subsection{The nucleosynthesis bound}\label{NSsect}

Nucleosynthesis successfully predicts the primordial abundances of
deuterium, $^3$He, $^4$He and $^7$Li in terms of one
cosmological  parameter $\eta$,
 the baryon to photon ratio. In the prediction enter also
parameters of the underlying particle theory, which are therefore
constrained in order not to spoil the agreement. In particular, the
prediction is  sensitive to the effective number of species at
time of nucleosynthesis, $g_*=g(T\simeq {\rm MeV})$. 
With some simplifications, 
the dependence on $g_*$ can be understood as follows. 
A crucial parameter in the computations of nucleosynthesis is
the ratio of  the number
density of neutrons, $n_n$, to the number density  of protons, $n_p$, 
As long as thermal equilibrium is mantained we have (for
non-relativistic nucleons, as appropriate at $T\sim$ MeV, when
nucleosynthesis takes place)
$n_n/n_p=\exp (-Q/T)$ where $Q=m_n-m_p\simeq 1.3$ MeV.
Equilibrium is mantained by the process $pe\leftrightarrow n\nu$, with
width $\Gamma_{pe\ra n\nu}$, as long as $\Gamma_{pe\ra n\nu}>H$. When
the rate drops below the Hubble constant $H$, the process cannot
compete anymore
with the expansion of the Universe and, apart from
occasional weak processes, dominated by 
the decay of free neutrons, the ratio $n_n/n_p$
remains frozen at the value $\exp (-Q/T_{\rm f})$, where
$T_{\rm f}$ is the value of the temperature at time of freeze-out.  This
number therefore determines the density of neutrons available for
nucleosynthesis, and since practically all neutrons available will
eventually form $^4$He, the final primordial abundance of $^4$He is
exponentially sensitive to the freeze-out 
temperature $T_{\rm f}$. Let us take for
simplicity $\Gamma_{pe\ra n\nu}\sim G_F^2T^5$ (which is really
appropriate only in the limit $T\gg Q$). The Hubble constant
is given by  $H^2=(8\pi /3)G\rho$, where $\rho$ includes all form of
energy density at time of nucleosynthesis, and therefore also the
contribution of primordial GWs. As usual, it is convenient to  write
the total energy density $\rho$ in terms of
$g_*$, see eq.~(\ref{g}), as $\rho =(\pi^2/30)g_*T^4$. 
We recall that, for
gravitons, the quantity $T_i$ entering eq.~(\ref{g}) is defined by
 $\rho_{\rm gw}=2(\pi^2/30)T_i^4$, and this does not imply
 a thermal spectrum.
Then the freeze-out temperature 
$T_{\rm f}$ is determined by the condition
\be
G_F^2T_{\rm f}^5\simeq \left(\frac{8\pi^3g_*}{90}\right)^{1/2}
\frac{T_{\rm f}^2}{\mpl}\, .
\ee
This shows that  $T_{\rm f}\sim g_*^{1/6}$, 
at least with the approximation that we used for 
$\Gamma_{pe\ra n\nu}$. A large energy density in
relic gravitons gives a large contribution to the total density $\rho$
and therefore to $g_*$. This results in a larger freeze-out
temperature, more available neutrons
and then in overproduction of $^4$He. This is the idea
behind the nucleosynthesis bound~\cite{Sch}. 
More precisely, since the density of $^4$He increases also with 
the baryon to photon ratio $\eta$, we could compensate an increase in
$g_*$ with a decrease in $\eta$, and therefore
we also need a lower limit on $\eta$, which is provided by the
comparison with the abundance of deuterium and $^3$He.

Rather than 
$g_*$, it is often used an `effective number of neutrino species'
$N_{\nu}$ defined as follows. In the standard model, at $T\sim $ 
a few MeV,
the active degrees of freedom are the photon, $e^{\pm}$, neutrinos and
antineutrinos, and they have the same temperature, $T_i=T$.  
Then,  for $N_{\nu}$ families of light neutrinos,
$g_*(N_{\nu})=2+(7/8)(4+2N_{\nu})$, where the factor
of 2 comes from the two elicity states of the photon, 4 from
$e^{\pm}$ in the two elicity states, and $2N_{\nu}$ counts the
$N_{\nu}$ neutrinos and the $N_{\nu}$ antineutrinos, each
with their single elicity state; the factor (7/8) holds for fermions. 
In the Standard Model with $N_{\nu}=3$,
 $g_*=43/4$. So
we can define an `effective number
of neutrino species' $N_{\nu}$ from
\be
\frac{43}{4}+
\sum_{i=\rm extra\, bosons}g_i\left(\frac{T_i}{T}\right)^4+
\frac{7}{8}\sum_{i=\rm extra\,
  fermions}g_i\left(\frac{T_i}{T}\right)^4
=2+\frac{7}{8}(4+2N_{\nu})
\, ,
\ee
or
\be
\sum_{i=\rm extra\, bosons}g_i\left(\frac{T_i}{T}\right)^4+
\frac{7}{8}\sum_{i=\rm extra\,
  fermions}g_i\left(\frac{T_i}{T}\right)^4
= \frac{7}{4}(N_{\nu}-3)\, .
\ee
One extra species of light neutrino, at the same temperature as
the photons, would contribute one unit to $N_{\nu}$, but all species,
weighted with their energy density, contribute to $N_{\nu}$, which of
course in general is not an integer.
For $i=$ gravitons, we have $g_i=2$ and
$(T_i/T)^4=\rho_{\rm gw}/\rho_{\gamma}$, where
$\rho_{\gamma}=2(\pi^2/30)T^4$ is the photon
energy density.
If gravitational waves 
give the only extra contribution to $N_{\nu}$, compared to the
standard model with $N_{\nu}=3$, then
\be
\sum_{i=\rm extra\, bosons}g_i\left(\frac{T_i}{T}\right)^4=
2\frac{\rho\sub{gw}}{\rho_{\gamma}}\, ,
\ee
and therefore
\be\label{NS}
\left(\frac{\rho_{\rm gw}}{\rho_{\gamma}}\right)_{\rm NS}
=\frac{7}{8}(N_{\nu}-3)\, ,
\ee
where the subscript NS reminds that this equality holds at time of
nucleosynthesis.
If more extra species, not included in the standard
model, contribute to $g_*(N_{\nu})$, then the equal sign in the above
equation is replaced by lower or equal.\footnote{To
compare with eq.~(56) of ref.~\cite{All}, note that 
while we use $\rho_{\gamma}$, in ref.~\cite{All} the bound is
written in terms of
the total energy density in radiation at time of nucleosynthesis,
which includes also the contribution of $e^{\pm}$, neutrinos and
antineutrinos,
$(\rho_{\rm rad})_{\rm NS}=[1+(7/8)(2+3)](\rho_{\gamma})_{\rm NS}$.}
The same happens if there is a contribution
from any other form of energy  present at the
time of nucleosynthesis and not included in the energy 
density of radiation, like, e.g., primordial black holes.

To obtain a bound on the energy density
at the present time, we note that from the time of
nucleosynthesis to the present time $\rho_{\rm gw}$ scaled as $1/a^4$,
while the photon temperature evolved following $g_S(T)T^3a^3=$
const. (i.e. constant entropy)
and $\rho_{\gamma}\sim T^4\sim 1/(a^4 g_S^{4/3})$. Therefore
\be
\left(\frac{\rho_{\rm gw}}{\rho_{\gamma}}\right)_0=
\left(\frac{\rho_{\rm gw}}{\rho_{\gamma}}\right)_{\rm NS}
\left( \frac{g_S(T_0)}{g_S({\rm 1\, MeV})}\right)^{4/3}=
\left(\frac{\rho_{\rm gw}}{\rho_{\gamma}}\right)_{\rm NS}
\left( \frac{3.913}{10.75}\right)^{4/3}\, ,
\ee
where the subscript zero denotes present time.
Therefore we get the nucleosynthesis bound at the
present time,
\be\label{227}
\left(\frac{\rho_{\rm gw}}{\rho_{\gamma}}\right)_0
\leq 0.227 (N_{\nu}-3)\, .
\ee
Of course this bound holds only for GWs that were  already produced at
time of nucleosynthesis ($T\sim$ MeV, $t\sim$ sec).
It does not apply to any stochastic
background produced later, like backgrounds of astrophysical origin.
Note that this is a bound on the total energy density in gravitational
waves, integrated over all frequencies. Writing $\rho_{\rm gw}=\int 
d(\log  f)\,
d\rho_{\rm gw}/d\log f$, multiplying both $\rho_{\rm gw}$ and
$\rho_{\gamma}$ in eq.~(\ref{227})
by $h_0^2/\rho_{c}$  and inserting the numerical value
$h_0^2\rho_{\gamma}/\rho_c\simeq 2.481\times 10^{-5}$~\cite{PDG}, we
get 
\be\label{nsbound}
\int_{f=0}^{f=\infty}d(\log f)\,\, \hogw (f)\leq
5.6\times 10^{-6}  (N_{\nu}-3)\, .
\ee
The bound on $N_{\nu}$ from nucleosynthesis is subject to various
systematic errors in the analysis, which have to do mainly with the
issues of how much of the observed
$^4$He abundance is of primordial origin,
and of the nuclear processing of $^3$He in stars, and as a consequence
over the last five years  have been quoted limits on $N_{\nu}$ ranging
from 3.04 to around 5. 
The situation has been  reviewed in ref.~\cite{CST}. The
conclusions of ref.~\cite{CST} is that, until current astrophysical
uncertainties are clarified, $N_{\nu}<4$ is a conservative limit.
Using  extreme assumptions, a meaningful limit $N_{\nu}<5$ still
exists, showing the robustness of the argument.
Correspondingly, the right-hand side of eq.~(\ref{nsbound}) is,
conservatively, of order
$5\times 10^{-6}$ and anyway cannot exceed $10^{-5}$.

If the integral cannot exceed these values, also its positive definite
integrand $\hogw (f)$ cannot exceed it over an appreciable interval of
frequencies, $\Delta\log f\sim 1$. One might still have, in principle,
a very narrow peak in $\hogw (f)$ at some frequency $f$, with a peak
value larger than say $10^{-5}$, while still its contribution to the
integral could be small enough. But, apart from the fact that 
such a behaviour seems rather implausible, or at least is not
suggested by any
cosmological mechanism,  
in the detection at broadband detectors like VIRGO, one must also
take into account that, even if we gain in the height of the signal,
this is partially compensated by the fact that  we lose
because of the reduction of
the useful frequency band $\Delta f$, see eq.~(\ref{SNR2}).

It is sometimes  useful to think in terms of the dimensionless
amplitude $\hc$ instead of
$\hogw$. From \eq{rho5}, we see that a bound $\hogw (f)<5\times
10^{-6}$, valid independently of the frequency $f$, translates into a
frequency dependent bound on $\hc$,
\be
\hc < 2.82\times 10^{-21}\(\frac{{\rm 1\, Hz}}{f}\)\, ,
\ee
so that, for instance, $\hc <  10^{-23}$ around 200-300 Hz,
where ground-based interferometers have their highest sensitivity, and 
$\hc <  10^{-17}$ at the lower range accessible to LISA,
$f\sim 10^{-4}$ Hz.

\subsection{The COBE bound}\label{COBE}

Another important constraint comes from the COBE measurement of the
fluctuation of the temperature of the cosmic microwave background
radiation (CMBR). 
The basic idea is that
a strong background of GWs at very long wavelengths
produces a stochastic redshift on the frequencies of the photons of
the 2.7K radiation, and therefore a fluctuation 
in their temperature. This
is known as the Sachs-Wolfe effect~\cite{SW,KT}. We first give a
qualitative understanding of this bound, and then quote more precise
numbers. 

It is conventional to expand
the CMBR temperature fluctuations  on the sky in spherical
harmonics, 
\be
\frac{\delta T(\hat{\Omega})}{T}=\sum_{l=2}^{\infty}\sum_{m=-l}^l
a_{lm}({\bf r})Y_{lm}(\hat{\Omega})\, .
\ee
The  dipole term is dominated by the Earth motion, and
therefore is omitted. The multipole amplitudes $a_{lm}$ depends on
the observer position ${\bf r}$, and $|a_{lm}({\bf r})|^2$, averaged
over all available positions, is the {\em angular power spectrum}. 
The rms  fluctuations of the temperature, averaged over the
sky, are  given by
\be
\langle \(\frac{\delta T}{T}\)^2\rangle =\sum_{l=2}^{\infty}
\frac{2l+1}{4\pi}\langle |a_{lm}|^2\rangle W_l\, ,
\ee
where $W_l$ depends only on the detector. COBE measures a temperature
difference between two antennas 
separated by an angle $\theta$,  and 
\be
W_l=2[1-P_l(\cos\theta )]e^{-(l+1/2)^2\sigma^2}\, ,
\ee
where $P_l(\cos\theta )$ are the Legendre polynomials and
$\sigma$ is the finite width of the beam, which effectively cuts
the resolution at $l\sim 1/\sigma$.  
For the quadrupole moment, one defines
\be
Q=T\, \[\sum_{m=-2}^{2}\frac{\langle |a_{2m}|^2\rangle}{4\pi}\]^{1/2}\, .
\ee
Assuming a flat Harrison-Zeldovich spectrum, i.e. spectral index $n=1$
(see sect.~\ref{slowroll}), the COBE result is
\be
Q=18\pm 1.6\, \mu K
\ee
while fitting also $n$ the best fit gives  $n=1.2\pm 0.3$ and $Q=15.3\mu K$.
In any case, from the quadrupole alone we get 
\be\label{dT}
\frac{\delta T}{T}\simeq (5-6)\times 10^{-6}\, .
\ee
It is not difficult to see that the COBE bound on $\ogw$ has
qualitatively the form
\be\label{cob1}
\ogw (f)<
\(\frac{H_0}{f}\)^2\, \(\frac{\delta T}{T}\)^2\, .
\ee
and holds for
\be
3\times 10^{-18}\,{\rm Hz}<f<10^{-16}\,{\rm Hz}\, .
\ee
Let us first discuss why this is the relevant frequency range. 
The limitation $f>3\times 10^{-18}$ Hz is just the condition that
these waves are inside the horizon today. The upper limit instead comes
out as follows.
It is  important to distinguish between
large angle  and small angle fluctuations. The dividing line is given
by the angle $\theta\sub{LSS}$ such that fluctuations on angular
scales $\theta <\theta\sub{LSS}$ corresponds 
correspond to comoving lengths that were inside the horizon at
time of last scattering, while  $\theta > \theta\sub{LSS}$ corresponds to
super-horizon sized scales at time of last scattering. 
Small scale fluctuations are sensitive to microphysical processes taking
place at time of last scattering; 
large scale fluctuations are instead insensitive to microphysics,
because they refer to regions that were not in causal contact at that 
time, and on these angular scales the Sachs-Wolfe effect, due either
to scalar perturbations or to tensor perturbations, i.e.  GWs,
is the dominant contribution to the CMBR anisotropy. 

To compute $\theta\sub{LSS}$, one
first defines   conformal time today, $\eta_0$, and at the last 
scattering surface, $\eta\sub{LSS}$ as
\bees
\eta_0&=&\int_0^{t_0}\frac{dt}{a(t)}=2H_0^{-1}\, ,\nonumber\\
\eta\sub{LSS}&=&\int_0^{t\sub{LSS}}\frac{dt}{a(t)}\simeq
\frac{\eta_0}{\sqrt{1+z\sub{LSS}}}\, ,
\ees
where we assumed a flat matter-dominated Universe today. The
quantities
$c\eta_0$ and $c\eta\sub{LSS}$ are the comoving size of the horizon
today and at last scattering, respectively; the comoving distance to
the last scattering surface is 
\be
d\sub{LSS}=c\eta_0\( 1-\frac{1}{\sqrt{1+z\sub{LSS}}}\)\simeq 
c\eta_0\, .
\ee
In the standard picture, where recombination occurs at 
$z\sub{rec}\simeq 1300$,
then $z\sub{LSS}\simeq 1100$. In non-standard scenarios, 
$z\sub{LSS}$ could be as low as $O(100)$~\cite{TS}.
The angle subtended by the horizon at last scattering is~\cite{TWL}
\be\label{thetaLLS}
\theta\sub{LSS}\sim \frac{\eta\sub{LSS}}{\eta_0}\sim
\frac{1}{\sqrt{z\sub{LSS}}}
\, .
\ee
For a non-flat  Universe, $\Omega_0\neq 1$, this generalizes to 
$\theta\sub{LSS}\sim (\Omega_0/z\sub{LSS})^{1/2}$.
For $z\sub{LSS}\gg 100$
this is a good approximation to the exact formula. 
The deviations for smaller redshift can be
found in~\cite{TS}. 
For $z\sub{LSS}=1100$, \eq{thetaLLS} gives
$\theta\sub{LSS}\sim 2^o$.

The temperature fluctuations over
an angle $\theta$  are related to $a_{lm}$ by
\be
\(\frac{\delta T}{T}\)^2_{\theta}\sim l^2
\langle |a_{lm}|^2\rangle\, ,
\ee
for 
\be\label{l200}
l\sim \frac{200^o}{\theta}\, .  
\ee
Therefore multipoles up to, say, $l\sim 30$,
are due to large angle fluctuations, and the corresponding
anisotropies $a_{lm}$ are dominated by the Sachs-Wolfe effect, due
either to scalar perturbations  or to GWs.

The waves that were outside the horizon at the last scattering, today
have a frequency,
\be
f < \sqrt{z\sub{LSS}}\, H_0\simeq 10^{-16}\,{\rm Hz}\, ,
\ee
where in the last equality we have taken $z\sub{LSS}=O(10^3)$. For a
non-standard reionization scenario with $z\sub{LSS}\sim 100$ the 
bound becomes approximately $3\times 10^{-17}$ Hz.

Let us now explain how \eq{cob1} comes out. 
A GW with a wave number 
\be
k (\theta )\sim \(\frac{200^o}{\theta}\)\eta_0^{-1}
\ee
subtends an angle $\theta$ on the last scattering surface.
and creates  temperature fluctuation on
this angular scale
given approximately  by~\cite{TWL}
\be
\(\frac{\delta T}{T}\)_{\theta}\sim h[z\sub{LSS}, k(\theta )]
\ee
where $h[z\sub{LSS}, k(\theta )]$ is the characteristic amplitude
that this GW had at time
of last scattering.

We therefore must connect the value of the amplitude at time of last
scattering 
with the value of $\hc$ today.  One then  observes
that, as long as the GW is outside the horizon, its amplitude  is
constant, while, after it enters the horizon, it redshifts with the
FRW scale factor $a(t)$ as $1/a$. This can be derived  from
the fact that the energy density of the GW is $\rho\sub{gw}\sim f^2\hc^2$,
\eq{rho4}; $f$ redshifts as $1/a$ and, after entering the horizon, 
$\rho\sub{gw}\sim 1/a^4$, as for any relativistic particle, so that
$\hc\sim 1/a$. 
Therefore, denoting by $a\sub{hor}$ the value of $a$
at horizon crossing,
\be
\hc\sub{today} = \frac{a\sub{hor}}{a_0}\hc\sub{LSS}= 
          \frac{a\sub{hor}}{a\sub{LSS}}
\frac{1}{z\sub{LSS}}h\sub{LSS}\, .
\ee
During matter dominance, the scale factor depends on conformal time as
$a(\eta )\sim \eta^2$. Then,
for waves that enter the horizon during matter dominance, 
$a\sub{hor}/a\sub{LSS}\simeq (\eta\sub{LSS} f)^{-2}$. Using
$\eta\sub{LSS}\simeq
\eta_0/\sqrt{z\sub{LSS}}=2/(H_0\sqrt{z\sub{LSS}})$, 
one finally gets
\be\label{cobe1}
\hc\sub{today} \simeq \(\frac{H_0}{2 f}\)^2\hc\sub{LSS}\sim
 \(\frac{H_0}{2 f}\)^2\, \(\frac{\delta T}{T}\)\, .
\ee
Using \eq{ogwhc}, we therefore get
\be
\ogw (f)=\frac{2\pi^2}{3H_0^2} f^2h_c^2(f)\sim
\(\frac{H_0}{f}\)^2\, \(\frac{\delta T}{T}\)^2\, .
\ee
Of course, the value of $(\delta T/T)$ 
induced by GWs cannot
exceed the observed value, \eq{dT}, and this gives the bound on
$\hogw$. 

Using the more quantitative analysis of refs.~\cite{AKo,All}, where
the effect of multipoles with $2\leq l\leq 30$ is included, this bound reads
\be\label{cobe2}
\hogw (f)<
7\times 10^{-11}\(\frac{H_0}{f}\)^2\, ,
\hspace*{20mm}(3\times 10^{-18}\,{\rm Hz}<f<10^{-16}\,{\rm Hz})
\, .
\ee 
This bound  is  stronger at the upper edge of its range of
validity,  $f\sim 30 H_0\sim 10^{-16}$ Hz,
where it gives
\be\label{cobebound}
\hogw (f)< 10^{-14}\, ,\hspace*{20mm}(f \sim\, 10^{-16}\,\rm{Hz} )\, .
\ee
Another issue is whether GWs can saturate this bound, or whether 
instead the dominant contribution to $\delta T/T$ 
comes from scalar rather than tensor
perturbations; the answer is in
general model-dependent, and we will discuss examples in
sect.~\ref{slowroll}.

The important point that we can
draw is the following: from the explicit computations discussed below, 
we will find  cosmological spectra that extends from  very low
frequencies, 
$f\sim H_0\simeq 3\times 10^{-18} $ Hz,
 up to the frequencies accessible at GW
experiment and higher. This will be typically the case of the spectra
predicted by the amplification of vacuum fluctuations. The frequency
dependence can often be computed reliably. Then, in order for these
spectra to be observable in the 1Hz-1kHz region, they must grow fast
enough, so that  they can satisfy \eq{cobebound}, and still have a
sizeble value of $\hogw$ at higher frequencies. The requirement
becomes even more stringent if we combine it with the nucleosynthesis
bound: the spectra cannot keep growing until they reach the cutoff,
say at 1 GHz, because otherwise the NS bound on the integral of
$\hogw$ would be violated. Therefore it is necessary that there is at some
stage a change of regime, so that the spectrum flattens or starts
decreasing. 

Thus, it is clear that the combination of the NS bound and the 
COBE bound, together with the request of having
a sizeble value of $\hogw$ at, say, 1 kHz, 
restricts considerably  the  cosmological models that can produce an
interesting spectrum of vacuum fluctuations. Quite remarkably, there
is however one candidate, theoretically  well motivated, that passes
these severe restrictions; this is the string cosmology model of
Gasperini and Veneziano, and we will discuss it in
sect.~\ref{prebigbang}. 

\subsection{Pulsars as GW detectors: msec pulsars, 
binary pulsars  and pulsar arrays}
Pulsars are rapidly rotating, highly magnetised neutron stars formed
during the supernova explosion of stars with 5 to 10 solar masses
(for a  recent review see ref.~\cite{Lor}).
Soon after their discovery, it became clear that pulsars are excellent
clocks. Actually, single pulses have a rather herratic profile in
time, but after integrating over a number  of pulses one finds an
integrated profile that is extremely stable, and can be used as a
clock. Most of the known  pulsars have a period  of the order of one
second, while a growing fraction of this sample has periods in the
range 1.5 to 30 msec. These
millisecond pulsar are an extremely impressive source of high
precision measurements. For instance, the  observations of the
first msec pulsar discovered, B1937+21, after 9 yr of data, 
give a period of  
$1.557\, 806\, 468\, 819\, 794\, 5\pm
 0.000\, 000\, 000\, 000\, 000\, 4$ ms. The speed up of the orbital period 
of the  double
neutron star system  B1913+16 provided experimental evidence for the
existence of GWs
and a test of General Relativity at the level of 1\%~\cite{HT,Tay,TW,DD}.
This phenomenal  stability makes pulsar competitive with atomic clocks
(for instance PSR B1855+09 becomes competitive with atomic clocks
after 3 yr of observations), and has even allowed to detect orbiting
bodies smaller than the Earth from the radial acceleration that they
induce on the pulsar~\cite{Lor}. 

Pulsars are also a natural  detector of
GWs~\cite{Saz,Det,RT,HD,BCR,BNR,SRTR,KTR}.
In fact, a GW passing between us and the pulsar 
causes a fluctuation in the time of arrival of the pulse,
proportional to the GW amplitute $\hc$. Consider a GW traveling
along the $z$ axis, and  let  $\theta$ be the polar angle of the
Earth-pulsar direction, measured  from the $z$ axis. Then the Doppler
shift at time $t$ induced by the GW is~\cite{Det,FB}
\be\label{pul1}
\frac{\Delta\nu (t)}{\nu}=\frac{1}{2}(1-\cos\theta )\[
\cos 2\psi\,  h_{+}(t-\frac{s}{c})+\sin 2\psi\, 
h_{\times}(t-\frac{s}{c})\]\, ,
\ee
where, as in sect.~3, 
$\psi$ is a rotation in the plane orthogonal to the direction of
propagation of the wave (in this case the $(x,y)$ plane), 
corresponding to a choice of the axes to which the
$+$ and $\times$ polarization are referred. 
The factor $(1/2)(1-\cos\theta )$ 
is the standard angular dependence of the Doppler shift, and $s$ is
the distance along the path, so that $t-s/c$ is the time when the GW
crossed the Earth-pulsar direction.

If the uncertainty in the time of
arrival of the pulse is $\epsilon$ and the total observation time is
$T$, this `detector' would be sensitive to $\hc\sim \epsilon/T$, for
frequencies $f\sim 1/T$. The highest sensitivities can then
be reached for a continuous source, as a stochastic background,  after
one or more years of integration, and therefore for $f\sim
10^{-9}-10^{-8}$ Hz. Based on the data from PSR B1855+09 and
correcting an error from a previous analysis, ref.~\cite{TD}
gives a limit, at $f\equiv f_*=4.4\times 10^{-9}$ Hz,
\be
\hogw (f_*)<  1.0\times 10^{-8}\, ,
\hspace{20mm} (95\% \, \rm{c.l.})\, ,
\ee
or
\be
\hogw (f_*)<  4.8\times 10^{-9}\, ,
\hspace{20mm} (90\% \, \rm{c.l.})\, .
\ee
Since the error on $\hc$ is proportional to $1/T$ and $\hogw\sim
h_c^2$, the bound for $f>f_*$ is (at 90\% c.l.)
\be
\hogw (f)<  4.8\times 10^{-9}\, \(\frac{f}{f_*}\)^2\, ,
\ee
and therefore it is quite significant (better than the NS bound)
also for $f\sim 10^2 f_*$. For $f<f_*$, instead, the pulsar provides
no limit at all.

These limits are likely to continue to improve since a number of
recently discovered msec pulsars appear to have a timing stability
comparable or better than PSR B1855+09. These limits are especially
significant for cosmic strings, as we will see in
sect.~(\ref{cosmicstrings}).

Another important limit comes from  pulsars in binary systems,
e.g. with a white dwarf or neutron star companion. These 
systems have a second clock, which is the binary orbit itself, and
are very clean. The change in the orbital period $P$ can be
computed from General Relativity and in many cases is neglegibly
small. This has the advantage that it not necessary to make a fit to
an unknown derivative of the orbital period, $\dot{P}$, and
subtract it to isolate the contribution of GWs.
This gives a limit that holds for a large range of frequencies, 
\be
\frac{c}{D}<f<\frac{1}{T}\, ,
\ee
where $T$ is the observation time and $D$ the distance to the
pulsar. A detailed analysis of this limit has been performed
in \cite{Kop}, correcting previous estimates.  In particular 
the limit  from B1855+09 turns out to be the most stringent
and gives~\cite{Kop}
\be
\hogw <  2.7\times 10^{-4}\, ,
\ee
in the frequency range $10^{-11}<f<4.4\times 10^{-9}$ Hz.

Finally, one can observe that these limits have been set using a
pulsar as a single detector. From the analysis of sect.~3 it is clear
that if one can use two or more pulsars as coincident detectors one
could obtain a much better sensitivity \cite{HD,FB,Lor}. 
From \eq{pul1} we see that the Doppler shift from the i-th
pulsar has the general form
\be
\frac{\Delta\nu_i (t)}{\nu_i}=\alpha_i s(t)+n_i(t)\, ,
\ee
where $s(t)$ is the GW signal, $\alpha_i$ a geometrical factor
depending of the line-of-sight direction, and $n_i(t)$ contains the
noise intrinsic to the pulsar timing. The cross-correlation between
two pulsars then gives
\be
\alpha_i\alpha_j\langle s^2(t)\rangle +\alpha_i\langle sn_i\rangle
+\alpha_j\langle sn_j\rangle +\langle n_in_j\rangle\, ,
\ee
and, since the noises from different pulsars are uncorrelated, 
increasing the observation time this
tends to $\alpha_i\alpha_j\langle s^2(t)\rangle$. Furthermore, the
correlation between many pulsars allows  to subtract the errors due to
terrestrial clocks, since there is now the possibility of timing
multiple pulsars against each other. Furthermore, precise timing also
require to subtract the effect of the Earth motion,  and the precision
of pulsar timing measurement is now  near the limit of models of the
Earth motion. Again, multiple pulsar timing allows to subtract this
effect. 

Performing pulsar coincidences
using the present database of long-term timing observation does not
improve on the limits obtained with single pulsars~\cite{Lor}. The
basic reason is that, in order to effectively subtract the error due
to the Earth motion, one needs  three pulsars widely separated across the
sky. A fourth pulsar is needed as a clock, and a fifth to measure the
GW amplitude~\cite{FB}. However, the galactic msec pulsars known
around 1990, for which we now have  many years of data taking, were
all concentrated in the same region of the sky (as a consequence of
the fact that they were all found at Arecibo); in this situation one
cannot subtract effectively the error due to the Earth motion.
However, the msec pulsars known today
are distributed much more  uniformly across the sky, and
therefore the continued timing of these pulsar in the next few years
will greatly improve the sensitivity of pulsar arrays as GW detectors.

Finally, we observe that the limit on $\hogw$ from pulsar timing holds
for all sort of GW backgrounds, not necessarily of primordial origin,
contrarily to the nucleosynthesis bound discussed in
sect.~\ref{NSsect} that of course holds only for GWs that were already
present at time of nucleosynthesis.

\subsection{Light deflection by GWs}
We finally mention that there has been  quite some activity in trying
to obtain bounds on $\hogw$ from the effect that a stochastic
background of GWs would have on the propagation of light from distant
sources~\cite{ZB,BKPN}. 
Examining the proper
motion of quasars, ref.~\cite{GEPBM} sets a limit
\be
\int_{-\infty}^{f_*}df\,\hogw <0.11\, ,
\ee
where $f_*\sim 2\times 10^{-9}$ Hz is the inverse of the observation
time, and states that considerable improvement in this limit should be
possible in the next decade. Note that the limit is weaker
than the one given by nucleosynthesis, but it is not restricted to GWs
present before nucleosynthesis.

A number of other mechanisms related to light deflection by GWs
have been investigated. In particular, the apparent
clustering in the galaxy-galaxy correlation function which, according
to ref.~\cite{Linder}, provides a bound $\hogw <10^{-3}$ for
wavelength between a few tens of kpc to a few hunderds Mpc, i.e. for
$10^{-14}$ Hz $<f<10^{-11}$ Hz. 
This bound however has been reanalyzed
in ref.~\cite{KJ} where (confirming older computations~\cite{ZB})
it has been found that, to linear order in the metric perturbation,
the effect due to GWs does not grows linearly with the distance to the
source, contrarily to previous claims, but only logarithmically, and
therefore does not cause an appreciable galaxy clustering.
Similarly, no useful bound on $\hogw$ can be set from
lateral displacement of
caustic networks or weak-lensing of distant objects.
This result is valid for  stochastic GWs, and  does not apply
immediately  to the
deflection of light by GWs from localized sources. However, in this
case the analysis of ref.~\cite{DE} confirms the  conclusion
that these effects are  too small to be of observational interest.

\section{Production of relic GWs: a general orientation}\label{gen}
Theoretical predictions of relic GW backgrounds
are  necessarily subject to large 
uncertainties, due either to the fact that  we must use physics
beyond the Standard Model, or to uncertainties in the details of
the cosmological mechanisms.
It is therefore important to understand
what features of a given result are
relatively general, and what are specific to a given model. 

In this section we  discuss results on the characteristic
frequency, the form of the spectrum
and the  typical intensity, that we will
consider  as a sort of `benchmark'. The results from various
specific computations can than be better understood  from the
comparison with these benchmarks, and in particular one can get a
feeling of what results are relatively general and model independent,
and what results are really peculiar to a specific computation.

\subsection{The characteristic frequency}\label{charfreq}
First of all, we  discuss what is the typical frequency where we
can expect to find today a  signal of cosmological origin. 
Obviously, part of the
answer depends on the dynamics of the production mechanism, and
therefore is model-dependent, and part is kinematical, depending on
the redshift from the production era. First of all, it is useful to
separate the kinematics  from the dynamics.

To specify the kinematics, we need a cosmological model. We 
consider  the standard Friedmann-Robertson-Walker (FRW)
cosmological model, consisting of a radiation-dominated (RD) phase
followed by the present matter-dominated (MD) phase, and we call
$a(t)$ the FRW scale factor. The RD phase goes backward in time until
some new regime sets in. This could be an inflationary epoch, e.g. at
the grand unification scale, or 
 the RD phase could go back in time until Planckian energies are
reached and quantum gravity sets in, i.e., until $t\sim t_{\rm
Pl}\simeq 5\times 10^{-44}$~s.

A graviton produced with a frequency $f_*$, 
at a time $t=t_*$ within the RD phase, has
today ($t=t_0$) a redshifted frequency $f_0$ given by
$f_0=f_* a(t_*)/a(t_0)$. To compute the ratio $a(t_*)/a(t_0)$ one uses
the fact that during the standard RD and MD phases the Universe
expands adiabatically. 
The entropy per unit comoving volume is 
\be
S={\rm const.} g_S(T)a^3(t)T^3\, ,
\ee
where $g_S(T)$  is defined by~\cite{KT}. 
\be\label{gS}
g_S(T)=\sum_{i={\rm bosons}}g_i\left(\frac{T_i}{T}\right)^3+
\frac{7}{8}\sum_{i={\rm fermions}}g_i\left(\frac{T_i}{T}\right)^3\, .
\ee
Here $g_i$ counts the internal degrees of freedom of the i-th particle
(spin, color, etc.).

Thus $g_S(T)$ is a measure of the effective
number of degrees of freedom at temperature $T$,  as far as
the entropy is concerned. Another useful quantity is
\be\label{g}
g(T)=\sum_{i={\rm bosons}}g_i\left(\frac{T_i}{T}\right)^4+
\frac{7}{8}\sum_{i={\rm fermions}}g_i\left(\frac{T_i}{T}\right)^4
\ee
which counts the effective
number of degrees of freedom at temperature $T$, as far as
the energy is concerned, and that entered our discussion of
nucleosynthesis, sect.~\ref{NSsect}.
In the early Universe most particles
had a common temperature, and $g(T)$ and $g_S(T)$ are indistinguishable.
In the Standard Model, at
$T\gsim 300$ GeV, they become constant and have the value
$g_S=g=106.75$,
while today, assuming 3 neutrino species, $g_S(T_0)\simeq 3.91$
and $g_S(T_0)\simeq 3.36$~\cite{KT}. 

The sum in \eqs{gS}{g} 
runs over relativistic species. This holds if a
species is in thermal equilibrium at the temperature $T_i$. If instead
it does not have a thermal spectrum (which in general
is the case for gravitons) we can still use the above equation, where
for this species $T_i$ does not represent a temperature but is defined
(for bosons) 
by $\rho_i=g_i(\pi^2/30)T_i^4$, where $\rho_i$ is the energy density
of this species.
Using
\be\label{adia}
g_S(T_*)a^3(t_*)T_*^3= g_S(T_0)a^3(t_0)T_0^3\, ,
\ee
and
$T_0=2.728\pm 0.002 $K~\cite{COBE}
one finds~\cite{KKT}
\be\label{f0}
f_0=f_*\frac{a(t_*)}{a(t_0)}\simeq  8.0\times 10^{-14}f_*
\left(\frac{100}{g_S(T_*)}\right)^{1/3}
\left(\frac{1{\rm GeV}}{T_*}\right)\, .
\ee
We must now ask what is the characteristic value of
the frequency $f_*$ of a graviton
produced at time $t_*$, when the temperature was
$T_*$.  Here of course the dynamics enters, but still some general
discussion is possible. One of the relevant
parameters in this estimate 
is certainly  the Hubble parameter at time of production,
$H(t_*)\equiv H_*$. This comes 
from the fact that $H_*^{-1}$ is the size
of the horizon at time $t_*$. The horizon size, physically, is the
length scale beyond which causal microphysics cannot operate (see
e.g.~\cite{KT}, ch.~8.4), and
therefore, for causality reasons, we  expect that the characteristic
wavelength 
of gravitons or any other particles produced  at time $t_*$ will
be of order  $H_*^{-1}$ or smaller.\footnote{On 
a more technical side, the deeper reason has
really to do with the invariance of general relativity
under coordinate transformations, combined with the expansion over a
fixed, non-uniform,
background. Consider for instance a scalar field $\phi (x)$ and
expand it around a given classical configuration, $\phi (x)=
\phi_0(x)+\delta\phi (x)$. Under a general coordinate transformation
$x\ra x'$, by definition  a scalar field transforms as
$\phi (x)\ra\phi '(x')=
\phi (x)$. However, when we expand around a given background, we keep
its functional form fixed and therefore under $x\ra x'$,
$\phi_0(x)\ra\phi_0(x')$, which for a non-constant field configuration,
is different from $\phi_0(x)$. It follows that 
the perturbation $\delta\phi (x)$  is not a scalar under general
coordinate transformations, even if $\phi (x)$ was a scalar. 
The effect becomes important for the Fourier components of $\delta\phi
(x)$ with a wavelength comparable or greater than the variation scale
of the background $\phi_0(x)$. (We are discussing a scalar field for
notational simplicity, but of course the same holds for the metric
tensor $g_{\mu\nu}$). In a homogeneous FRW background the only
variation is temporal, and its timescale is given by the
$H^{-1}$. Therefore modes with wavelength greater than $H^{-1}$ are in
general plagued by gauge artefacts. This problem 
manifests itself, for instance, when computing density fluctuations in
the early Universe. In this case one finds spurious modes which can be
removed with an appropriate gauge choice, see e.g. ref.~\cite{KT},
sect.~9.3.6 or ref.~\cite{Muk}.} 

Therefore, we write 
\be
\lambda_*=\epsilon H_*^{-1}\, . 
\ee
The above argument
suggests $\epsilon\leq 1$.
During RD, $H_*^2=(8\pi /3)G\rho_{\rm rad}$. Then
\be\label{H*}
H_*^2=\frac{8\pi^3 g_*T_*^4}{90\mpl^2}\, ,
\ee
and, using $f_*\equiv H_*/\epsilon$, 
eq.~(\ref{f0}) can be written as~\cite{KKT}
\be\label{f1}
f_0\simeq 1.65\times 10^{-7}\frac{1}{\epsilon}
\left(\frac{T_*}{1\rm GeV}\right)
\left(\frac{g_*}{100}\right)^{1/6}\,{\rm Hz}\, .
\ee
This simple equation allows to understand a number of important points
concerning the energy scales that can be probed in GW experiments. 
Basically, the effects of the dynamics have been isolated into the
parameter $\epsilon$.

The relation between time and temperature during the RD phase then
tells us
how far back in time are we exploring the Universe, when we observe a
graviton produced at  temperature $T_*$,
\be
t_*\simeq \frac{2.42}{g_*^{1/2}}
\left(\frac{\rm MeV}{T_*}\right)^2\, {\rm sec}\, ,
\ee
and therefore, detecting a GW that today has  a frequency $f_0$, we
are looking back at the Universe  at time
\be
t_*\simeq 6.6\times 10^{-21}\frac{1}{\epsilon^2}\,
\(\frac{1\,{\rm Hz}}{f_0}\)^2
\(\frac{100}{g_*}\)^{1/6}\, {\rm sec}\, .
\ee
The simplest estimate  are obtained taking $\epsilon =1$ in
eq.~(\ref{f1}). Table~3 gives some representative values for 
the production time $t_*$ and the temperature of the Universe $T_*$
corresponding to  frequencies relevant to LISA and 
 to ground based interferometers, setting $\epsilon =1$.

\begin{table}
\begin{center}
\begin{tabular}{|l|l|l|}\hline
detection frequency & production time  & production temperature   
\\ \hline\hline
$10^{-4}$ Hz    & $7\times 10^{-13}$ sec & 600 GeV            \\ \hline
$1$ Hz          & $7\times 10^{-21}$ sec & $6\times 10^6$ GeV \\ \hline
$100$ Hz        & $7\times 10^{-25}$ sec & $6\times 10^8$ GeV \\ \hline
$10^3$ Hz       & $7\times 10^{-27}$ sec & $6\times 10^9$ GeV \\ \hline
\end{tabular}
\caption{The production time $t_*$ and the production temperature
  $T_*$ for GWs observed today at frequency $f_0$, if at time
of production they had
 a wavelength of the order of the horizon length.} 
\end{center}
\label{tabf*}
\end{table}

However, one should keep in mind that
the estimate  $\lambda_*\sim H_*^{-1}$,  i.e. $\epsilon\sim 1$,
can sometimes be incorrect,
even as an order of magnitude estimate. Below we will
illustrate this point
with some  specific examples. The values in table~3 are more
correctly interpreted as a starting point for understanding how the
result is affected by the dynamics of the specific production mechanism.

An important question is whether GW experiments can explore the
very high energy regime, when the Universe had temperatures of the
order of a typical grand unification scale,
or of the order of the Planck scale. 

The scale of quantum gravity is given by the Planck mass, 
related to Newton constant by $G=1/\mpl^2$. More
precisely, since in the gravitational action enters the combination
$8\pi G$, we expect that the relevant scale is  the reduced
Planck mass $\mpl /(8\pi )^{1/2}\simeq 2.44\times 10^{18}$ GeV.
Using eq.~(\ref{f1}) with $T_*=\mpl /(8\pi )^{1/2}$ and $\epsilon =1$ gives
\be\label{400}
f_0\sim 400\left(\frac{g_*}{100}\right)^{1/6}\, {\rm GHz}\, ,
\hspace{30mm}( T_*= \mpl /\sqrt{8\pi})\, .
\ee
The dependence on $g_*$ is rather weak because of the power 1/6 in
eq.~(\ref{f1}). For $g_*=1000$, $f_0$ increases by a factor $\sim
1.5$ relative to $g_*=100$. Instead, 
for $T_*=\mgut\sim 10^{16}$ GeV, and $g_*\simeq 220$ (likely
values for a supersymmetric unification),
\be
f_0\sim 2\,{\rm GHz}\, ,\hspace{30mm}(T_*= M_{\rm GUT})\, .
\ee
Using instead 
the typical scale of string theory, $M_S$,
the characteristic frequency is between these two values, since $M_S$
is expected to be approximately
in the range $10^{-2}\lsim M_S/\mpl\lsim 10^{-1}$ \cite{Kap}.

Thus, the expected peak frequency of  GW spectra produced at these
very early epochs is in the GHz region. In this region
 no present or future
experiment can operate, and the reason is easily seen from \eq{hrms}:
considering that the maximum value of $\hogw$ is
fixed by nucleosynthesis, the required sensitivity in the GW amplitude
goes as $f^{-3/2}$. At $f=$ 1~GHz one needs a value of
$\tilde{h}_f$ smaller by a factor $10^9$, compared to a value that
would give the same sensitivity in $\hogw$ at  $f=$ 1kHz.

However, one should not hurry toward pessimistic conclusions. 
The characteristic
frequency that we have discussed is the value of the cutoff frequency
in the graviton spectrum. Above this frequency, the spectrum decreases
exponentially, 
and no signal can be detected. Below this frequency,
however, the form of the spectrum is not fixed by general arguments. 
We discuss the issue in the next section.

\subsection{The form of the spectrum}
For a thermal spectrum, the occupation number per cell of the phase
space,
$n_f$, is given by the Bose-Einstein
distribition and therefore, below the characteristic frequency,
\be
n_f=\frac{1}{e^{2\pi f/kT}-1}\sim \frac{kT}{2\pi f}
\ee
Then, from \eq{37} we then find that at low frequencies
a thermal spectrum corresponds to
\be
\ogw (f)\sim f^3\, .
\ee
Of course, if we have a thermal spectrum with a cutoff frequency in
the GHz region, the value of $\hogw (f)$ at, say, $f=1$ kHz is utterly
neglegible. 
However, we have seen in the Introduction that
below the Planck scale  gravitons interact too weakly 
to thermalize, and therefore there is
no a priori  reason for a $\sim f^3$ dependence. The gravitons will
retain the form of the spectrum that they had at time of production,
and this is a very model dependent feature. However, from a number of
explicit examples and general arguments
that we will discuss below, we learn that spectra
 flat or almost flat over a large  range of frequencies seem
to be not at all unusual in cosmology. 
 In particular:

\begin{enumerate}
\item The amplification of vacuum fluctuations always give spectra that
  extend over a huge range of frequency, from $\sim 3\times 10^{-18}$
  Hz, corresponding to wavelength equal to the present size of the
  horizon, up to a cutoff in the GHz region. The spectrum found in
  some cases in string cosmology is particularly interesting from this
  point of view, since it can satisfy the COBE and nucleosysnthesis
  bound, and still have a large intensity at the frequency accessible
  to experiments. The basic reason for this behaviour is the fact that
  in the problem there is only one relevant lengthscale, given by the
  value of the Hubble parameter.

\item Another example of relic GW spectrum which is almost flat over a 
very large
range of frequencies is provided by GWs produced by the decay of
cosmic strings, discussed in sect.~\ref{cosmicstrings}.  
In this case the spectrum has a peak
around $f\sim 10^{-12}$ Hz, where $\hogw$ can reach
a few times $10^{-6}$,
and then it is almost flat
from $f\sim 10^{-8}$ Hz to
the GHz region, where it has the cutoff (fixed by the
arguments previously discussed, considering that the relevant scale
for cosmic string is $\mgut$).  The basic reason for such a behavior
is the scaling property of the string network, which says that a
single lengthscale, the Hubble length, characterizes all properties of
the string network~\cite{Vil,All}. The network of strings evolves
toward a self-similar configuration, with small loops being chooped
off very long strings, and the typical radiation emitted by a single
loop has a wavelength related to the length of the loop.

\item Phase transitions can also give spectra that extend over a large
  range of  frequencies; in this case, as we will see in
  sect.~(\ref{relaxsection}), the  reason is  that in an
  expanding Universe a field cannot have a correlation length larger
  than the horizon size, basically because of causality.

\end{enumerate}

These facts have potentially important
consequencies. They suggest that, {\em 
even if a spectrum of gravitons produced
during the Planck era has a cutoff at frequencies much larger
than the range of frequencies accessible to interferometers, 
still in this range we can hope to observe 
 the low-frequency part of these spectra.}

\subsection{Characteristic intensity}

The numbers discussed in sect.~(\ref{NSsect})
give a first  idea of what can be considered an
interesting detection level for $\hogw (f)$, which should be at least
\be
\hogw\,\lsim\, 5\times 10^{-6}\, ,
\ee
especially considering that the
bound~(\ref{nsbound}) refers not only to
gravitational waves, but to all possible sources
of energy which have not been included, 
like particles beyond the standard model, primordial black
holes, etc.

The next question is whether it is reasonable to expect that some
cosmological production mechanism saturates the nucleosynthesis bound. 
This of course depends on the production mechanism, but some
relatively general
considerations are possible. 

First of all, it is clear that, if GWs are
produced at the Planck scale by collisions and decays
together with the photons that we observe today in the CMBR,
and there is not an inflationary phase at later time,
we expect roughly $\rho_{\rm gw}\sim
\rho_{\gamma}$. Since
$h_0^2\rho_{\gamma}/\rho_c\simeq 2.481\times 10^{-5}$, in this case
 the bound~(\ref{NS}) is approximately
saturated. More precisely,
if at some time $t=t_*$ both the photons that we observe in the CMBR
and gravitons were produced,
and  with 
a total fraction of the energy density in gravity waves $\Omega_*$,
then,
taking into account the different redshifts
of GWs and photons due to the fact that GWs decouple immediately,  
at the present
time we would have~\cite{KKT}
\be
\hogw\simeq 1.67\times 10^{-5}\(\frac{100}{g_*}\)^{1/3}\Omega_*\, .
\ee
If instead the mechanism  that produces GWs is different from the mechanism
that produces the photons in the CMBR, often it is still  possible 
to relate the respective energy densities, simply because
the scales of the two processes are  related.  
A few examples will illustrate the point. 

\begin{enumerate}
\item In sect.~\ref{prebigbang} we will discuss the spectrum produced in
string cosmology by the amplification of vacuum fluctuations. However,
the typical value of the characteristic frequency and the peak value
of $\hogw$ can be understood using only general arguments, following
ref.~\cite{peak}.

The  production temperature is fixed by the
criterium $H_*\sim M_S$, since in this model the Hubble parameter is
stabilized by stringy corrections, and therefore at the string scale,
see sect.~\ref{prebigbang}. 
Using $H_*\sim T_*^2/\mpl$, this
corresponds to an
effective  `temperature' at time of
production $T_*\sim (M_S\mpl )^{1/2}=(M_S/\mpl)^{1/2}\mpl$. 
Redshifting $T_*$ we get a characteristic
frequency today 
\be
f_0\sim \(\frac{M_S}{\mpl}\)^{1/2}T_0\, ,
\ee
and a peak energy density 
\be\label{pe}
\frac{d\rho\sub{gw}}{d\log f}\sim f_0^4 \sim
\left(\frac{M_S}{\mpl}\right)^2\\
\frac{d\rho_{\gamma}}{d\log f}\, .
\ee
Therefore   $\rho_{\rm gw}$ is related to $\rho_{\gamma}$, although
with a suppression factor $(M_S/\mpl )^2$ which is expected to range
between $10^{-4}$ and $10^{-2}$, and numerical factors.

\item Another mechanism that we will discuss is 
the production of GWs through bubble
collisions when inflation terminates with a first order phase
transition. In this case there is only one energy scale, 
the vacuum energy  density  $M$
during inflation (so that $M^4$ is the false-vacuum energy density). 
The value of $M$ fixes the reheating temperature, and therefore
$\rho_{\gamma}$, from
$(\pi^2g_*/30)T_{\rm rh}^4\simeq M^4$, and 
fixes also the energy liberated in
gravitational waves.  Therefore
the energy density in photons and in
gravitons are related. Writing, as in sect.~\ref{charfreq},
$\lambda_* =\epsilon H_*^{-1}$, where $\lambda_*, H_*$ are
the typical wavelength produced and the 
Hubble parameter at time of production, 
the computation of sect.~\ref{bubble} shows that, in a strongly first
order phase transition,
\be\label{tw1}
\rho_{\rm gw}\sim\epsilon^2\rho_{\gamma}\, ,
\ee
so that again $\rho_{\rm gw}$ and $\rho_{\gamma}$ are related.
\end{enumerate}

From these explicit examples we see that, independently of the
production mechanism, an equation of the form
eq.~(\ref{tw1}) is quite general. The scale for
the intensity of a relic GW background is indeed fixed in many cases by
$\rho_{\gamma}$, which therefore gives a first order of magnitude
estimate of the effect. However, there are also suppression factors,
like the factor $\epsilon^2$ in eq.~(\ref{tw}) or
$(M_S/\mpl)^{2}$ in eq.~(\ref{pe}), that, together with the exact
numerical coefficients, are crucial for the detection at present
experiments. A value $\epsilon\sim 0.1$ would allow detection at the
level $\hogw\sim 10^{-7}$ while $\epsilon\sim 10^{-2}$ would require
$\hogw\sim 10^{-9}$, which is beyond the possibilities of first
generation experiments.

Furthermore, another  suppression factor is present 
if we have a spectrum of relic GWs with a total energy density
$\rho_{\rm gw}\sim \rho_{\gamma}$, and
 a cutoff in the GHz, and we
want to observe it in the kHz region. In this case, as we discussed,
we must hope that the spectrum
is practically flat between the kHz and the GHz. While we have seen
that there are various examples of such a behaviour, the fact
that the energy density  is spread over such a wide interval of
frequencies diminuishes  the maximum allowed
value of $\hogw (f)$ at a given
frequency, since the nucleosynthesis bound is a limit on the integral
of $\hogw (f)$ over all frequencies. 
The nucleosynthesis bound~(\ref{nsbound}) 
then gives a maximum value at 1kHz 
\be\label{ns1}
\hogw (1 {\rm kHz})\leq \frac{5.6\times 10^{-6}(N_{\nu} -3)}
{\log\left(\frac{1 {\rm GHz}}{1 {\rm kHz}}\right)}\simeq 4\times 10^{-7}
(N_{\nu}-3)\, .
\ee
If the spectrum extends further toward lower frequencies, the maximum
value of $\hogw$ decreases accordingly, with a factor $\log (1{\rm
GHz}/ f_{\rm min})$ instead of $\log (1{\rm GHz}/ 1{\rm kHz})$.

\section{Amplification of vacuum fluctuations}

\subsection{The computation of Bogoliubov coefficients}\label{Bogo}
The amplification of vacuum fluctuations 
is a very general  mechanisms, first discussed in a cosmological setting
in refs.~\cite{Gri,Sta} and then in many other papers, 
see e.g.~\cite{RSV,FaP,AW1,AW,AH,All2,SS,KW,Gri1}.
Specializing immediately to a FRW metric, it is  convenient to
introduce conformal time $\eta$, related to cosming time $t$ by
$d\eta =dt/a(t)$, so that
\be
ds^2=a^2(\eta )\( -d\eta^2 +d{\bf x}^2\)\, .
\ee
(In the following we set $c=1$).
In the presence of  a classical GW in a FRW background, we write 
\be 
g_{\mu\nu}= a^2(\eta )\(\eta_{\mu\nu} +h_{\mu\nu}\)\, ,
\ee
with $\eta_{\mu\nu}=(-,+,+,+)$ and, in the TT gauge,
\be\label{155}
h_{ab}(\eta ,{\bf x})=\sqrt{8\pi G_N}\,\sum_{A=+,\times}\sum_{{\bf k}}\,
 \phi^A_{\bf k}(\eta ) e^{i{\bf k\cdot x}}e_{ab}^A(\hat{\Omega})\, ,
\ee
Note that ${\bf x}$ are the comoving coordinates; the physical
coordinates are ${\bf x}\sub{phys}=a(t){\bf x}$. Correspondingly,
${\bf k}$ is a comoving momentum, and the physical momentum is
${\bf k}\sub{phys}={\bf k}/a(t)$.

The  Einstein equations, linearized
over the FRW background, give an equation for 
$\phi^A_{\bf k}$, 
\be\label{phi''}
\phi_{\bf k}''+2\frac{a'}{a}\phi_{\bf k}'+k^2\phi_{\bf k} =0\, ,
\ee
where the prime is the derivative with respect to $\eta$,
and we omitted the index $A$.
This equation is just the Klein-Gordon equation for a homogeneous
scalar field in a FRW background,
\be
\Box\phi \equiv \frac{1}{\sqrt{-g}}\partial_{\mu}
\(\sqrt{-g}g^{\mu\nu}\partial_{\nu}\)\phi =0\, .
\ee
The factor $\sqrt{8\pi G_N}$ in \eq{155}
is a convenient choice of normalization,
such that the action of  $\phi^A$ derived from the linearization of
the Einstein-Hilbert action corresponds to the action of two
($A=+,\times $)
canonically normalized scalar fields.
Therefore one can use the standard technique of Bogoliubov
coefficients for a scalar field~\cite{BD}. One considers the situation
in which, over a timescale $\Delta T$, there is a change of regime in
the cosmological evolution, between two regimes that we denote  I
and II
(for instance from an inflationary regime to a
standard radiation dominated era). Let $t_*$ be  the time of transition
between the two regimes, and
consider a mode whose physical frequency, at time $t_*$, is
$f_*$.  For values of $f_*$ such that
$2\pi f_*\Delta T\ll 1$, the change of regime can be considered abrupt,
while if $2\pi f_*\Delta T\gg 1$ the change of regime is
adiabatic.
It is convenient to define a new variable
\be
\psi_{\bf k}(\eta )=\frac{1}{a}\phi_{\bf k} (\eta )\, .
\ee
In terms of $\phi_{\bf k}$, \eq{phi''} becomes
\be\label{psi''}
\psi_{\bf k}''+\( k^2-\frac{a''}{a}\)\psi_{\bf k}=0\, .
\ee
Let us denote by $f_{\bf k}(\eta )$ the solution  of this equation
with the scale factor of phase I, and $F_{\bf k}(\eta )$ for phase II.
Then the mode expansion of $h_{ab}$ in the phase I is
\be\label{I}
h_{ab}=\sqrt{8\pi G_N}\sum_A\int \frac{d^3k}{(2\pi )^3\sqrt{2k}}\,
\frac{1}{a(\eta )}\left[
a_A({\bf k}) f_{\bf k}(\eta ) e^{i{\bf k\cdot x}}+
a_A^{\dagger}({\bf k})
f^*_{\bf k}(\eta )e^{-i{\bf k\cdot x}}
\right]
e_{ab}^A(\hat{\Omega})
\ee
(we have used the fact that our $e_{ab}^A$ are real, \eq{6});
$a_A({\bf k}),a^{\dagger}_A({\bf k})$ are the creation and
annihilation operators in this phase, so that the vacuum in phase I is
defined by
\be
a_{+}({\bf k})|0\rangle_I=a_{\times}({\bf k})|0\rangle_I=0\, ,
\ee
and we can construct the Fock space appropriate to this phase acting
with $a^{\dagger}({\bf k})$. Similarly, in phase II, one has the
expansion 
\be\label{II}
h_{ab}=\sqrt{8\pi G_N}\sum_A\int \frac{d^3k}{(2\pi )^3\sqrt{2k}}\,
\frac{1}{a(\eta )}\left[
A_A({\bf k}) F_{\bf k}(\eta ) e^{i{\bf k\cdot x}}+
A^{\dagger}_A({\bf k})
F^*_{\bf k}(\eta )e^{-i{\bf k\cdot x}}
\right]
e_{ab}^A(\hat{\Omega})
\ee
and a new vacuum state $|0\rangle_{II}$ such that
\be
A_{+}({\bf k})|0\rangle_{II}=A_{\times}({\bf k})|0\rangle_{II}=0\, .
\ee
Since both the $(f_{\bf k},f^*_{\bf k})$ and the 
$(F_{\bf k},F^*_{\bf k})$ are a complete set, we
can express one in terms of the others; of course, since the 
$f_{\bf k}$ alone are not a complete set, the expansion of 
$F_{\bf k}$ will involve both $f_{\bf k}$ and $f_{\bf k}^*$, and
therefore there will be a mixing of positive and negative frequency
modes. The relation between $F_{\bf k}$ and $(f_{\bf k},f^*_{\bf k})$
is known as a Bogoliubov transformation,
\be
F_{\bf k}=\sum_{\bf k'}\( \alpha_{{\bf k}{\bf k'}} f_{\bf k'}
+\beta_{{\bf k}{\bf k'}} f^*_{\bf k'}\)\, .
\ee
Inserting this relation into \eq{II} and comparing with \eq{I} one
finds the relation between creation and annihilation operators
(we omit hereafter the index $A=+,\times $),
\be\label{bog1}
a({\bf k})=\sum_{\bf k'}\[ \alpha_{{\bf k'}{\bf k}}A({\bf k'})
+\beta^*_{{\bf k'}{\bf k}}A^{\dagger}({\bf k'})\]
\ee
and 
\be\label{bog2}
A({\bf k})=\sum_{\bf k'}\[ \alpha^*_{{\bf k}{\bf k'}}a({\bf k'})
-\beta^*_{{\bf k}{\bf k'}}a^{\dagger}({\bf k'})\]\, .
\ee
The Bogoliubov coefficients satisfy 
\bees
\sum_{{\bf k}}\[\alpha_{{\bf k_1}{\bf k}}\alpha^*_{{\bf k_2}{\bf k}}-
\beta_{{\bf k_1}{\bf k}}\beta^*_{{\bf k_2}{\bf k}}\] &=&
\delta_{{\bf k_1}{\bf k_2}}\, ,\\
\sum_{{\bf k}}\[\alpha_{{\bf k_1}{\bf k}}\beta_{{\bf k_2}{\bf k}}-
\beta_{{\bf k_1}{\bf k}}\alpha_{{\bf k_2}{\bf k}}\] &=&0
\ees
If the cosmological background metric is time-dependent but isotropic
and spatially
homogeneous, the gravitational field can give energy to the particles,
but not momentum, so that 
\bees
\alpha_{{\bf k}{\bf k'}}&=&\alpha_f\delta_{{\bf k}{\bf k'}}\nonumber\\
\beta_{{\bf k}{\bf k'}}&=&\beta_f\delta_{{\bf k}{\bf k'}}\, .
\ees
Eqs.~(\ref{bog1},\ref{bog2}) give the formal relation between the Fock
spaces constructed in phase I and II. Now, the basic physical
idea is the following: suppose that just before the transition
the quantum state was $|s\rangle$. For instance, we can express 
$|s\rangle$ in terms of the occupation numbers $n_f$, relative  to the
creation operators $a^{\dagger}_fa_f$. 
Then, for modes such that the transition is sudden,
that is $2\pi f_*\Delta T\ll 1$, the physical state does not have time
to change during the transition, so that it is still 
given by $|s\rangle =|\{ n_f\}\rangle$
just after the transition. However, we now must express 
the occupation numbers with respect to $N_f=A^{\dagger}_fA_f$. Using
the Bogoliubov transformation, one immediately finds
\be\label{N_f}
N_f=n_f+2|\beta_f|^2\left( n_f+\frac{1}{2}\right)\, .
\ee
This shows, first of all, that any preexisting value of $n_f$ is
amplified, with an amplification factor  $1+2|\beta_f|^2$. And
furthermore, even the `half-quantum' due to vacuum fluctuations is
amplified: due to the mixing between positive and negative
frequencies, even the vacuum state of phase I is a
multiparticle state in the Fock space appropriate to phase II.

The above discussion refers to  frequencies $f_*$ such that 
$2\pi f_*\Delta T\ll 1$. In the opposite limit $2\pi f_*\Delta T\gg 1$,
the mode whose physical frequency, at $t=t_*$, is
$f_*$,  sees the transition between the two
regimes as adiabatic, and the quantum state has the time to follow smoothly the
evolution of the scale factor. In this case, therefore, there is no
particle production, or more precisely the amplification factor
approaches one exponentially~\cite{BD}. Therefore, for any practical
purpose, there is a cutoff in the spectrum produced by the
amplification of vacuum fluctuations, at
\be
f\su{max}_*\sim\frac{1}{2\pi\Delta T}\, .
\ee
To obtain the value of the cutoff frequency today, one then redshifts
$f\su{max}_*$ from time of production $t=t_*$ to the present time $t_0$.
In a cosmological setting, the scale of time variation is given by the
Hubble constant, so that $\Delta T\sim H^{-1}$; then the condition 
$2\pi f_*\Delta T\gg 1$ is met when the reduced
physical wavelength $\lambda_*/(2\pi )$ is
smaller than the horizon; thus, modes that at time of transition where
outside the horizon are  amplified. Instead, no
amplification occurs for sub-horizon sized modes.

It is also interesting to note, from \eq{N_f},  that
even if we have an inflationary phase, the occupation numbers $N_f$
{\em after} inflation are very sensitive to the occupation numbers
$n_f$ {\em before} inflation. This is in contrast  with what happens
in most cases after an inflationary phase: usually 
inflation practically
erases the  information about a previous era. This does not happen for
the amplification of vacuum fluctuations, and
the technical reason behind this is that the quantity which is
amplified is a number of particle per unit cell of the phase space.
The volume of a cell of the phase space, $d^3xd^3k/(2\pi )^3$,
is unaffected by the expansion of the Universe,
contrarily to a physical spatial volume $d^3x\sim a^3(t)$.

\subsection{Amplification of vacuum fluctuations in inflationary models}
We now compute the Bogoliubov coefficients for the transition between
an inflationary phase and the radiation-dominated
(RD) phase, followed by the matter-dominated (MD) phase.
We start from the simplest case of exact De~Sitter inflation and then
we compute the modifications for slow-roll inflation.

The amplification of vacuum flucutations at the inflation-RD
transition is the most studied example of this general 
phenomenon~\cite{Gri,Sta,RSV,FP,AW,AH,All2,AKo,Sah,Gri1,Gri2},
but, as we will see, it is by no means the most promising from the
point of view of the detection at interferometers or resonant
masses. Still, we start with this example because is the simpler
setting to understand the computational technique.
 
\subsubsection{De~Sitter inflation}\label{DeSitter}
We start with a FRW cosmological model with 
$ds^2=dt^2-a^2(t)d{\bf  x}^2$, and we concentrate for the moment
on the particle production at the
De~Sitter-RD transition. The scale factors are
$a(t)\sim e^{Ht}$ for $-\infty <t<t_1$, and 
$a(t)\sim t^{1/2}$ for $t_1 <t<t\sub{eq}$, where
and $t\sub{eq}$ is the time when MD sets in.
In terms of
conformal time $\eta$,
\be
a(\eta )= - \frac{1}{H\eta}\, 
\ee
for $-\infty <\eta <\eta_1$, with $\eta_1<0$, and
\be
a(\eta )=  \frac{1}{H\eta_1^2}(\eta -2\eta_1)\, ,
\ee
for $\eta_1 <\eta <\eta\sub{eq}$.
We have fixed the constants so that $a(\eta )$ and $a'(\eta )$
are continuous across the transition.
Here we are considering an
istantaneous transition at $t=t_1$. More realistically, we should
consider a transition that takes place in a time $\Delta t$. 
As already discussed in the previous section, as long
as we consider frequencies such that $2\pi f_*\Delta t\ll 1$, the
approximation of an istantaneous transition gives neglegible
errors. When $2\pi f_*\Delta t\sim 1$, the approximation breaks down, and
there is an exponential suppression
of the particle production.

The solution of \eq{psi''} with these scale factors is easily found.
In general the solution is a superposition of a term that oscillates
as $e^{-ik\eta}$ and a term  $e^{ik\eta}$.
For $\eta <\eta_1$, we impose  the boundary condition that 
only the positive energy solution, i.e. the term $\sim e^{-ik\eta}$,
is present. This is not an assumption on the state of the system at 
$\eta <\eta_1$, but simply a trick to compute the Bogoliubov
coefficient, which can be defined by matching a solution with only
positive frequencies at $\eta <\eta_1$ with the most general solution
at $\eta >\eta_1$. Then, the solutions are
\be
\phi_k(\eta )=\( 1-\frac{i}{k\eta}\) e^{-ik\eta}
\ee
for $-\infty <\eta <\eta_1$, with $\eta_1<0$, and
\be
\phi_k(\eta )=\( \alpha_k e^{-ik\eta}+\beta_k e^{ik\eta}\) 
\ee
for $\eta_1 <\eta <\eta\sub{eq}$.
Requiring that the $\phi_k,\phi'_k$ are continuous across the
transition gives
\be
\alpha_k =1-\frac{i}{k\eta_1}-\frac{1}{2k^2\eta_1^2}\, ,
\hspace{10mm}
\beta_k=\frac{1}{2k^2\eta_1^2}\, .
\ee
Note that $|\alpha_k|^2-|\beta_k|^2=1$. 
The number of particles
produced at the De~Sitter-RD transition, per cell of the phase space,
is given by \eq{N_f}; in particular, if before the transition the
system was
in the vacuum state, $n_k=0$, after the transition 
\be
N_k=|\beta_k|^2=\frac{1}{4k^4\eta_1^4}\, . 
\ee
As discussed before,  $k$ is the comoving momentum and is
related to the physical momentum $k\sub{phys}$ by 
$k\sub{phys}=k/a(t)$. We must now express $N_k$ in terms of
physical quantities observed today, taking into account the redshift
during the RD and MD phases. Of course, during the RD-MD transition
there will be a further creation of quanta; we will discuss it below,
and for the moment we express the spectrum of particles produced at
the De~Sitter-RD transition
in terms of physical quantities measured today.

We reserve the notation $f$ for the {\em physical} 
frequency observed today, so that
\be
2\pi f=(k\sub{phys})_0=\frac{k}{a(t_0)}\, ,
\ee
where $t_0$
is the present value of cosmic time. Therefore, 
\be\label{eta1}
k|\eta_1|=2\pi fa(t_0)|\eta_1|=
\frac{2\pi f}{H}\, \frac{a(t_0)}{a(t_1)}=
\frac{2\pi f}{H}\left (\frac{t_0}{t_{\rm eq}}\right )^{2/3}
\left (\frac{t_{\rm eq}}{t_1}\right )^{1/2}\, ,
\ee
where in the second equality we have used
$a(t_1)=1/(H|\eta_1|)$; we now use~\cite{KT}
\be
\left (\frac{t_0}{t_{\rm eq}}\right )^{2/3}\equiv
1+z\sub{eq}=2.32\times 10^4\, (\Omega_0h_0^2) T_{2.75}^{-4}
\simeq 2.40\times 10^4\, (\Omega_0h_0^2)\, ,
\ee
$\Omega_0$ is the total density of the Universe in units of the
critical density, and we have taken $T=2.728$~K.  For the time of
matter-radiation equilibrium we have~\cite{KT}
$t_{\rm eq}=4.3608\times 10^{10}
(\Omega_0h_0^2)^{-2}(T_{2.75})^6\simeq
 4.1\cdot 10^{10}\,\Omega_0^{-2}h_0^{-4}s$.

From \eq{eta1} we see that
the parameter $\eta_1$ can be traded for a more meaningful 
parameter $f_1$ defined by
\be\label{freq1}
k|\eta_1|=\frac{f}{f_1}\, ,
\ee
so that
\be
f_1 = \frac{H}{2\pi z\sub{eq}}
\left (\frac{t_1}{t_{\rm eq}}\right )^{1/2}\, .
\ee
Note that here $t_1$ is not a free parameter: during RD,
$H(t)=1/(2t)$, so that $H(t_1)=1/(2t_1)$, since $t_1$ is the time when
RD  begin;
and since the scale factor and its derivative are
continuous across the transition, $H(t_1)$ is equal to the constant
value $H$ during the  De~Sitter phase, so that $t_1=1/(2H)$.
Then, inserting the numerical values,
\be
f_1\simeq 10^{9} 
\left( \frac{H}{10^{-4} \mpl} \right)^{1/2}\,\, {\rm Hz}\, .
\ee
Note that both $h_0$ and $\Omega_0$ cancel in $f_1$.
We have chosen $10^{-4}\mpl$ as a useful
reference value for $H$, for reasons
that will be clear below.
In terms of the physical frequency observed
today, $f$, and of the parameter $f_1$, 
\be
N_f=\frac{1}{4}\(\frac{f_1}{f}\)^4\, ,
\ee
if before the transition $n_f=0$.  Inserting this value into
\eq{37} we see that $f$ cancels and we get a flat spectrum,
\be\label{spe1}
\hogw (f)\simeq  10^{-13}\(\frac{H}{10^{-4}\mpl}\)^2\, .
\ee
This is the contribution of the De~Sitter-RD transition. A further
amplification comes from the RD-MD transition. The mode that,
at the time of the transition between RD and MD, had a physical
frequency much bigger than the typical timescale of the transition,
are not subject to further amplification. The typical timescale of the
transition is given by the inverse of the Hubble constant at time of
equilibrium, which is
$H(t\sub{eq})=4(\sqrt{2}-1)/(3t\sub{eq})$ (see~\cite{KT},
pg.~60). A mode that, at
$t=t\sub{eq}$, had a physical momentum  $k\sub{phys}=H(t\sub{eq})$, 
today has a physical frequency $f\sub{eq}$ given by
\be
f\sub{eq}=\frac{1}{2\pi}\,\frac{4(\sqrt{2}-1)}{3t\sub{eq}}\,
\frac{1}{z\sub{eq}}\simeq
10^{-16}(\Omega_0h_0^2)\, {\rm Hz}\, .
\ee
The modes that today have a frequency  $f>f\sub{eq}$ 
have not been further amplified at the RD-MD transition. 
Another useful way to express this result is the statement
that modes that, at the time of transition, had
a physical wavelength much smaller than the horizon are not amplified.
Therefore, the spectrum at $f>f\sub{eq}$  is given by
\eq{spe1}.

\begin{figure}
\centering
\includegraphics[width=\linewidth,angle=270]{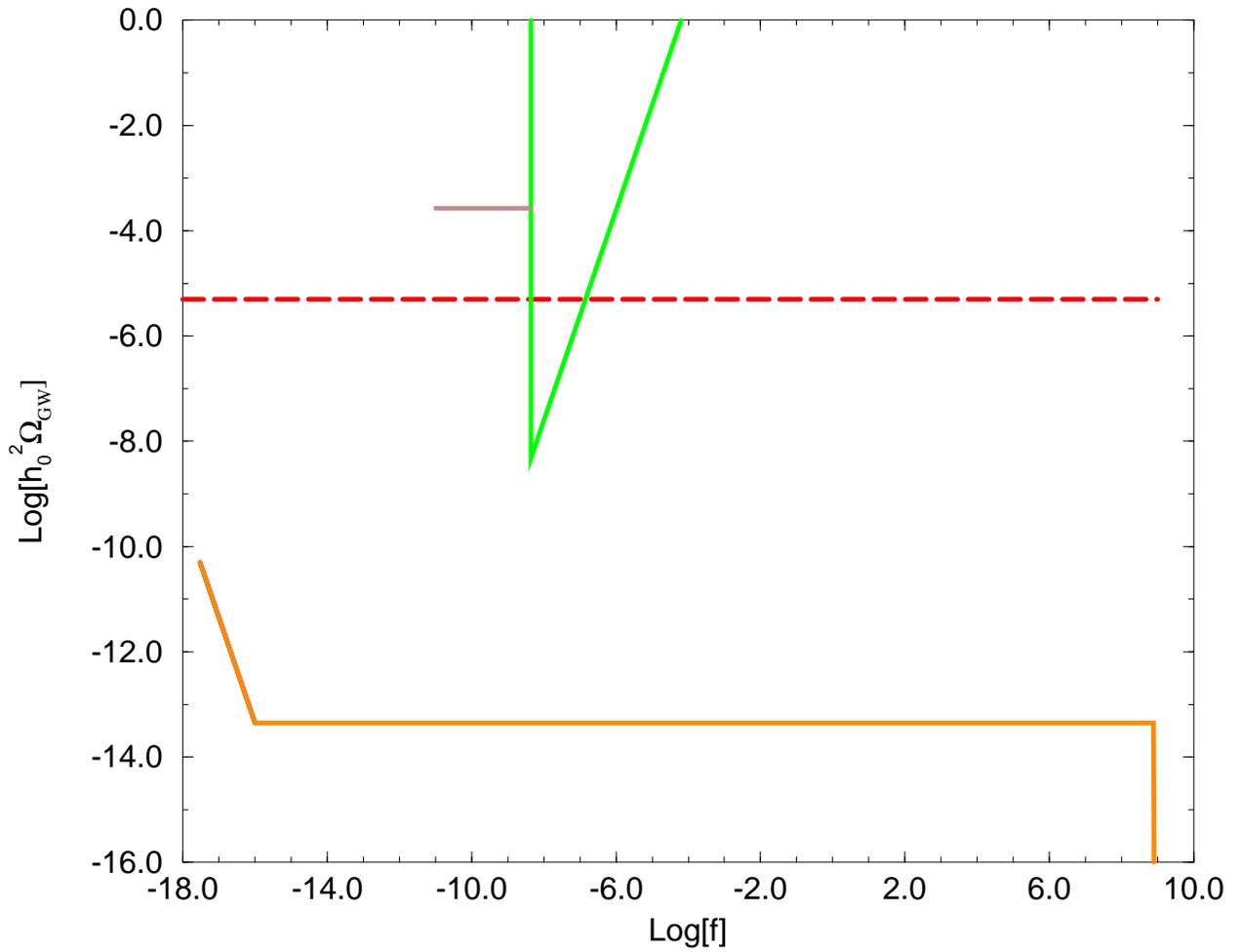}
\caption{The spectrum of amplification of vacuum fluctuations produced
  by a phase of De~Sitter inflation (solid line), with a value of $H$ that
  saturates the COBE bound. The nucleosynthesis bound (dotted line) and
  the pulsar bound (triangle shaped) of fig.~\ref{bounds} are also
  shown for comparison.}
\label{infl}
\end{figure}

For $3\times 10^{-18}$ Hz $<f<f\sub{eq}$ Hz, instead, we must
compute the Bogoliubov coefficients for the RD-MD transition, 
and the total Bogoliubov transformation is obtained with the
composition of the De~Sitter-RD and RD-MD transformations.
As always, the condition $f>3\times 10^{-18}$ Hz comes from requiring
that the mode is inside the horizon today. Thus, these are the modes
that were still outside the horizon at the RD-MD transition and re-entered
the horizon before the present time.

A straightforward computation, similar to the one performed above, 
shows that
in this frequency range the spectrum now goes like $1/f^2$ rather than
being flat. So the final result is
\be
\label{1/f^2}
\hogw (f)\simeq  10^{-13}\(\frac{f\sub{eq}}{f}\)^2\, 
\(\frac{H}{10^{-4}\mpl}\)^2 \, ,
\ee
if $3\times 10^{-18}\,{\rm Hz}<f<f\sub{eq}\simeq 10^{-16}$ Hz, and
\be\label{flat}
\hogw (f)\simeq   10^{-13}\(\frac{H}{10^{-4}\mpl}\)^2\, ,
\ee
for $f\sub{eq}<f<f_1$.
Finally, from the definition of $f_1$, \eq{freq1}, we see that a mode
that today has $f>f_1$,
at the time of transition between De~Sitter and RD had a
wavelength smaller than the horizon; therefore these modes are never
amplified, so that $f_1$ is the upper cutoff  of the spectrum. 
The spectrum is shown in fig.~\ref{infl}.

Note that the behaviour $1/f^2$ in the region
$3\times 10^{-18}$ Hz $<f< 10^{-16}$ Hz 
is  the same as the COBE bound,
and is in fact due to the same physical reason, i.e. to the further
amplification of modes that are inside the horizon at time of 
matter-radiation equilibrium. The comparison of \eq{1/f^2} with the COBE
bound, \eq{cobe2},
then gives an upper limit on the Hubble constant during De~Sitter
inflation~\cite{SS}; 
from a precise statistical analysis, the limit at 95\% c.l.
is~\cite{KW} 
\be
H< 6\times 10^{-5}\mpl\, .
\ee

\subsubsection{Slow-roll inflation}\label{slowroll}
In the above section we have considered a simplified model with an
exact De~Sitter inflation. If inflation is driven by a scalar field
slowly rolling in a potential $V(\phi )$, the Hubble parameter during inflation
is not exactly constant, and this results in a small tilt in the
spectrum. Furthermore, scalar perturbations will also contribute to
the COBE anisotropy, and the ratio of tensor to scalar perturbation is
fixed by the choice of $V(\phi )$. 
Then, in the region $f\sub{eq}<f<f_1$, one finds that
$\hogw (f)$ is not flat, as
in \eq{flat}, but has a frequency dependence
\be
\hogw (f)\sim f^{n_T}\, ,
\ee
where $n_T$ is called the spectral index for tensor perturbations. If 
$V_*$ is the value of the inflationary potential when the present
horizon scale ($k=H_0$) crossed the horizon during inflation, and
$V_*'$ is the first derivative of the potential at that point, then,
to lowest order in the deviation from scale invariance, and for single
field models with a smooth potential~\cite{LT,Tur2,LR}
\be\label{nT}
n_T=-\frac{\mpl^2}{8\pi}\,\(\frac{V_*'}{V_*}\)^2\, .
\ee
Therefore $n_T<0$, and the spectrum is decreasing rather than
flat. Since the slow-roll condition is just 
$|\frac{\mpl^2}{8\pi}\,\(\frac{V_*'}{V_*}\)^2|\ll 1$, then 
$|n_T|\ll 1$.
It is convenient to define the scalar ($S$) and tensor ($T$)
contributions to the
quadrupole anisotropy~\cite{CKLL},
\bees\label{quadS}
S&\equiv& \frac{5\langle |a_{2m}^S|^2\rangle}{4\pi}=
\frac{2.2(V_*/\mpl^4)}{(\mpl V_*'/V_*)^2}\, ,\\
\label{quadT}
T&\equiv& \frac{5\langle |a_{2m}^T|^2\rangle}{4\pi}=
0.61 (V_*/\mpl^4)\, ,
\ees
where the second equalities hold for slow-roll inflation.
From \eqss{nT}{quadS}{quadT} it follows that
\be\label{1/7}
n_T=-\frac{1}{7}\, \frac{T}{S}\, .
\ee
To obtain the precise frequency dependence, we must also take 
into account  the fact that the behaviours
 $\hogw\sim 1/f^2$
and $\hogw\sim$ const. in \eqs{1/f^2}{flat} 
hold only in the limits $f\ll f\sub{eq}$ and $f\gg
f\sub{eq}$. Performing accurately the matching at
$f=f\sub{eq}$, taking into account the tilt $n_T$ 
and defining $k\sub{eq}=2\pi f\sub{eq}$, one finds the
spectrum~\cite{Tur2}
\be
\frac{d\ogw}{d\log k}=
\frac{\Omega_0^2(V_*/\mpl^4)}{(k/H_0)^{2-n_T}}
\[ 1+\frac{4}{3}\frac{k}{k\sub{eq}}+\frac{5}{2}
\(\frac{k}{k\sub{eq}}\)^2\]\, .
\ee
Defining $H_*^2=(8\pi /3)(V_*/\mpl^2 )$ and inserting the numerical
values, we can rewrite it as
\bees
\hogw (f)&\simeq &0.8\times 10^{-13}\(\frac{H_*}{10^{-4}\mpl}\)^2
\,\(\frac{5\times 10^{-17}h_0\,{\rm Hz}}{f}\)^{-n_T}\times\nonumber\\
& &\[ 1+\(\frac{5\times 10^{-17} (h_0^2\Omega_0)\,{\rm Hz}}{f}\)
+\(\frac{6\times 10^{-17} (h_0^2\Omega_0)\,{\rm Hz}}{f}\)^2 \]\, .
\ees
The relation between $n_T, T$ and $S$, \eq{1/7},
goes in the wrong direction for the detection at
interferometers: the higher the contribution of tensor perturbations
at COBE, and therefore the maximum allowed value
of the GW spectrum at, say, 
$f=10^{-16}$ Hz, the steeper is the decrease in the spectrum with the
frequency. Furthermore, when extrapolating through a very large range
of frequencies, 
one must also take into account that $n_T$ itself is not constant, but
rather~\cite{KoT2} 
\be
\frac{dn_T}{d\log k}=-n_T\frac{\mpl^2}{4\pi}\(
\frac{V_*'}{V_*}\)'\, .
\ee
Combining these effects, one finds that, at a fixed
frequency $f=10^{-4}$ Hz, relevant for LISA, the spectrum is maximized
for $n_T\simeq -0.025$~\cite{Tur2}, or $T/S\simeq 0.175$. 
Values of this order are realized
in various models of inflation, although there are as well models that
predict $T/S\sim 10^{-3}$.
As discussed in sect.~\ref{COBE}, 
GWs can contribute to the anisotropy only on angular scales larger
than approximately $1^o$. Combining the results of COBE with 
 other observations on smaller angular scales, one can place upper
limits on $T/S$, and the analysis of ref.~\cite{MSSV,ZSW} gives
$T/S<0.5$, at 95\% c.l.

Even with the value of $n_T$ that maximizes the signal at $f=10^{-4}$ Hz,
the expected
signal at LISA is of order $10^{-15}$, therefore 
various order of magnitudes below the expected sensitivity.

\subsection{Pre-big-bang cosmology}\label{prebigbang}

\subsubsection{The model}
In recent years, a cosmological model derived from the low-energy
effective action of string theory has been proposed~\cite{Ven,GV}
(see ref.~\cite{LWC} for a recent review). 
The simplest starting point is the effective action in the metric-dilaton
sector;
at lowest order in the derivatives and in $e^{\phi}$ it is given by
\be\label{Seff}
S\sub{eff}\sim\int d^Dx\, \sqrt{-g}\, \left [ e^{-\phi}
\left( R+\partial_{\mu}\phi\partial^{\mu}\phi\right)
\right ]\, ,
\ee
where $\phi$ is the dilaton field.
This action receives two kinds of perturbative corrections:
corrections parametrized by $e^{\phi}$, which are just higher loops in
the field theory sense. And corrections proportional to $\alpha '$,
where  $\lambda_s=\sqrt{2\alpha '}$ 
is the string length,
which are genuinely string effects, and take into account the finite
size of the string. Since $\alpha '$ has dimension of (length)$^2$, it
is associated with higher derivative operators as for instance 
the square of the
Riemann tensor, $R_{\mu\nu\rho\sigma}^2$. The remarkable fact observed
in refs.~\cite{Ven,GV} is that  there  exits a regime
in the early Universe 
evolution of this model,  where the perturbative approach is well
justified, and in a first approximation one can use \eq{Seff},
neglecting the corrections. This regime occurs if we take as initial
condition a Universe at weak coupling and low curvatures. 
Then, a generic inhomogeneous string vacuum shows a gravitational
instability~\cite{BDV}. Specializing for simplicity
to a homogeneous model,
the solution of the equations of motion derived from the action
(\ref{Seff}) reads, for  generic anisotropic FRW scale factors
$a_i(t)$, $i=1,\ldots D-1$,
\bees
a_i(t)&=&(-t)^{c_i}\, ,\nonumber\\
\label{kasner}
\phi_i(t)&=&\phi_0+c_0\log (-t)\, ,
\ees
with
\be\label{kas2}
\sum_{i=1}^{D-1}c_i^2=1\, ,\hspace{10mm}
\sum_{i=1}^{D-1}c_i= 1+ c_0 \, .
\ee
This is just a generalization in the presence of the dilaton of the Kasner
solution of general relativity. Considering for simplicity an
isotropic model, one has in particular a  solution with
$c_i=-1/\sqrt{D-1}$, $c_0=-1-\sqrt{D-1}$.
This corresponds to a superinflationary evolution:  the Hubble
parameter $H=\dot{a}/a$ grows until it reaches the string scale
$\sim 1/\lambda_s$, and finally formally diverges at a singularity.
When $H\sim 1/\lambda_s$, of course,
the description  based on the lowest
order action is not anymore adequate. Much work has been devoted
recently to understanding how string theory cures the apparent
singularity, and if it is possible to match this `pre-big-bang' phase
to standard post-big-bang cosmology
\cite{BS,BV,KMO,EMW,CLW2,GMV,MM,FMS0,BM,FMS2}. Apart from the phenomenological
aspects concerning GWs that we will discuss, it is clear that the
problem has great conceptual interest, since it a theoretically well
motivated attempt to
cure the Big-Bang singularity.

\begin{figure}
\centering
\includegraphics[width=0.7\linewidth,angle=270]{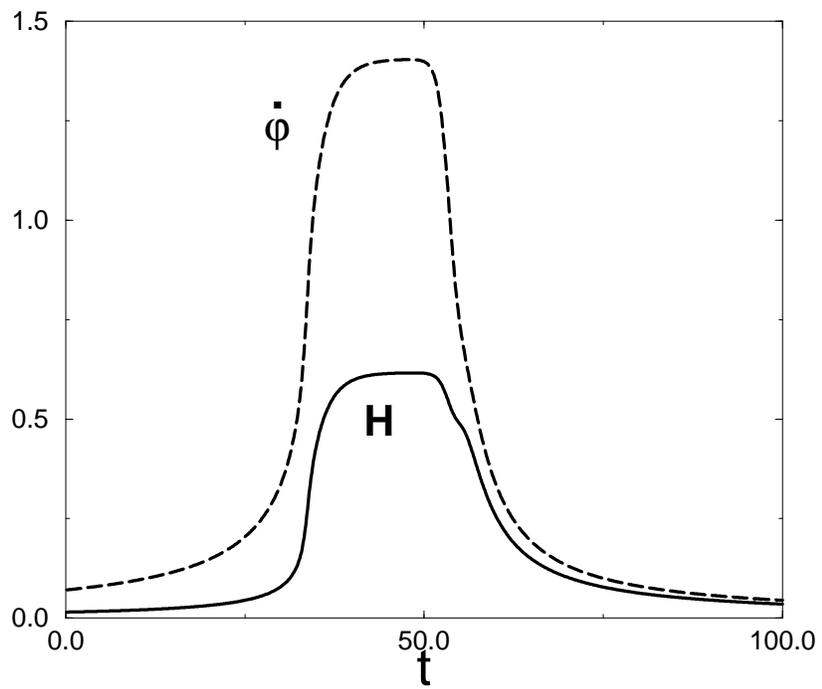}
\caption{The evolution of the Hubble parameter $H$ and of $\dot{\phi}$
  in string cosmology with $\alpha '$ and loop corrections (from
  ref.~\cite{FMS2}).}
\label{hf2}
\end{figure}

One can then investigate whether
 pertubative corrections succeed in turning the regime of 
pre-big-bang accelerated expansion into the standard 
decelerated expansion. Indeed, the inclusion of $\alpha '$ corrections can
turn the unbounded growth of the curvature into a
De~Sitter phase with linearly growing dilaton~\cite{GMV}. Since 
in this new regime the
dilaton keeps growing linearly, finally also $e^{\phi}$ gets large,
 even if the evolution started at very weak coupling; at
this stage loop corrections are important, and they can trigger the
exit from the De~Sitter phase \cite{BM,FMS2}. An example of this
behaviour, using loop corrections derived from orbifold
compactifications of string theory, has been found in
ref.~\cite{FMS2}, and is shown in fig.~\ref{hf2}.

These results are encouraging,  but various technical limitations in
the analysis have also indicated
that most probably
the true resolution of the singularity problem lies outside the domain
of validity of the perturbative approach~\cite{FMS2}.
In this case one must take into 
account that at strong coupling and large curvature
new light states appear
and then the approach based on the
effective supergravity action plus string corrections breaks down.
The light modes are now different and one must  turn to a new
effective action, written in terms of the new relevant degrees of
freedom. In particular, at strong coupling  $D$-branes~\cite{Pol} 
are expected to play an important
role, since their mass scales like the inverse of the string coupling,
$\sim 1/g$ and they are copiously produced by gravitational
fields~\cite{MR}.

For the purpose of computing the amplification of vacuum fluctuations
in this model, the  approach that can be taken is to consider a model
with three phases: first an isotropic 
 superinflationary evolution described by
\eq{kasner}, with the value of the constants $c_i,c_0$
determined by \eq{kas2}. Then this phase is matched to a De~Sitter
phase with linearly growing dilaton, which may be considered as
representative of a typical solution in the large curvature regime;
the constant values of $H,\dot{\phi}$ in this phase are taken as free
parameters, of the order of the string scale. 
Finally, we match this phase to the standard RD era. 
Of course, this model can only be taken as indicative of possible
behaviours in string cosmology, since what happens in the large
curvature phase is certainly much more complicated. However,
we will see that some important 
characteristics of the GW spectrum depend only on the
low curvature regime, where we can use the lowest order effective
action and therefore   is under good theoretical
control. 

Variants of this model can of course be constructed. In particular,
ref.~\cite{ML} considers the effect of a second burst of inflation, 
or  of the possibility that the Universe becomes temporarily dominated
by long-lived massive particles, as for instance the moduli fields of
string theory, which then decay, restoring a standard RD
phase~\cite{GLR}.

\subsubsection{Production of GWs}
We can now discuss the production of GWs through amplification of vacuum 
fluctuations in this model \cite{GV2,GG,BGGV}
(see also refs. \cite{BMU,peak,Bru,AB,MM2,MS,Gas1,Gas2,Gas3,CLLW,Hwa}). We
specialize for illustration to a isotropic model compactified to four
dimensions. Writing the scale factors and the dilaton in terms of
conformal time, from \eq{kasner},  we find that
for $-\infty <\eta <\eta_s$ (with $\eta_s <0$), we have
 a dilaton-dominated  regime with
\bees\label{dil}
a(\eta )&=&-\frac{1}{H_s\eta_s}\left( 
\frac{\eta-(1-\alpha )\eta_s}{\alpha\eta_s}\right)^{-\alpha}\\
\phi (\eta )&=&\phi_s-\gamma\log
\frac{\eta-(1-\alpha )\eta_s}{\eta_s}\, .
\ees
and the Kasner conditions \eq{kas2} give 
$\alpha = 1/(1+\sqrt{3}),\gamma =\sqrt{3}$.

At a value $\eta =\eta_s$ the curvature becomes of the order of the
string scale, and we assume a De~Sitter expansion 
with linearly growing dilaton; in terms of conformal time,
this means
\be\label{str}
a(\eta )=-\frac{1}{H_s\eta}\, ,\hspace{8mm}
\phi (\eta )=\phi_s-2\beta\log\frac{\eta}{\eta_s}\, .
\ee
The parameter $H_s$ is of the order of the string scale, while $\beta$
is a free parameter of the model. Note that $2\beta =\dot{\phi}/H_s$,
where $\dot{\phi}$ is the derivative with respect to cosmic time.

The stringy  phase lasts  for $\eta_s<\eta <\eta_1$ (again 
$\eta_1 <0$), and at 
$\eta_1$ we match this phase to a standard RD era.
This gives, for $\eta_1 <\eta <\eta_r$, (with $\eta_r>0$)
\be\label{RD}
a(\eta )=\frac{1}{H_s\eta_1^2}\,(\eta -2\eta_1)\, ,\hspace{8mm}
\phi =\phi_0\, .
\ee
After that, the standard matter dominated era takes place.
We have chosen the additive and multiplicative
constants in $a(\eta )$ in such a way that 
$a(\eta)$ and $da/d\eta$ (and therefore also $da/dt$)
are continuous across the transitions.

The equation for the Fourier modes of
metric tensor perturbations for the two physical polarizations 
in the transverse traceless gauge is~\cite{GG}
\be\label{psi}
\frac{d^2\psi_k}{d\eta^2}+\left[ k^2-V(\eta )\right]\psi_k =0\, ,
\ee
\be\label{pot}
V(\eta )=\frac{1}{a}\,e^{\phi /2}\,\frac{d^2}{d\eta^2}\(
a\,e^{-\phi /2}\)\, .
\ee
Note that, for constant $\phi$, $V(\eta )$ reduces to \eq{psi''}.
Inserting the expressions~(\ref{dil}-\ref{RD}), the potential is
\bees
&&V(\eta )= \frac{1}{4}\, (4\nu^2-1)\left(\eta-(1-\alpha)\eta_s
\right)^{-2},
\hspace*{10mm} -\infty <\eta <\eta_s \nonumber \\
&&V(\eta )= \frac{1}{4}\, (4\mu^2-1)\eta^{-2},\hspace*{40mm}
\eta_s <\eta<\eta_1\\
&&V(\eta )= 0\,,\hspace*{64mm}\eta_1 <\eta<\eta_r \nonumber 
\ees
where $2\mu =|2\beta -3|,2\nu =|2\alpha-\gamma+1|$. From
$\alpha =1/(1+\sqrt{3})$ and $\gamma =\sqrt{3}$, we get $\nu =0$.
 The exact
solutions of eq.~(\ref{psi}) in the three regions are
\bees\label{sol}
&& \psi_k(\eta )=\sqrt{|\eta -(1-\alpha )\eta_s|}\, C
\,H_{0}^{(2)}(k|\eta -(1-\alpha )\eta_s|)\,,
\hspace*{2mm}-\infty <\eta<\eta_s\nonumber \\
&& \psi_k(\eta )=\sqrt{|\eta|}\,\left[ A_+ \,H_{\mu}^{(2)}(k|\eta|)
+A_- \,H_{\mu}^{(1)}(k|\eta|)\right]\,,
\hspace*{13mm}\eta_s <\eta<\eta_1\\
&& \psi_k(\eta )=i\,\sqrt{\frac{2}{\pi k}}\,\left[ B_+\,e^{ik\eta}
-B_-\, e^{-ik\eta}\right]\,, \hspace*{30mm}\eta_1 <\eta<\eta_r 
\nonumber 
\ees
where $H_{\nu}^{(1,2)}$ are Hankel's functions. 
As we have seen in sect.~\ref{DeSitter}, it is convenient to trade 
$\eta_s,\eta_1$ for more physically meaningful quantities. 
As before, we then write
\be
k|\eta_1|=2\pi fa(t_0)|\eta_1|=
\frac{2\pi f}{H_s}\, \frac{a(t_0)}{a(t_1)}=
\frac{2\pi f}{H_s}\left (\frac{t_0}{t_{\rm eq}}\right )^{2/3}
\left (\frac{t_{\rm eq}}{t_1}\right )^{1/2}\, ,
\ee
and  therefore the parameter $\eta_1$ can be traded for a
parameter $f_1$ defined by
\be\label{f1def}
k|\eta_1|=\frac{f}{f_1}\, ,\hspace{10mm}f_1\simeq 
4.3\cdot 10^{10} {\rm Hz}\, \,
\left(\frac{H_s}{0.15\, M_{\rm pl}}\right)\,
\left(\frac{t_1}{\lambda_s}\right)^{1/2}
\, .
\ee
where we have chosen a reference value $H_s$ of the order of the
string scale, and $t_1\simeq\lambda_s$.
Similarly we can introduce a parameter $f_s$ instead of $\eta_s$,
from $k|\eta_s|={f}/{f_s}$. This parameter depends on the duration
of the string phase, and  therefore, contrarily to $f_1$,
is totally unknown, even as an
order of magnitude. However, since $|\eta_1|<|\eta_s|$, we have
$f_s<f_1$.

To summarize, the model has a  parameter $f_s$ with dimensions of a
frequency,  which
can have any value in the range
$0<f_s <f_1$, with $f_1 \sim 1-100 $ GHz,
and a dimensionless parameter $\mu \ge 0$
(or equivalently $\beta$ with $2\mu=|2\beta -3|$).

Performing the matching between the three phases one finally gets the
spectrum. The exact form is~\cite{BMU}
\bees\label{res}
& & \Omega_{\rm gw}(f)=a(\mu )\,\frac{(2\pi f_s)^4}{H_0^2M_{\rm pl}^2}\,
\left(\frac{f_1}{f_s}\right)^{2\mu +1}\,
\left(\frac{f}{f_s}\right)^{5-2\mu}\,
 \left| H_0^{(2)}\left(\frac{\alpha f}{f_s}\right)\,
J_{\mu}'\left(\frac{f}{f_s}\right)\right. +\nonumber\\
& &\left. +H_1^{(2)}\left(\frac{\alpha f}{f_s}\right)
\,J_{\mu}\left(\frac{f}{f_s}\right)
 -\frac{(1-\alpha)}{2\alpha}\,\frac{f_s}{f}\,
H_{0}^{(2)}\left(\frac{\alpha f}{f_s}\right)\,
J_{\mu}\left(\frac{f}{f_s}\right)
\right|^2\, ,
\ees
where $a(\mu )$ is a number, which depends on $\mu$.
Expanding this expression for small values of $f/f_s$ one  gets
a result of the form
\be\label{low}
\Omega_{\rm gw}(f)\sim
\frac{(2\pi f_s)^4}{H_0^2M_{\rm pl}^2}\left(\frac{f_1}{f_s}
\right)^{2\mu +1}
\left(\frac{f}{f_s}\right)^3
\left\{  a_1 +
\left[
\left( a_2 \log\frac{\alpha f}{2f_s}+a_3\right)
\right]^2
\right\}\, ,
\ee
where $a_1,a_2,a_3$ depend on $\mu$.

\begin{figure}
\centering
\includegraphics[width=0.6\linewidth,angle=270]{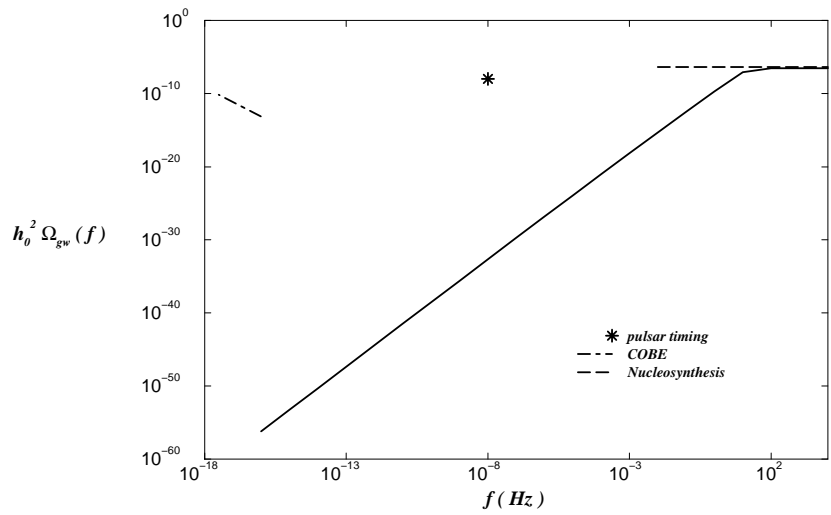}
\caption{The spectrum of GWs due to amplification of vacuum
  fluctuations in string cosmology, compared with the COBE bound
  (dot-dashed), pulsar bound (asterisk) and nucleosynthesis (dotted),
  for a choice of the parameters $f_s=10$ Hz, $\mu =1.5$
(from ref.~\cite{BMU}).}
\label{bmu1}
\end{figure}

\begin{figure}
\centering
\includegraphics[width=0.6\linewidth,angle=270]{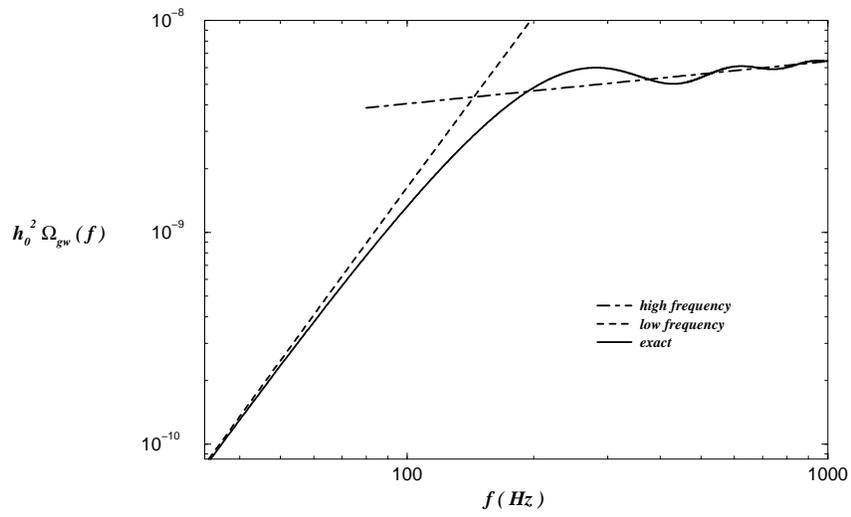}
\caption{The GW spectrum, enlarging the region between
  the Hz and the kHz, for $f_s=100$ Hz,
$\mu =1.4$. The dashed and dot-dashed lines are the low- and
high-frequency limits (from  ref.~\cite{BMU}).}
\label{bmu2}
\end{figure}

In the opposite limit
 $f\gg f_s$ (but still $f$ smaller than the cutoff $f_1$),
one has instead
\be\label{large}
\Omega_{\rm gw}(f)\sim
\frac{(2\pi f_s)^4}{H_0^2M_{\rm pl}^2}
\,\left(\frac{f_1}{f_s}\right)^{2\mu +1}
\,\left(\frac{f}{f_s}\right)^{3-2\mu}
= \frac{(2\pi f_1)^4}{H_0^2M_{\rm pl}^2}\,
\left(\frac{f}{f_1}\right)^{3-2\mu}\, .
\ee
Both limits  reproduces  the frequency dependence first found
in ref.~\cite{GG,BGGV}. It is important to stress that in the high
frequency limit the unknown parameter $f_s$  cancels.

This spectrum has quite interesting properties, depending on the two
free parameters $f_s$ and $\mu$. First of all, it goes like
$f^3$ at low frequencies. This result is a consequence of the
behaviour of the cosmological model in the superinflationary 
pre-big-bang phase, where everything is under good theoretical
control, so this is a rather general prediction of the model.
This means that $\hogw (f)$ can evade the COBE bound
very easily, and yet be large in the range accessible to
interferometers. This is very different from slow-roll inflation,
where instead the spectrum is  decreasing. 

If however the spectrum would
behave as $f^3$ for all frequencies, up to the cutoff $f_1$, 
it would be totally neglegible at,
say, 1 kHz, since the value of $\hogw (f)$ at the cutoff frequency $f_1$
 is bounded by the nucleosynthesis limit discussed in
sect.~\ref{NSsect}.  However, above the unknown frequency $f_s$, the
spectrum predicted by this model
changes  and behaves as $f^{3-2\mu}$; this reflects the
fact that the Universe changed regime and entered a phase of De~Sitter
inflation with linearly growing dilaton. The parameter $\mu$ measures
the growth of the dilaton in units of the Hubble constant $H_s$ during
the string phase, and for an almost constant dilaton,
$3-2\mu\simeq 0$. 

This means that, in some range for the parameters $f_s$ and $\mu$, it
is possible to have the optimal situation from the point of view of
interferometers: a spectrum that grows at very low frequencies, so
that it can evade the COBE bound, and then flattens, at the maximum
level allowed by the nucleosynthesis limit.

The resulting spectrum is shown in fig.~\ref{bmu1}, compared to the
COBE, pulsar and nucleosynthesis  bounds discussed in
sect.~\ref{bounds}. Fig.~\ref{bmu2} enlarges the frequency range
relevant for ground-based interferometers, displaying some features
of the spectrum which might be a signature of this background.

Finally, it is clear that in this model GWs cannot be responsible for
the COBE anisotropies, since the $\sim f^3$ behaviour makes $\hogw$
neglegible at very low frequencies. In the pre-big-bang model, a
possible seed for the CMBR anisotropies can be provided by
pseudo-scalar fields~\cite{CEW,CLW,DGSV,BMUV,BH,MVDV}.

\section{Other production mechanisms}\label{other}

\subsection{Phase transitions}\label{phase}

In the history of the Universe are expected a number of phase
transition. 
In particular, the QCD phase transition takes place at $T_*\sim 150$
MeV. Above this temperature quarks and gluons are deconfined and form
a quark-gluon plasma, while below they are in the confined
phase. Recent lattice calculations with three quark flavors and with the
physical mass for the strange quark,  using Wilson
fermions, suggest that the  phase transition is  first
order~\cite{Iwa}, but the issue is not yet settled (see also
\cite{SSW} for a detailed discussion of cosmological consequencies of
the QCD phase transition).  
Around $T_*\sim 100$ GeV we expect the electroweak
phase transition, when $SU(2)\times U(1)$ breaks to $
U(1)\sub{em}$. Further phase transitions could occur even earlier, at
the grand unified scale. 
These phase transitions are dramatic events in the history of the
Universe, and are good places to look for GW production.

Furthermore, phase transition will in general modify previously
produced spectra~\cite{DSch}. This follows from the fact that during
the phase transition one can have a large drop in the number $g_S$ of
effective species (e.g. by a factor $\sim 3$ during the QCD phase
transition). Since the entropy  $S\sim g_Sa^3T^3$ is constant, this
results in a smaller value of the growth rate of $H^{-1}$ and
therefore in a different rate at which modes cross into the
horizon. For the QCD transition this results in a step in the spectrum
at a frequency $f\sim 10^{-7}$ Hz 
(determined by the value of the Hubble parmeter
at the transition), so that the spectrum above $10^{-7}$ Hz is
diminuished by approximately 30\%~\cite{DSch}. Similar drops happen at
$e^+e^-$ annihilation, with a step at $f\sim 10^{-10}$ Hz, and at the
GUT phase transition, with a step in the GHz region.

\subsubsection{Bubble collisions}\label{bubble}

In first order phase transitions the Universe finds itself in a
metastable state. True vacuum bubbles are then nucleated via quantum
tunneling. In strongly first order phase transition the subsequent
bubble dynamics is relatively simple: once the bubble are nucleated,
if they are smaller than a critical size their volume energy cannot
overcome the shrinking effect of the surface tension, and they
disappear. However, as the temperature drops below the critical
temperature $T_c$, it becomes possible to nucleate bubble that are
larger  than this critical size. These `critical bubbles'
 start expanding, until
their wall move at a speed close to the speed of light. The energy
gained in the transition from the metastable state to the ground state
is transferred to the kinetic energy of the bubble wall.
As the bubble expands, more and more regions of space convert to the
ground state, and the wall  becomes more and more energetic. At the
same time, it also becomes thinner, and therefore the energy
density stored in the wall increases very fast. 
As long as we have a single spherical bubble, this large energy of
course cannot be converted in GWs. But when two bubbles collide we have
the condition for the liberation of a large amount of energy into
GWs. In particular, there are two possible `combustion' modes for the
two bubbles: detonation~\cite{Ste}, that basically takes place when the
boundaries propagate faster than the speed of sound, and deflagration,
when instead they move slower. In the first case there is a large
production of GWs~\cite{Hog,Wit,KKT}.

The characteristic frequency, as seen today, of these GWs is given by
\eq{f1}, with $T_*$ equal to the temperature of the phase transition. 
First of all, we must estimate the parameter $\epsilon$ that enters
\eq{f1}.
 Plausible values of $\epsilon$ have been discussed in
detail by Hogan~\cite{Hog} and by Witten (ref.~\cite{Wit},
app.~B). In the case of the QCD and electroweak transitions
$\epsilon\sim 1$ is excluded because otherwise
in the bubble  collisions there would be overproduction of primordial
black holes. Production of primordial black holes
 is severely constrained because the energy
density of primordial black holes scales like $1/a^3$ in the RD phase,
and therefore they would come to dominate the energy density of
the Universe. In particular, the Universe would not be radiation
dominated   during nucleosynthesis. Actually, primordial black holes
can evaporate through emission of Hawking radiation.  However, in
order to evaporate before nucleosynthesis, the scale at which
the phase transition takes place must be higher than $10^{11}$
GeV~\cite{TuWi}. Therefore $\epsilon$ could be of order one in phase
transitions which take place at these temperatures, but not in the QCD
or electroweak phase transition.

To estimate $\epsilon$ it is necessary to make
assumptions about how the phase transition is nucleated. If it is
nucleated by thermal fluctuations, ref.~\cite{Hog} suggests an upper
bound $\epsilon \lsim 10^{-2}$, but recent lattice simulations 
(restricted however to  pure QCD without fermions) give
a much smaller value $\epsilon\sim 10^{-6}$, which would make the GWs
from the QCD transition neglegible, as will be clear below.
For nucleation of bubbles via quantum
tunneling a detailed analysis has been done in ref.~\cite{TWW}. 
The relevant parameter is the nucleation rate; the phase transition
can be modelled assuming an exponential bubble nucleation rate per
unit volume,
\be
\Gamma =\Gamma_0e^{\beta t}\, .
\ee
The parameter $\beta$ sets the scale for the typical time
variation, and therefore the frequency at time of production, $ f_*$,
is determined by $\beta$. In fact, ref.~\cite{KKT} founds that the
spectrum peaks at 
 $2\pi f_*\simeq 2\beta$. Then
\be
\frac{1}{\epsilon} \equiv \frac{f_*}{H_*}\simeq
\frac{\beta}{\pi H_*}\, .
\ee
Of course, $\beta$
is a model-dependent quantity,
but for a wide class of models one typically finds $\epsilon\sim
10^{-3}-10^{-2}$~\cite{Hog1,Hog,TWW,KoT,KKT}.  

If however the transition is nucleated
by impurities (which is most often the case, except in very pure and
homogeneous samples) the issue is much more complicated. For instance,
the impurities could be given by turbolent motion of the cosmic fluid
generated prior to the QCD epoch; they would depend on the detailed
spectrum of the fluid motion, or in general on the characteristic
distance between impurities, and in this case $\epsilon$ is basically
impossible to estimate, even as an order of magnitude~\cite{Wit}.

Coming back to the simplest case of nucleation by quantum or thermal
fluctuations, 
for the electroweak phase transition, ref.~\cite{KKT}
finds
\be
\frac{\beta}{H_*} \simeq 1.3\times 10^{-3}
\ee
and therefore
\be
f_0\simeq 4.1\times 10^{-3}\, {\rm Hz}\, .
\ee
This phase transition is therefore very relevant for LISA. 
With the same value for $\epsilon$, the QCD phase transition would
instead peak around $f_0\sim 4\times 10^{-6}$ Hz. A smaller value of
$\epsilon$ is required to  bring it into the frequency window of LISA.

A first estimate of
the value of $\hogw$, that as we will see really holds only for
strongly first order phase transitions, can be given as follows~\cite{TuWi}.
The typical wavelength $\lambda_*$
produced in the collision of two bubbles will
be of the order of the radius of the bubbles when they collide, which
in turn is a fraction of the horizon scale $H_*^{-1}$.
 The energy liberated in GWs in the
collision of two bubbles is of order 
\be
E_{GW}\sim \frac{GM_B^2}{\lambda_*}\, ,
\ee
where $M_B\sim \rho\sub{vac}\lambda_*^3$ 
is the energy of a typical bubble and $\rho\sub{vac}$ the difference
in energy density from the metastable state to the ground state. The
fraction of the false vacuum energy that
goes into GWs instead of going into $\rho_{\gamma}$
is therefore $E\sub{GW}/M_B$. Since
\be
\frac{E\sub{GW}}{M_B}\sim \frac{GM_B}{\lambda_*}\sim
G  \rho\sub{vac}\lambda_*^2
\ee
and $H_*^2\sim G \rho\sub{vac}$, we get,  at time of
production 
\be\label{tw}
\rho_{\rm gw}\sim \lambda_*^2H_*^2\rho_{\gamma}=
\epsilon^2\rho_{\gamma}\, .
\ee
Therefore $\rho_{\rm gw}$ is related to $\rho_{\gamma}$, and the basic
reason is that there is essentially only one dimensionful parameter
that enters in the estimate.  
If we neglect the variation in $g(T)$ from time of
production to the present value, 
 $\rho\sub{gw}/\rho\sub{\gamma}$ is
almost constant, and since today 
$h_0^2\rho\sub{\gamma}/\rho_c\simeq 2.5\times 10^{-5}$, this quick 
order of magnitude estimate gives $\hogw (f_0)\sim
10^{-5}\epsilon^2$. However, further suppression factors appears in
the accurate computation of ref.~\cite{KKT}, which  gives
\be
\hogw (f_0)\simeq 1.1\times 10^{-6}\kappa^2
\(\frac{H_*}{\beta}\)^2\(\frac{\alpha}{1+\alpha}\)^2
\(\frac{v^3}{0.24+v^3}\)\(\frac{100}{g_*}\)^{1/3}\, .
\ee
Here $\alpha$ is the ratio of vacuum energy to thermal energy in the
symmetric phase. It characterizes the strength of the phase
transition; for $\alpha\ra 0$ we have a very weak first order and for
$\alpha\ra\infty$ a very strong first order phase transition.
For the electroweak phase transition, ref.~\cite{KKT} gives
\be
\alpha =1.4\times 10^{-3}\, .
\ee
$\kappa$ is an efficiency factor quantifying the fraction of
vacuum energy which goes into bubble wall kinetic energy, and in the
electroweak case is estimated to be $\kappa \simeq 7.8\times 10^{-3}$.
Finally, $v$ is the velocity of the 
bubble wall, and is approximately equal to the speed of sound during
RD, $v\simeq 1/\sqrt{3}$. Altogether, these suppression factors give
a disappointing 
\be
\hogw\sim 10^{-22}\, .
\ee
These suppression factors are a
consequence of the fact that the electroweak phase transition is very
weakly first order, if first order at all. 
It is clear that strongly first order phase
transitions are needed for a detectable signal.

Quite interestingly, the requirement of strongly first order phase
transition is the same that it is necessary to make possible electroweak
baryogenesis  (see e.g. ref.~\cite{RTr} for a recent
review). In particular, the ways in which baryons can be produced have
been separated into `local baryogenesis', when both
baryon number violating
and CP violating processes 
occur near the bubble wall, and `nonlocal baryogenesis', when only CP
violating processes take place near the wall. The condition for local
baryogenesis to dominate is that the speed of the wall be greater
than the speed of sound, which is the same condition for detonation to
dominate the bubble-bubble collisions. 

Partly motivated by the importance for baryogenesis, there have been
many recent investigations on the strength of the phase transition in
extensions of the Standard Model, and in particular in the Minimal
Supersymmetric Standard Model (MSSM)~\cite{Giu,Myi,EQZ,BEQZ}.
Both analytical~\cite{CQW,DGGW,Esp,BJLS,Los,FL,dCE}
and lattice~\cite{Lai,LaR,CK} computations have shown that in the MSSM
the phase transition can be strongly first order if the top squark is
lighter then the top quark. 
The observation of a relic stochastic background at LISA could
therefore  give us a hint for physics beyond the Standard Model.

\subsubsection{Turbolence}

For weakly first order phase transitions the situation is actually
more complex. Bubble nucleation occurs both via quantum
tunneling and with thermal fluctuations, and the subsequent evolution
of the bubble  depends on the interaction of the wall with the
surrounding plasma; part of the energy
gained in the transition from the metastable state to the ground state
is used to heat up the plasma, and another fraction  is converted into
bulk motion of the fluid.  If the Reynolds number of the Universe at
the phase transition is large enough, then this results in the onset of
turbolence in the plasma, and this is another powerful
source of GWs. 

The estimates for the characteristic frequency and for $\hogw$ turn
out to be~\cite{KKT}
\be
f_0\simeq 2.6\times 10^{-8}\,{\rm Hz}\,  \frac{v_0}{v}\(\frac{\beta}{H_*}\)
\(\frac{T_*}{1\,{\rm GeV}}\)\(\frac{g_*}{100}\)^{1/6}\, ,
\ee
\be
\hogw\simeq 10^{-5}\(\frac{H_*}{\beta}\)^2vv_0^6
\(\frac{100}{g_*}\)^{1/3}\, ,
\ee
where $v_0$ is the fluid velocity on the largest length scales on
which the turbolence is being driven, and $v$ the velocity of the
bubble wall. 
These estimates indicate that fully developed turbolence can be
comparable, or even more potent than bubble collisions in generating
GWs. 

\subsubsection{Scalar field relaxation}\label{relaxsection}

Consider a
global phase transition in the early Universe, 
associated with some scalar field $\phi$
that, below a critical temperature, gets a vacuum expectation value
$\langle\phi\rangle$. For causality reasons,
in an expanding Universe the scalar field
cannot have a correlation length larger than the horizon size. So,
even if the configuration which minimizes the energy is a constant
field, the field will be constant only over a horizon distance.
During the RD phase the horizon expands, and the
field will relax to a spatially uniform configuration within
the new horizon distance. This relaxation process will in general
produce gravitational waves, since there is no reason
to expect  $\langle T_{\mu\nu}\rangle$ for the classical field
just  entering the horizon to be spherically symmetric~\cite{Kra}.
A simple estimate of the spectrum produced makes use of the fact that the
characteristic scale of spatial or temporal variation of the field is
given by the inverse of the
Hubble constant, $H^{-1}$, at the value of time under consideration. Then,
the energy density of the field is $\rho\sim (\partial\phi )^2
\sim \langle\phi\rangle^2H^2$. The total energy 
in a Hubble volume is $\rho H^{-3}
\sim \langle\phi\rangle^2H^{-1}$ and
the corresponding quadrupole moment is  $Q\sim (\rho H^{-3}) H^{-2}\sim
\langle\phi\rangle^2H^{-3}$, times a constant smaller than one, which
measures the non-sphericity of the configuration.
The energy liberated in GWs in a horizon
time is then given by~\cite{Kra}
\be
\Delta E\sim H^{-1}\times {\rm Luminosity}\sim
H^{-1}G\left(\frac{d^3Q}{dt^3}\right)^2\sim 
G\langle\phi\rangle^4H^{-1}\, ,
\ee
and the energy density of GW produced is
\be\label{kr}
\(\frac{d\rho_{\rm gw}}{d\log f}\)_*
\sim \frac{\Delta E}{H^{-3}}\sim G\langle\phi\rangle^4H^{2}\, .
\ee
This is the energy density  at a frequency $f\sim H$,  at a
value of time $t=t_*(f)$ which is the time  when the mode with frequency
$f$ enters the horizon. The asterisk in  $(d\rho/d\log f)_*$  reminds that
this quantity is
evaluated at horizon crossing, therefore
at different values of time for different
frequencies. However, in the RD phase, the energy density
scales like $\rho\sim
1/a^4(t)\sim t^{-2}$ and $H\sim t^{-1}$ so that 
$\rho$ scales as $H^2$. Therefore, eq.~(\ref{kr}) means that, if we
evaluate $d\rho/d\log f$ at the same value of time, for all
frequencies which at this value of time are inside the horizon, the
spectrum is flat, $\hogw (f)\sim$ const.
This is a rather nice example of how a flat spectrum can follow simply
from the requirement of causality and simple dimensional estimates.
The simplest way to redshift \eq{kr} to the present time is to 
use $H^2=(8\pi /3)G\rho_{\gamma}$, so that \eq{kr} can be written as
\be
\(\frac{d\rho_{\rm gw}}{d\log f}\)_*\sim \frac{8\pi}{3}
G^2\langle\phi\rangle^4 (\rho_{\gamma})_*\, ,
\ee
where both sides are evaluated
at horizon crossing. Then $\rho_{\rm gw}$ redshifts as $1/a^4$,
$\rho_{\gamma}$ adiabatically, \eq{adia}, and $\rho\sub{rad}=
g(T)\rho_{\gamma}$, so that
$(\ogw )_0=(g_0/g_*)^{1/3}
(\ogw )_*$, and therefore \eq{kr} gives, for the energy density today, 
\be\label{relaxation}
\hogw \sim 5\times 10^{-5}\(\frac{\langle\phi\rangle}{\mpl}\)^4
\(\frac{100}{g_*}\)^{1/3}\, .
\ee
A quite  similar mechanism has been studied in ref.~\cite{Hog2},
considering a theory with a multicomponent scalar field 
$\vec{\phi}$ and an effective potential $V(\vec{\phi})$ with
degenerate minima at a set of points with $|\vec{\phi}|=\phi_0\gg 1$. 
The direction in field space with $\partial V/\partial\vec{\phi}$
define massless Goldstone bosons. Cooling down from a state with
higher symmetry generates spatial gradients in $\vec{\phi}$ and therefore
excites these Goldstone bosons. The modes that enter the horizon then
produce GWs with the same mechanism described above, and again one
finds~\eq{relaxation}.

Clearly, the effect is significant only if
$\langle\phi\rangle$ is not too far from the Planck mass.
Since the mechanism is completely 
general, it is well conceivable that it might take place in string theory,
which has a multitude of 
scalar fields, like the dilaton and the moduli of the
compactification. At tree level these fields correspond to flat
directions, but non-perturbative effects will  lift the degeneracy; it
is then very natural to expect that at the same time they produce a
number of degenerate or almost degenerate vacua.
In this case this effect could give an extremely interesting signal
both for LISA and for ground based interferometers. 

Of course, if we want to avoid the COBE bound, it is necessary to
suppress the spectrum at low frequencies. As suggested in
ref.~\cite{Hog2}, this might happen if the Goldstone modes are coupled to
fields that strongly damp their oscillations after a certain epoch. For
example there could be a second phase transition that removes the
degeneracy between the minima of the potential; the field would then
relax everywhere to a single minimum, suppressing further GW
production. 

Finally, another related production mechanism is the collapse of
solitons or of 
unstable domain walls~\cite{Gle,GR}. In this case the spectrum is quite
peaked, at a frequency determined by the expectation value of the scalar field
$\phi_0$, and for $\phi_0$ of the order of the electroweak scale the
signal would be relevant for LISA.

\subsection{Cosmic strings}\label{cosmicstrings}

Cosmic strings are topological defect that might have formed
during phase transitions in the early Universe
(see e.g.~\cite{Vil,VS} and references
therein). Interest in cosmic strings has also been fueled by the
fact that they were considered a candidate for seeding structure
formations. The most recent comparison with CMB anisotropies show that
in models with  vanishing cosmological constant they do not reproduce
the dependence of the anisotropy from the multipole
moment~\cite{PST,ABR}. However, with a non-vanishing cosmological
constant they can still be a viable option~\cite{ASWA,BRA}.

The spectrum of GWs produced by cosmic strings is discussed in detail
in refs.~\cite{Vil,VS,CA,All,CBS,BCS}. Gravitational waves are produced by the
relativistic oscillations of a cosmic string, in a cosmic string
network. The dynamics of this string network  has been studied in
detail, and it has been found that it obeys a scaling property, so
that the only relevant scale is given by the Hubble length. Small
loops oscillate producing GWs and disappear, while more small loops
are chooped of very long strings, replacing the loops that
disappear, and  the network has 
a self-similar configuration, with loops of
all lengthscales. Since the wavelength of the GWs emitted is fixed by
the length of the loop, the spectrum of the cosmic string network
extends across a very large frequency band.

The spectrum has two main features: an almost flat region that extends
from $f\sim 10^{-8}$ Hz up to $f\sim 10^{10}$ Hz, and a peak in the
region $f\sim 10^{-12}$ Hz.

At LIGO/VIRGO frequencies we are in the flat part of the spectrum, 
and the typical 
estimate of the intensity is of order 
\be
\hogw\sim 10^{-8}-10^{-7}\, .
\ee
(Similar results are obtained from hybrid topological
defects~\cite{XMV}). 
The scale for these numbers is given first of all
by the combination $(G\mu )^2$, where $\mu$ is the mass per unit
length of the string. It comes from the fact that a loop radiates with
a power $P\sim \gamma G\mu^2$, ($\gamma$ is a dimensionless constant)
while, evaluating $\hogw$, another factor of $G$ comes from $1/\rho_c$.
For strings created at the GUT scale $G\mu\sim 10^{-6}$. Then, there
are also large 
numbers related to the number of strings per horizon volume,
which is thought to be of order 50, to the fact that 
also $\gamma\sim 50$,
and to the size of the loops at formation time~\cite{All}, that finally
give a value $\hogw\sim 10^{-8}-10^{-7}$.

For the cosmic string spectrum the most relevant bound comes from the
pulsars, discussed in sect.~\ref{bounds}. Pulsar timing implies a
bound on the mass per unit length of the string~\cite{BCS}
\be
\frac{G\mu}{c^2}< (5.4\pm 1.1) \times 10^{-6}\, .
\ee
For comparison, 
the value of $\mu$ obtained fitting  CMBR anisotropies in an open
Universe is $(G\mu )/c^2\sim 1.7\times 10^{-6}$ for $\Omega_0=0.2$ 
\cite{ACM}. The improvement in pulsar timing expected in the next few
years, see sect.~\ref{bounds}, 
will therefore be crucial for testing the cosmic string scenario.

\subsection{Production during reheating}
If we  have an inflationary phase below the Planck scale, there
is no gravitational analog of the 2.7K radiation. At the Planck scale
photons and gravitons can be produced simply by thermal collisions,
with probably similar production rates. However, the 
particles produced in this way at the
Planck (or string) scale  are exponentially diluted 
by the subsequent inflationary phase, and the photons that we see
today in the cosmic microwave background radiation (CMBR)
have been produced during the reheating era which terminated the
inflationary phase. 

Since the reheating temperature $T_{\rm rh}<M_{\rm
infl}\ll \mpl$, thermal collisions are by now
unable to produce a substantial amount of
gravitons. However, in this case
relic GWs can  be produced  through
some  non-equilibrium  phenomenon connected with
the reheating process. 
For a process that takes place at the reheating temperature $T_{\rm
rh}$ the estimate of the characteristic frequency  is given by 
eq.~(\ref{f1}) with $T_*=T_{\rm rh}$. In principle the reheating
temperature can be between a minimum value at the  
TeV scale (since this
is the last chance for baryogenesis, via the anomaly in the
electroweak theory) and a maximum value $T_{\rm rh}\sim M_{\rm infl}
\lsim 3 \times 10^{16}$ GeV. In supersymmetric theories, the gravitino
problem give a further constraint $T_{\rm rh}\lsim 10^9$
GeV~\cite{KT,KM}. More precisely, ref.~\cite{KM} gives a bound on
$T_{\rm rh}\sim 10^6-10^9$ GeV for a gravitino mass 100 GeV--1~TeV,
and a bound $10^{11}-10^{12}$ GeV independent of the gravitino mass.

So, phenomena occurring at reheating can
manifest themselves with cutoff frequencies between $10^{-4}$  and
$10^9$ Hz,  if we set $\epsilon =1$ in eq.~(\ref{f1}).
In typical non-supersymmetric models the reheating temperature is
large, say $10^{14}$ GeV, corresponding to $f_0\sim (1/\epsilon )
10^7$ Hz, and in this case interferometers or resonant masses
could only look for their low-frequency tails. 
One should  however keep in mind
the possibility that bubble nucleation  occurs before the end of
inflation, and then the cutoff frequency would be redshifted by
the subsequent inflationary evolution toward lower values, possibly
within the VIRGO/LIGO frequency range~\cite{BAFO}.

If instead
the reheating temperature is $T_{\rm rh}\sim 10^6-10^9$ GeV, non-equilibrium
phenomena taking place during reheating would give a signal just in
the LIGO/VIRGO frequency band. A reheating 
temperature $T\sim$ TeV would instead be relevant for LISA. 

A  mechanism for GW generation during reheating is  bubbles
collision, in the case when inflation terminates with a first order
phase transition~\cite{TuWi,KKT}. We have  already discussed it  in
sect.~\ref{bubble}. 

Another possibility is that  reheating occurs through the decay
of the inflaton field. In this case, in a very general
class of models, there is first an explosive stage called preheating,
where the inflaton field decays through a non-perturbative process
due to  parametric resonance~\cite{KLS}, and at this stage there are 
mechanisms that can produce GWs, see~\cite{param,Bas}. 
We believe that in these cases a reliable estimate of the
characteristic frequency is really very difficult to obtain, since the
relevant parameter, which is
the value of the Hubble constant at time of production, depends
on the complicated dynamics of preheating and on the specific
inflationary model considered. Ref.~\cite{param} examines models 
where the characteristic frequency turns out to be
in the region between a few tens and a few
hundreds  kHz.

\section{Stochastic backgrounds of astrophysical origin}\label{astro}

The emission of GWs from a large number of unresolved astrophysical
sources can create a stochastic background of GWs. This background
would give very interesting informations on the state of the Universe at
redshifts  $z\sim 2-5$. It will provide a  probe 
of star formation rates, supernova rates, branching ratios between
black hole and neutron star formations, mass distribution of black
holes births, angular momentum distributions and black hole growth
mechanisms~\cite{BJ2}. At the same time, however, from the point of
view of cosmological backgrounds produced in the primordial Universe,
the astrophysical background is a `noise', which can mask the relic
cosmological signal. 

In this section we investigate a number of astrophysical mechanisms
that can produce a stochastic background of GWs. 
As for a general orientation into the typical intensity and frequency,
we should note that 
GWs of astrophysical origin of course are not subject to the nucleosynthesis
bound, since they were created much later, so that one of our main
`benchmarks' disappears in this case. The estimates of the typical
values of $\hogw$ is strictly dependent on the mechanisms, and we will
see examples below. 

Concerning the frequency, 
a first observation is that there is a maximum frequency at which 
astrophysical sources can radiate. This
comes from the fact that a source of mass $M$, even if very compact,
will be at least as large as its gravitational radius $2GM$, the bound
being saturated by black holes. Even if its
surface were rotating at the speed
of light, its rotation period would be at least $4\pi GM$, and the
source cannot emit waves with a period much shorter than that. 
Therefore we have a maximum frequency~\cite{Tho2},
\be
f\lsim \frac{1}{4\pi GM}\sim 
10^4\frac{M_{\odot}}{M}{\rm Hz}\, .
\ee
To emit near this maximum frequency an object must presumably
have a mass of the order of the
 Chandrasekhar limit $\sim 1.2 M_{\odot}$, which gives a maximum
frequency of order 10~kHz~\cite{Tho2}, and this limit can be saturated
 only by very compact  objects
 (see ref.~\cite{FP} for a recent review of GWs emitted in the
gravitational collapse to black holes, with typical frequencies 
$f\lsim 5$ kHz).
The same numbers, apart from factors of order one, can be obtained
using the fact that for a self-gravitating Newtonian system 
with density $\rho$, radius $R$ and mass $M=\rho (4/3)\pi R^3$,
there is a natural
dynamical frequency~\cite{Schu} 
\be
f_{\rm dyn}=\frac{1}{2\pi}(\pi G\rho )^{1/2}=
\left(\frac{3GM}{16\pi^2 R^3}\right)^{1/2}\, .
\ee
With $R\geq 2GM$ we recover the same order of magnitude estimate
apart from a factor $(3/8)^{1/2}\simeq 0.6$.
This is  an interesting  result, because it
shows that the natural frequency domains of cosmological and
astrophysical sources can be very different. In particular,
a GW signal detected  above,  say, 10 kHz, would be
unambiguosly of cosmological origin. However, as we have seen, it is
not easy to build sensitive detectors operating at such high
frequencies.

In discussing a stochastic background 
of astrophysical origin, apart
from the characteristic frequency and characteristic  intensity, there
is a third important parameter, which, 
for burst sources,  is the duty cycle $D$. 
This is given by the ratio of the typical duration of the signal
(e.g. 1~ms for supernovae) to the average distance between successive
bursts. If $D\ll 1$, the signal is not stochastic. For intermediate
values, say $D\simeq 0.1$, we have a so called `popcorn' noise,
or shot noise. As
$D\ra 1$, we get a continuous, stochastic background. 
The value of $D$ is important because, if $D\ll 1$, we can distinguish
a signal from the very early Universe form an astrophysical signal,
even when the latter corresponds to higher peak amplitude.

\subsection{Supernovae collapse to black-hole}\label{core}
Sufficiently massive stars, at the final stage of their evolution,
collapse to form a black hole or a neutron star. In this explosive
supernova event GWs are liberated. 
Depending on the rate of supernova
events, this can give origin to a stochastic background.
The typical timescale of supernovae is the millisecond, and therefore
the typical frequency of the GWs produced is around the kHz. These
processes are therefore relevant for ground based experiments.
 
Earlier
estimates~\cite{BJ1,BJ2} suggested  a duty cycle $D\sim 0.3$ and
a value $\hogw$ possibly even of order $10^{-6}$ at the kHz, but
acknowledged the need for better collapse models and duty cycle
informations.

Recently, a careful study has been done in ref.~\cite{FMSa}. 
The crucial point is the determination of the star formation
rate. However, in the last  few years the understanding of the origin
and evolution of galaxies has greatly improved, thanks to the
spectacular data from the Hubble Space Telescope, Keck, and other
large telescopes, that have allowed to investigate the Universe up to
redshifts $z\sim 4-5$. This makes possible a determination of the star
formation rate density evolution, based on observations. Denote by 
$\dot{\rho}_*(z)$ the comoving star formation rate density, i.e. the
mass of gas that goes into stars, per unit time and comoving volume
element, at redshift $z$. Then  the rate of core-collapse
supernovae event $R\sub{SN}(z)$, i.e. the number of events per unit
time in a comoving volume, at redshift $z$, is~\cite{FMSa}
\be
R\sub{SN}(z)=\int_0^z dz'\, \frac{dV}{dz'} \,
\frac{\dot{\rho}_*(z')}{z'+1}
\int_{M_p}^{M_u}\Phi (M) dM\, ,
\ee
where $\Phi (M)$ is the Salpeter initial mass function
and $dV/dz$ the comoving volume element. The lower
limit $M_p$ depends on the specific nature of supernovae considered:
the smallest progenitor mass which is expected to lead, after
collapse,  to black hole is in the range $(18-30)\, M_{\odot}$, and
the reference value $25M_{\odot}$ has been used in~\cite{FMSa}, while $M_u=
125M_{\odot}$. With these values,  adopting a value $h_0=0.5$ for
the Hubble constant and assuming a flat cosmology, $\Omega_0=1$, with
vanishing cosmological constant, the total number of supernovae
explosions per unit time leading to black hole formation 
is~\cite{FMSa} 
\be
R\sub{BH}=4.74\,\,{\rm events/s}\, .
\ee
The duty cycle due to all sources up to redshift $z$ is defined as
\be
D(z)=\int_0^{z}dR\sub{BH}\, {{\Delta\tau}}\sub{GW}(1+z)\, ,
\ee
where ${{\Delta\tau}}\sub{GW}$ is the average time duration of
single bursts at the emission, and is of order 1ms. Then the total
duty cycle turns out to be~\cite{FMSa}
\be
D=1.57\times 10^{-2}\, .
\ee
This implies that this GW background is not stochastic, but rather a
sequence of bursts, with typical duration 1ms and a much larger
typical separation $\sim 0.2$~s.  This is  a good new from the
perspective of primordial Universe cosmology, since this astrophysical
background can be distinguished from relic GWs produced at much higher
redshifts. 

\begin{figure}
\centering
\includegraphics[width=0.6\linewidth,angle=270]{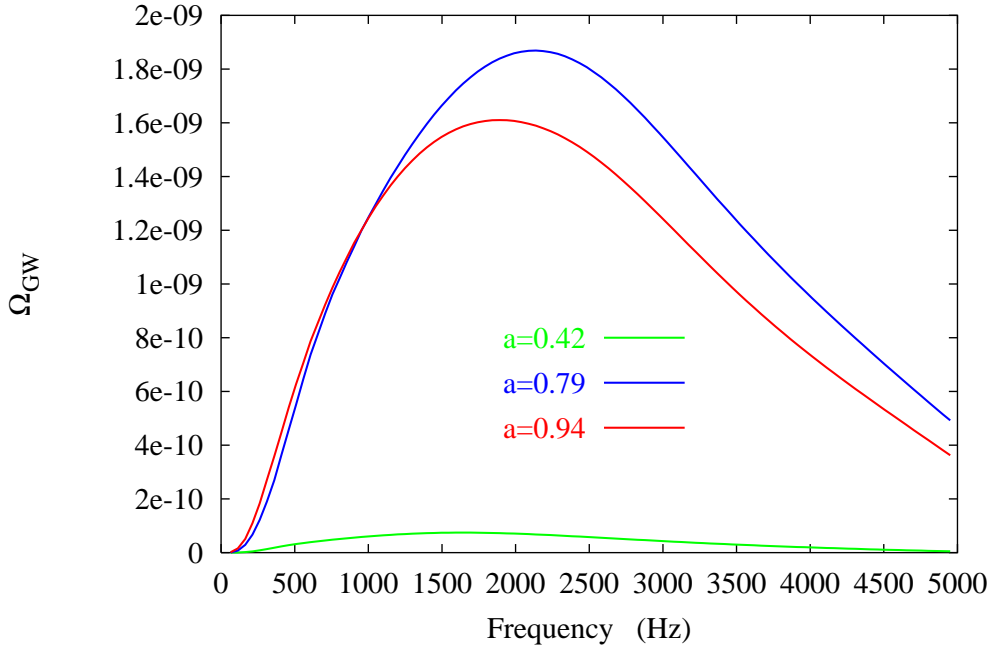}
\caption{$\ogw$ against frequency, for GWs emitted in the supernovae
  collapse to black holes, from ref.~\cite{FMSa}.}
\label{fmsfig1}
\end{figure}

The value of $\hogw$ has been computed in ref.~\cite{FMSa} and is of order
$10^{-10}$, peaked around 2 kHz.  It is shown in fig.~\ref{fmsfig1}
(provided by the authors of ref.~\cite{FMSa}, and including a recent 
update of the star formation rate at large $z$) 
for three different values of the rotation parameter
$a\equiv Jc/(GM^2\sub{core})$, where $J$ is the angular momentum.  A
black hole forms only if $a$ is smaller than a critical value
estimated in the range $a\sub{crit}\sim 0.8-1.2$; otherwise rotational
energy dominates, the star bounces, and no collapse occurs. For
comparison with the other plots of $\hogw$ shown in this report, note
that fig.~\ref{fmsfig1} shows $\ogw$ rather than $\hogw$, and that
ref.~\cite{FMSa} uses $h_0=0.5$, so the value of $\hogw$ is obtained
from fig.~\ref{fmsfig1} dividing by 4. Note also that both axes are on
a linear scale, rather than a log-log scale, as the other plots that
we show. 

From the figure, we see that
this background  is out of reach for
first generation experiments, while it could  be detectable with
advanced ground based interferometers. 

It is also important to observe that the resulting GW signal is
insensitive to  the uncertainties that are present in the star
formation rate in the high
redshift region $z\sim 4-5$, since the result turns out to be
dominated by low-to-intermediate redshift sources.

\subsection{GWs from hydrodynamic waves in  rotating neutron
  stars}\label{rmode} 

In rapidly rotating neutron stars
there is a whole class of instabilities driven  by the emission of
GWs, called CFS instabilities after Chandrasekhar~\cite{Cha}, who
discovered them, and Friedman and Schutz~\cite{FS}, who showed that
they are generic to rotating neutron stars. Basically, they form when
there is a mode that is forward-going, as seen from a distant
observer, but backward going with respect to the rotation of the
star. In this case, when this mode radiates away angular momentum, the
star can find a rotation state with lower energy, than fueling further
growth of the mode and therefore an instability occurs. Viscous forces,
instead, tend to damp this instability.

Recent studies of rapidly rotating relativistic stars have revealed
the existence of a particularly intersting
class of modes (r-modes) that are unstable due to
the emission of GWs~\cite{And,FM,LOM}. As the star spins down, 
for these modes an
energy of order of 1\% of a solar mass is emitted in GWs, making the
process very interesting for GW detection~\cite{OLCSVA,AKS}.
Ref.~\cite{OLCSVA} give a rough estimate of the spectrum assuming a
comoving number density of neutron star births constant in the range
$0<z<4$, and zero at higher $z$. A more detailed analysis has been
done in ref.~\cite{FMSb,FMSc}, using  again the star formation rate
determined by observation, mentioned in the previous section. 
The analysis is performed for three different cosmological models, a
flat Universe with vanishing cosmological constant
($\Omega_M=1,\Omega_{\Lambda}=0$) and $h_0=0.5$,
a flat, low-density
model ($\Omega_M=0.3,\Omega_{\Lambda}=0.7, h_0=0.6$),
and an open model
with $\Omega_M=0.4,\Omega_{\Lambda}=0, h_0=0.6$. Similarly to the
computation of $R\sub{NS}$ in the previous section, one can now
compute the number of neutron stars formed per unit time within a
comoving volume out at redshift $z$. Depending on the cosmological
model, and on the  upper cutoff in the progenitor mass, it turns out that
\be
R\sub{NS}\simeq (15-30)\,{\rm events/s}\, .
\ee
Comparing with the collapse to black hole, with the same parameters,
one finds that $R\sub{NS}\sim 4R\sub{BH}$.

\begin{figure}
\centering
\includegraphics[width=0.6\linewidth,angle=270]{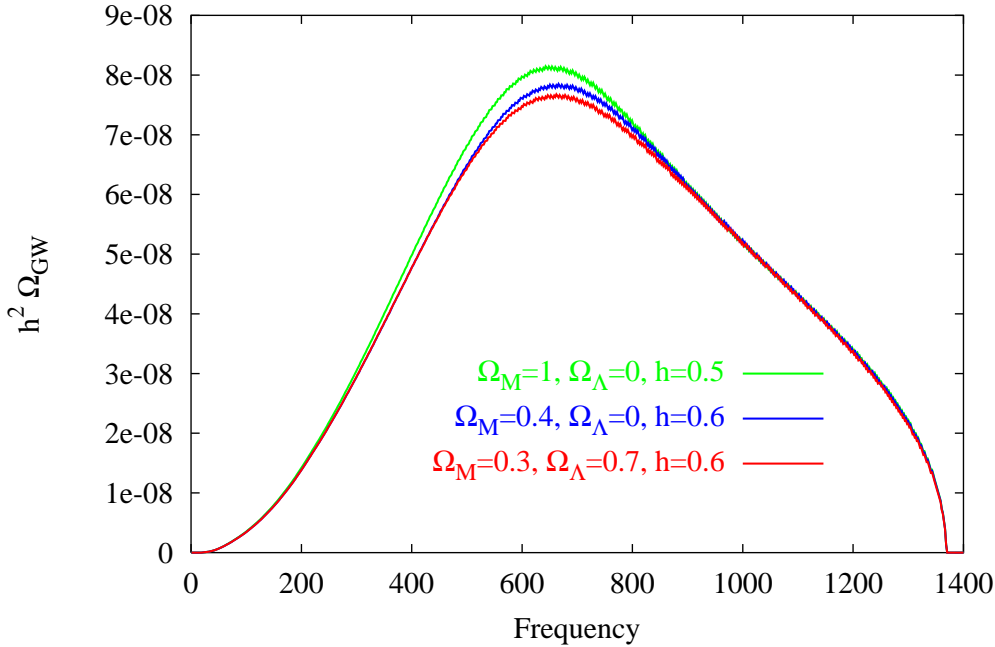}
\caption{$\hogw$ against frequency, for GWs produced by the r-mode of
rotating  neutron stars (from ref.~\cite{FMSb}).}
\label{fmsfig2}
\end{figure}

The mechanism of GW emission is discussed in 
refs.~\cite{LOM,OLCSVA,AKS,AKS2}.  The neutron star is modeled
as a fluid with a polytropic equation of state $p=k\rho^2$, with $k$
chosen so that the mass of the neutron star is $1.4 M_{\odot}$ and the
radius is $R=12.53$ km. 
The r-mode excitation starts as a
small perturbation of the velocity field. As long as 
the amplitude $\alpha$ of the r-mode us small,
the effect of the instability is such that $\alpha$ grows exponentially
while the angular velocity $\Omega$ of the neutron star is nearly
constant. After approximately 500s, $\alpha$ becomes of order one, and
we  reach a  regime where non-linear hydrodynamic effects are
important. The non-linear effects saturate the growth of $\alpha$,
which stays approximately constant. This phase lasts  approximately
for a time
$\bar{\tau}\sub{NS}\sim$ 1~yr, 
during which the star loses approximately 2/3 of its rotational
energy by emission of GWs. Finally, the angular velocity becomes
sufficiently low so that the r-mode ceases to be unstable. 

Since GWs are emitted continuously over approximately 1yr, the duty
cycle for this process is~\cite{FMSb}
\be
D=\int_0^{\infty}dR\sub{NS}\,{\bar{\tau}}\sub{NS}(1+z)\sim 10^9\, ,
\ee
and therefore we definitely have a stochastic background. 

The minimum and maximum values of the frequency radiated are estimated
as follows~\cite{OLCSVA}. 
It is known that the maximum possible angular velocity of a
rotating star is
\be
\Omega_K\sim 7.8\times 10^3\, {\rm Hz} \[
\frac{M}{M_{\odot}}\( \frac{10\,{\rm km}}{R}\)^3\]^{1/2}\, ,
\ee
where $M,R$ are the mass and radius of the corresponding non-rotating
star. Assuming that most neutron stars are born with $\Omega$ close to
$\Omega_K$, one can write the initial rotational energy of the
star. Using the fact that about 2/3 of it is radiated into GWs, 
one finds
\be
f\sub{max}\simeq \frac{2}{3\pi}\Omega_K\, .
\ee
The lower cutoff in $f$ is associated to the fact that the r-mode
instability becomes ineffective below some critical rotational
frequency, and it can be estimated that
\be
f\sub{max}\simeq 120\, {\rm Hz}\, .
\ee
The resulting spectrum of GWs is shown in fig.~\ref{fmsfig2}, 
(provided by the authors of ref.~\cite{FMSb}, and including a recent 
update of the star formation rate at large $z$ and other technical
improvements) 
for the three
different cosmological models defined above. We see that
the three curves are very similar, and therefore the result
is quite solid against changes in the cosmological model, 
and~\cite{FMSb}
\be
\hogw\simeq 8\times 10^{-8}
\ee
in the frequency range (500-700) Hz.
This value is quite interesting, but from the discussion in
sect.~\ref{5} it is clear that it is still below the sensitivities that can
be obtained with correlations  of first
generation interferometers. It will be instead quite accessible to
second generation experiments. 

\subsection{GWs from mass multipoles of  rotating neutron stars}

In the previous subsection we examined the case where GWs are emitted
by rotating neutron stars through their coupling to the current
multipoles associated to hydrodynamic waves on the star surface.
We now discuss the emission by the simplest mechanism, which is the
coupling of the GWs to the multipoles of the mass distribution of a
rotating neutron star.

The stochastic background from mass multipoles of
rotating neutron stars has been
discussed in ref.~\cite{Pos}. The main uncertainty 
comes from the estimate of the typical ellipticity $\epsilon$ of the
neutron star, which measures its deviation from sphericity. 
An upper bound on $\epsilon$ can be obtained assuming
that the observed slowing down of the period of known pulsars
  is interely due to
the emission of gravitational radiation. This is almost certainly a
gross overestimate, since most of the spin down is probably due to
electromagnetic losses, at least for Crab-like pulsars. 

The typical  frequency  for these GWs can be in the range of
ground-based interferometers.
With realistic estimates for $\epsilon$, ref.~\cite{Pos} gives, at
$f=100$ Hz, a value of $h_c(f)\sim 5\times 10^{-28}$, 
that, using eq.~(\ref{rho5}), corresponds to 
\be
\hogw (100 {\rm Hz}) \sim 10^{-15}\, . 
\ee
This is very far from the sensitivity of even the
advanced experiments. An absolute upper bound can be obtained assuming
that the spin down is due only to gravitational losses, and this gives
$\hogw (100 {\rm Hz}) \sim 10^{-7}$, but again this value is probably a
gross overestimate.

Techniques for the detection of this background with a single
interferometer using the fact that it is not isotropic and exploiting
the sideral modulation of the signal have been
discussed in ref.~\cite{GBG}.

\subsection{Unresolved galactic and extragalactic binaries}

These backgrounds will be relevant 
in the LISA frequency 
band.
For frequencies below a few mHz, one expects a stochastic
background due to a large number of galactic
white dwarf binaries~\cite{EIS,BenH,Tho2,Schu,Pos,LISA}. 

In sections~(\ref{rmode},\ref{core}),
in order to determine whether the superposition of burst signals from
supernovae formed a stochastic background, we used the duty cycle. For
a continuous signal as that produced by binaries, instead, whether we
have a stochastic background or not depends on our frequency
resolution.  After a
time $T$ of observation, we can resolve $\Delta f=1/T$; for $T=$ 1~yr, 
$\Delta f\sim 3\times 10^{-8}$ Hz, and the number of resolvable
frequencies in the LISA frequency band is of order 
$(1\, {\rm Hz})/(3\times 10^{-8}\, {\rm Hz})=3\times 10^7$.
It is
estimated that most frequency bins below a critical frequency of the order
of 1 mHz will contains signal from more than one galactic binary, and
will therefore form a confusion-limited background. 
Above this critical frequency, instead, individual signals will be
resolved.

In particular, it is expected a contribution from compact white dwarf
binaries. 
The computation of the intensity
of this background depends on the rate of white dwarf mergers, which
is uncertain. A possible estimate is shown in
fig.~\ref{lisa}~\cite{LISA}. We see that, below a few mHz, it would
cover a stochastic background of cosmological origin at the level
$\hogw\sim 10^{-11}$.

It should be observed that, even if an astrophysical background is
present, and masks a relic background, not all hopes 
of observing the cosmological signal are lost. If we
understand well enough the astrophysical background, we can subtract
it, and  the relic background would still be observable if it is much larger
than the uncertainty that we have on the 
subtraction of the astrophysical background.  In
fact, LISA should be able to subtract the background due to white
dwarf binaries,
since there is a large number of binaries close enough to be
individually resolvable~\cite{LISA}. 
This should allow to predict with some
accuracy the space density of white dwarf binaries in other parts of
the Galaxy, and therefore to compute the stochastic background that
they produce. Furthermore, any background of galactic origin is likely
to be concentrated near the galactic plane, and this is another handle
for its identification and subtraction. The situation is more
uncertain for  the contribution of extragalactic  binaries, which
again can be relevant at LISA frequencies.
The uncertainty  in the merging rate is such that it cannot be
predicted reliably, but it is believed to be lower than the galactic
background~\cite{Pos}. In this case the only handle for the
subtraction  would be the form of the spectrum. In fact,
even if the strength is quite uncertain, the form of the spectrum may
be known quite well~\cite{LISA}. 

Finally,  another possible astrophysical background
can be expected if MACHO's are black holes of mass $\sim
0.5M_{\odot}$~\cite{NSTT}. In this case we can expect that a large
number of them will be in binary systems, and they could create a
stochastic background relevant both for LISA and for ground-based
interferometers~\cite{ITN}.

\section{Conclusions}
Present GW experiments have not been designed especially for the
detection of  GW backgrounds of cosmological origin. Nevertheless,
there are chances that in their frequency window there  might be a
cosmological signal. The most naive estimate
of the frequency range for signals from the
very early Universe singles out the
GHz region, very far from the region accessible to 
ground based interferometers,
$f<$ a few kHz, or to resonant masses. 
To have a signal in the accessible region, one of these two
conditons should be met: either we find a spectrum with
a long low-frequency tail, that extends from the GHz down to the kHz
region, or we have some explosive production mechanism much below the
Planck scale. As we have discussed, both situations seem to be not at
all unusual, at least in the examples that have been worked out to
date. The crucial point is
 the value of the intensity of the background.

With a very optimistic attitude, one could hope for a signal,
with the maximum  intensity
compatible with the nucleosynthesis bound,  $\hogw\sim$ a few $\times
10^{-6}$ (or even $10^{-5}$, stretching all parameters to the maximum
limit). 
Such an option is not excluded, and the fact that such a
background is not predicted by the mechanisms that have been investigated
to date is probably  not a very strong
objection, given our theoretical ignorance of physics at the Planck scale
and the rate at which new production mechanisms have been
proposed in recent years, see the reference list. 
However, with more realistic estimates, on general grounds
it appears difficult to  predict  a
background that, either  in the  region between a few Hz and a few
kHz, relevant for 
ground based interferometers, or in the region $(10^{-4}-1)$ Hz
relevant to LISA, exceeds the level 
$\hogw\sim$ a few $\times 10^{-7}$, independently of the production
mechanism. This should be considered the minimum detection level 
where a significant search can start. Such a level is beyond the
sensitivity of first generation ground based
detectors, but it is accessible both to advanced interferometers and
to the space interferometer LISA.

The difficulties of such a detection are clear, but 
 the payoff of a positive result would be enormous,
opening up a window in the Universe and in fundamental high-energy
physics that will never be reached with particle physics experiments.

\vspace*{5mm}

{\bf Acknowledgments.}
I am  grateful to Adalberto Giazotto for many interesting
discussions and  questions, which stimulated me to write down
this report.
I  thank for useful discussions or comments on the
manuscript 
Danilo Babusci,  Stefano Braccini, Maura Brunetti, Alessandra Buonanno, 
Massimo Cerdonio, Eugenio Coccia, Viviana Fafone, Valeria Ferrari,
Stefano Foffa, Maurizio Gasperini,  Alberto Nicolis, 
Emilio Picasso, Raffaella Schneider, Riccardo Sturani, Carlo Ungarelli,
Gabriele Veneziano and  Andrea Vicer\`e.

I thank Barry Barish and Alberto Lazzarini for providing  
figs.~\ref{4km_dhs} and \ref{ligoII},  Karsten Danzmann and  Roland
Schilling for providing figs.~(\ref{lisa},\ref{lisa2}).
I  thank Danilo Babusci for computing the overlap reduction
functions and producing figs.~(\ref{dad1},\ref{dad2}),
Viviana Fafone for providing fig.~\ref{10_8_98}, and
Raffaella Schneider and Valeria Ferrari for
figs.~(\ref{fmsfig1},\ref{fmsfig2}).

I am grateful to Carl Gwinn for useful comments on sect.~7.4.

\end{document}